\documentclass[%
 reprint,
superscriptaddress,
preprintnumbers,
nofootinbib,
 amsmath,amssymb,
 aps,
 prd,
 floatfix,
]{revtex4-2}

\usepackage{graphicx}

\usepackage{siunitx}

\usepackage{bm}
\usepackage{braket}
\usepackage{mathtools} 
\usepackage{slashed}
\usepackage{cancel} 
\usepackage{enumitem} 
\usepackage{placeins} 

\usepackage[pdftex]{color}
\usepackage{hyperref}
\hypersetup{
    colorlinks=true,     
    linkcolor=blue,      
    citecolor=blue,      
    filecolor=blue,      
    urlcolor=blue        
}

\usepackage{cleveref}
\crefname{equation}{Eq.}{Eqs.}
\crefname{figure}{Fig.}{Figs.}
\crefname{table}{Tab.}{Tabs.}
\crefname{section}{Sec.}{Secs.}
\crefname{appendix}{App.}{Apps.}

\renewcommand{\vec}[1]{{\mathbf{#1}}} 
\newcommand{\Tr}{{\mathrm{Tr}}}

\begin{document}

\preprint{FERMILAB-PUB-24-0407-T}

\title{Lanczos algorithm for lattice QCD matrix elements}

\author{Daniel C. Hackett}
 \affiliation{Fermi National Accelerator Laboratory, Batavia, IL 60510, U.S.A.}
\author{Michael L. Wagman}
 \affiliation{Fermi National Accelerator Laboratory, Batavia, IL 60510, U.S.A.}

\date{\today}

\begin{abstract}    
    Recent work~\cite{Wagman:2024rid} found that an analysis formalism based on the Lanczos algorithm allows energy levels to be extracted from Euclidean correlation functions with faster ground-state convergence than effective masses, convergent estimators for multiple states from a single correlator, and two-sided error bounds.
    After filtering out spurious eigenvalues and using outlier-robust estimators within a nested bootstrap framework, Lanczos estimators behave more like multi-state fit results than effective masses---but without involving statistical fitting.
    We extend this formalism to the determination of matrix elements from three-point correlation functions and provide a physical picture of ``spurious state filtering'' involving restriction to a Hermitian subspace.
    We demonstrate similar advantages for matrix elements as for spectroscopy through example applications to noiseless mock-data and (bare) forward matrix elements of the strange scalar current between both ground and excited states with the quantum numbers of the nucleon.
\end{abstract}

\maketitle

\section{Introduction}
\label{sec:intro}

By stochastically evaluating discretized path integrals, numerical lattice QCD calculations provide a first-principles approach to studying the dynamics of the strong force~\cite{Detmold:2019ghl, USQCD:2022mmc, Boyle:2022ncb, Achenbach:2023pba}, including hadron spectroscopy and scattering amplitudes~\cite{Briceno:2017max,Bulava:2022ovd,Hanlon:2024fjd} and various aspects of hadron structure~\cite{Cichy:2018mum,Ji:2020ect,Constantinou:2022yye,Gao:2024pia}.
Due to the stochastic nature of these calculations, statistical analysis of noisy Monte Carlo ``data'' is necessary.
Although there are well-established analysis methods, developing improved techniques is still an active topic of research~\cite{Fleming:2004hs,Beane:2009kya,Detmold:2014hla,Wagman:2016bam,Detmold:2018eqd,Fischer:2020bgv,Yunus:2022wuc}.
Better analysis tools can improve both statistical precision by alleviating signal-to-noise issues, as well as accuracy by providing more robust control of systematic uncertainties.

Lattice QCD calculations often involve the extraction of hadronic matrix elements from simultaneous analysis of hadronic two- and three-point correlation functions (correlators).
This analysis task underlies the calculation of many different quantities of physical interest including form factors, parton distribution functions (PDFs), and generalizations thereof like generalized parton distributions (GPDs) and transverse-momentum distribution PDFs (TMDs)~\cite{Cichy:2018mum,Ji:2020ect,Constantinou:2022yye,Gao:2024pia}.
However, presently standard methods may produce unreliable results~\cite{Jang:2019vkm,Gupta:2021ahb,Gupta:2024krt} due to a combination of excited-state contamination (ESC)~\cite{Gupta:2017dwj,RQCD:2019jai,Jang:2023zts,Gupta:2023cvo,Barca:2024sub} and exponentially decaying signal-to-noise ratios (SNR)~\cite{Parisi:1983ae,Lepage:1989hd}.
Analysis techniques which address these issues are a topic of active research~\cite{Jang:2019vkm,Gupta:2021ahb,Gupta:2024krt}, but obtaining full control over all sources of uncertainty remains challenging for many quantities of physical interest.

Spectroscopy---the extraction of finite-volume energy levels from analysis of hadronic two-point correlation functions---is hindered by the same issues as matrix-element extractions, i.e.~ESC and decaying SNR,
and methods to improve these issues can be useful in both contexts.
There has been extensive work to develop improved spectroscopy methods less susceptible to ESC and which offer bounds on systematic uncertainties---notably 
approaches based on generalized eigenvalue problems (GEVP) that provide one-sided variational bounds on energy level systematic uncertainties~\cite{Fox:1981xz,Michael:1982gb,Luscher:1990ck,Blossier:2009kd,Fleming:2023zml}, which have already been adapted to matrix-element calculations~\cite{Blossier:2009kd,Aubin:2011zz,HadStruc:2021wmh,Barca:2024sub,Bulava:2011yz,Barca:2022uhi,RBC:2023ynh,Green:2019zhh,Stokes:2018emx,Dragos:2016rtx,Shultz:2015pfa,Green:2014xba,Green:2013hja,Bulava:2011yz,Bennett:2024wda,Braun:2015axa,Bussone:2023kag}.
Recent work has shown that a novel formalism based on the Lanczos algorithm~\cite{Lanczos:1950zz} can provide qualitative and quantitative improvements for spectroscopy including faster convergence and two-sided bounds on systematic errors, even when ESC is large.
After filtering out spurious eigenvalues~\cite{Cullum:1981,Cullum:1985} and applying outlier-robust bootstrap-median estimators in a nested bootstrap framework, Lanczos results are more analogous to multi-state fit results than effective masses and exhibit 
stability of statistical estimates for large numbers of iterations~\cite{Wagman:2024rid}.

It is appealing that the uncertainties of Lanczos results---just like those of multi-state fits---are not increased by including additional noisy large-time data points; however, it is worth emphasizing that Lanczos analyses cannot extract more useful information from these large-time points than other methods. 
Large correlations appear between Lanczos results that differ only in how many noisy large-time data points are included~\cite{Wagman:2024rid,Ostmeyer:2024qgu,Chakraborty:2024scw,Hackett:2024nbe}, and therefore one cannot achieve significantly higher precision than correlator fits by attempting to fit large-time Lanczos results.
The stability of large-time Lanczos results does not mean that Lanczos methods avoid the problem that correlator SNR decays exponentially. Instead, it should be taken as an indication that once sufficiently large-time points are included all available statistical information has been extracted and Lanczos spectral estimates are unchanged by including additional large-time data points.
Further, this stability only indicates that the information available at finite statistics is exhausted and does not formally guarantee that results have converged to their infinite-statistics values within uncertainties; the application of Lanczos to noisy data can only strictly be interpreted within an oblique Lanczos framework in which periods of ``stagnation'' with slow convergence for many iterations are possible~\cite{Leyk:1997,Gaaf:2016,Saad:2011}.
The convergence of oblique Lanczos is the presence of statistical noise is being actively investigated~\cite{Abbott:2025yhm}.

In this work, we present a new approach to matrix element analyses using a simple extension of this Lanczos formalism.
The essence of the idea is to evaluate operator matrix elements between the Ritz vectors, the eigenstate approximations found by the Lanczos procedure of Ref.~\cite{Wagman:2024rid}.
As derived below, after constructing initial and final-state Ritz coefficients $P$ and $P'$ from the corresponding two-point functions $C(t)$ and $C'(t)$, extracting matrix elements from an arbitrary three-point function amounts to matrix multiplication, taking the form\footnote{Note that ground-state overlap factors appearing in the spectral representation of the three-point function are canceled by the Ritz coefficients and do not need to be explicitly included in this formula; see \cref{eq:lanczos:mx-elts} for more details.}
\begin{equation}
    \braket{f'|J|i} \approx
    \sum_{\sigma \tau} P^{\prime*}_{f \sigma} \frac{C^\mathrm{3pt}(\sigma, \tau)}{\sqrt{C'(0) C(0)}} P_{i \tau},
    \label{eq:intro:mx_elt}
\end{equation}
where $\tau$ is the source-operator separation and $\sigma$ is the operator-sink separation.
As explored below, this provides several important advantages over the previous state of the art, namely simplicity, direct and explicit computation of excited-state and transition matrix elements, and avoidance of inductive biases unavoidably introduced by implicit methods involving statistical modeling and fits of correlation functions.
The method involves only a few {analysis} hyperparameters associated with eigenstate identification and none specific to the operator or three-point function.
Furthermore, we find that the new method is dramatically less susceptible to yielding deceptive results when applied to three-point correlators with large ESC.
The data required is the same as for presently standard analyses, with the caveat that sparsely evaluated three-point functions limit the number of Lanczos iterations that can be evaluated.

The remainder of this paper proceeds as follows.
\Cref{sec:review} defines the analysis task, and reviews both the transfer matrix formalism necessary to understand the Lanczos approach as well as previously available methods used for comparison.
\Cref{sec:lanczos} derives the method.
\Cref{sec:noiseless} applies the method to a noiseless mock-data example and compares with previous approaches, demonstrating its improved convergence properties.
\Cref{sec:lattice} discusses how the method must be adapted in the presence of statistical noise and presents a calculation of bare matrix elements of the strange scalar current for the low-lying states in the nucleon spectrum using lattice data.
\Cref{sec:attack} subjects both the summation method and Lanczos to adversarial attacks; the results of these experiments suggest that Lanczos estimates are qualitatively more robust against excited-state contamination.
Finally, \cref{sec:conclusions} concludes and discusses opportunities for future work.

Note that throughout this paper, repeated indices do not imply summation and all sums are written explicitly.
We also use lattice units to simplify the notation, setting the lattice spacing $a=1$ throughout.
In these units, physical quantities like energies and matrix elements are dimensionless, and Euclidean times $t$ take on integer values such that they may be used interchangeably as arguments and indices.

\section{Background}
\label{sec:review}

We are interested in computing hadronic matrix elements of some operator $J$, i.e.,
\begin{equation}
    J_{fi} = \braket{f'|J|i},
\end{equation}
where $\ket{i}$ and $\ket{f'}$ are energy eigenstates with the quantum numbers of the initial (unprimed) and final (primed) state, which may in general be different.
For example, their momenta will differ for off-forward matrix elements, i.e.~when $J$ carries some nonzero momentum.
Hadronic matrix elements are not directly calculable using numerical lattice methods, and instead are typically extracted from simultaneous analysis of two- and three-point correlators.\footnote{The Feynman-Hellmann theorem and generalizations thereof provides a distinct approach; see~\cite{QCDSF:2012mkm, Batelaan:2023jqp, QCDSFUKQCDCSSM:2022ncb, Hannaford-Gunn:2022lez} for examples.}
The ones relevant to the calculation are, in Heisenberg picture,\footnote{Suppressed lattice spatial indices on each operator are assumed to be absorbed into these quantum numbers, e.g.~by projection to a definite momentum.}
\begin{equation}\begin{aligned}
    C(t) &= \braket{\psi(t) \, \overline{\psi}(0)}, \\
    C'(t) &= \braket{\psi'(t) \, \overline{\psi'}(0)}, \\
    C^{\mathrm{3pt}}(\sigma, \tau) &= \braket{ \psi'(\tau+\sigma) \, J(\tau) \, \overline{\psi}(0) } 
\end{aligned}
\label{eq:heis-corrs}
\end{equation}
where $\psi$ and $\psi'$ are interpolating operators (interpolators) with initial- and final-state quantum numbers.
Note that the arguments of $C^{\mathrm{3pt}}$ are more often defined with a different convention, with the sink time $t_f = \sigma + \tau$ as the first argument.
While not explicitly notated, we generally assume definitions such that all vacuum contributions are subtracted out.
These expectations may be evaluated stochastically by lattice Monte Carlo methods.

\subsection{Transfer matrices \& spectral expansions}

To see how these data constrain the matrix element of interest, we use the Schr\"odinger-picture transfer-matrix formalism~\cite{Kogut:1974ag} to derive their spectral expansions.
This exercise also serves to establish notation and as a review of this formalism, used throughout this work.

We begin with the assumption\footnote{This holds only approximately for many lattice actions in standard use~\cite{Luscher:1976ms,Luscher:1984is}.} that Euclidean time evolution can be described by iterative application of the transfer matrix $T = e^{-H}$, such that 
\begin{equation}
    T \ket{n} = \lambda_n \ket{n} \equiv e^{-E_n} \ket{n},
\end{equation}
where $\ket{n}$ are unit-normalized energy eigenstates such that $\braket{n|m} = \delta_{nm}$, $\lambda_n$ are transfer-matrix eigenvalues, and ${E_n = -\log \lambda_n}$ are the energies.
Note that states $\ket{\cdot}$ and operators including $J$ and $T$ are not directly accessible, but rather formal objects that live in the infinite-dimensional Hilbert space of states.
Applied to the vacuum state $\ket{\Omega}$, the adjoint interpolating operator $\overline{\psi}$ excites the state $\ket{\psi}$, which may be decomposed as
\begin{equation}
    \overline{\psi} \ket{\Omega} 
    \equiv \ket{\psi}
    = \sum_n \braket{n|\psi} \ket{n}
    \equiv \sum_n Z_n \ket{n},
\end{equation}
where $Z_n = \braket{n|\psi}$ are the overlap factors. 
The sum may be assumed to be restricted to eigenstates with the quantum numbers of $\overline{\psi}$, as $Z_n=0$ otherwise.
The transfer matrix acts nontrivially on $\ket{\psi}$, as
\begin{equation}
    T^t \ket{\psi} = \sum_n Z_n e^{-E_n t} \ket{n} ~.
\end{equation}
Under such Euclidean time evolution, the amplitudes of higher-energy eigenstates decay more quickly.
Taking $t \rightarrow \infty$, the ground eigenstate $\ket{0}$ dominates.
Importantly, this amounts to application of the power-iteration algorithm~\cite{VonMises:1929}, as discussed further below.

For the initial-state correlator $C(t)$ from \cref{eq:heis-corrs}, translating to Schr\"odinger picture using $O(t) = T^{-t} O T^{t}$ gives
\begin{equation}\begin{aligned}
    C(t) 
    &= \braket{\Omega | T^{-t} \psi T^t T^0 \overline{\psi} T^0 | \Omega} \\
    &= \braket{\psi | T^t |\psi} \\
    &= \sum_i |Z_i|^2 e^{-E_i t},
    \label{eq:Ci-spectral-decomp}
\end{aligned}\end{equation}
using $\bra{\Omega} T = \bra{\Omega}$ in the second equality.
An analogous expression holds for the final-state correlator,
\begin{equation}
    C'(t) = \sum_f |Z'_f|^2 e^{-E'_f t},
    \label{eq:Cf-spectral-decomp}
\end{equation}
where $Z'_f = \braket{f'|\psi'}$.
Similar manipulations produce the spectral expansion of the three-point correlator,
\begin{equation}\begin{aligned}
    C^\mathrm{3pt}(\sigma, \tau)
    &= \braket{\psi' | T^\sigma J T^\tau |\psi } \\
    &= \sum_{fi} Z^{\prime *}_f Z_i J_{fi} \, e^{-E'_f \sigma - E_i \tau} ~.
    \label{eq:C3pt-spectral-decomp}
\end{aligned}\end{equation}
Note that these definitions assume zero temperature.\footnote{Thermal effects are handled automatically in applications of Lanczos to two-point correlators as discussed in Sec.~A of Ref.~\cite{Wagman:2024rid}'s Supplemental Material.
The resulting isolation of thermal states automatically removes their effects from all matrix element results.
}
Comparing \cref{eq:Ci-spectral-decomp,eq:Cf-spectral-decomp,eq:C3pt-spectral-decomp}, we see that $C$ and $C'$ carry the necessary information to isolate $J_{fi}$ in $C^\mathrm{3pt}$.

A standard approach to extracting $J_{fi}$ is to use statistical inference, i.e.~simultaneously fitting the parameters $Z_i$, $Z'_f$, $E_i$, $E'_f$, and $J_{fi}$ of a truncated spectral expansion to \cref{eq:Ci-spectral-decomp}, \eqref{eq:Cf-spectral-decomp}, and \eqref{eq:C3pt-spectral-decomp}.
The resulting estimates converge to the underlying values given sufficiently high statistics, large Euclidean times, and states included in the models.
This has been used to much success, but has certain serious disadvantages.
Specifically, the black-box nature of statistical inference and large number of hyperparameters to vary (e.g.~number of states modeled, data subset included, choice of selection/averaging scheme) induce systematic uncertainties which can only be assessed with caution and experience.

\subsection{Power iteration \& other explicit methods}

In this work, we restrict our consideration to explicit methods that do not involve statistical modeling.
Familiar examples include effective masses\footnote{And generalizations thereof including GEVPs~\cite{Fox:1981xz,Michael:1982gb,Luscher:1990ck,Blossier:2009kd,Fleming:2023zml} and Prony's method~\cite{Fleming:2004hs,Beane:2009kya,Fischer:2020bgv}.} and ratios of correlation functions.
Here, we present these as examples of the power iteration algorithm, thereby motivating the use of Lanczos in \cref{sec:lanczos}.
We also review the summation method~\cite{Maiani:1987by,Dong:1997xr,Capitani:2012gj}, for later comparison with Lanczos results.

As already noted, using Euclidean time evolution to remove excited states may be thought of as applying the power iteration algorithm~\cite{VonMises:1929} to extract the ground eigenstate of the transfer matrix~\cite{Wagman:2024rid}.
With power iteration, increasingly high-quality approximations of the ground eigenstate, $\ket{b^{(m)}} \approx \ket{0}$, are obtained using the recursion
\begin{equation}
    \ket{b^{(m)}} 
    = \frac{ T \ket{b^{(m-1)}} }{ || T \ket{b^{(m-1)}} || }
    = \frac{ T \ket{b^{(m-1)}} }{ \sqrt{ \braket{ b^{(m-1)} | T^2 | b^{(m-1)}} } },
    \label{eq:power-iter-recursion}
\end{equation}
starting from
\begin{equation}
    \ket{b^{(0)}} 
    \equiv \frac{ \ket{\psi} }{ |\psi| } 
    \equiv \frac{ \ket{\psi} }{ \sqrt{\braket{\psi|\psi}} } 
    = \frac{ \ket{\psi} }{ \sqrt{C(0)} } ~.
\end{equation}
The resulting approximate eigenstates $\ket{b^{(m)}}$ can be used to compute ground-state matrix elements of different operators.
For example, using them to extract the ground-state eigenvalue of $T$ yields the usual effective energy
\begin{equation}\begin{aligned}
    E^\mathrm{eff}(2m) 
    &= - \log \braket{ b^{(m)} | T | b^{(m)} } \\
    &= - \log \frac{ \braket{\psi | T^{2m+1} | \psi} }{ \braket{\psi | T^{2m} | \psi} } \\
    &= - \log \frac{ C(2m+1) }{ C(2m) },
    \label{eq:meff}
\end{aligned}\end{equation}
where the second equality fully evaluates the recursion \cref{eq:power-iter-recursion}.
Note that while this power-iteration version of $E^\mathrm{eff}$ is defined for only even arguments $2m$, we generalize and evaluate it for all $t=2m$ as usual.

More relevantly to this work, power-iteration states may also be used to compute ground-state matrix elements of some operator $J$ as
\begin{equation}\begin{aligned}
    \braket{ {b'}^{(m)} | J | b^{(m)} } 
    &= \frac{ \braket{\psi' | T^m J T^m | \psi} }
        { \sqrt{ \braket{\psi'|T^{2m}|\psi'} \braket{\psi|T^{2m}|\psi} } }
    \\ &= \frac{ C^\mathrm{3pt}(m,m) }{ \sqrt{C'(2m) \, C(2m)} } ,
\end{aligned}
\label{eq:pow-ratio}
\end{equation}
where $\ket{{b'}^{(m)}}$ are from $\ket{\psi'}$.
We thus arrive at the usual approach of constructing ratios of three-point and two-point functions to isolate the desired matrix element.
It is straightforward to insert the spectral expansions \cref{eq:Ci-spectral-decomp,eq:Cf-spectral-decomp,eq:C3pt-spectral-decomp} and verify that $\braket{ {b'}^{(m)} | J | b^{(m)} } = J_{00}$ up to excited-state effects.
Thus, in analogy to the effective energy \cref{eq:meff}, $\braket{ {b'}^{(m)} | J | b^{(m)} }$ may be thought of as an ``effective matrix element'' expected to plateau to $J_{00}$ as $m \rightarrow \infty$.
The new method presented in \cref{sec:lanczos} simply applies this same idea with the improved approximations of the eigenstates afforded by the Lanczos algorithm, with the notable difference that approximate eigenstates are available for excited states as well.

The ratio \cref{eq:pow-ratio} derived by power iteration is not the one in standard use.
For easier comparison with other works, we instead employ the standard ratio
\begin{equation}\begin{aligned}
    R(\sigma, \tau) &= 
    \frac{ C^\mathrm{3pt}(\sigma, \tau) }{ C'(\sigma+\tau) }
    \sqrt{
    \frac{ C(\sigma) }{ C'(\sigma) }
    \frac{ C'(\sigma+\tau) }{ C(\sigma+\tau) }
    \frac{ C'(\tau) }{ C(\tau) }
    }
    \\
    &= J_{00} + (\text{excited states}) ~.
    \label{eq:std-ratio}
\end{aligned}\end{equation}
When $\psi = \psi'$, this reduces to $\braket{b^{(\sigma)} | J | b^{(\tau)}}$; additionally taking $\tau = \sigma = m$ reproduces \cref{eq:pow-ratio} exactly.
However, when $\psi \neq \psi'$ the two expressions are inequivalent.
We use this standard ratio to define a power-iteration-like effective matrix element for sink time $t$ as
\begin{equation}
    J^\mathrm{PI}(t) = \begin{cases}
        R(\frac{t}{2}, \frac{t}{2} ), & t\text{ even} \\
        \frac{1}{2} \left[ R(\frac{t+1}{2}, \frac{t-1}{2}) + R(\frac{t-1}{2}, \frac{t+1}{2}) \right], & t\text{ odd},
    \end{cases}
    \label{eq:power-iter-Jeff}
\end{equation}
similar to \cref{eq:pow-ratio} for even $t$ and averaging the two equivalently contaminated points for odd $t$.
This quantity is what is referred to as ``Power iteration'' in all plots below.

Further manipulation leads to the summation method~\cite{Maiani:1987by,Dong:1997xr,Capitani:2012gj}, presently in common use, which provides an effective matrix element\footnote{This differs superficially from the typical presentation of the summation method, which prescribes fitting $\Sigma(t_f)$ to extract the part linear in $t_f$. Linear fits to $\Sigma(t_f)$ are identical to constant fits to $J^\mathrm{eff}(t_f)$ if their covariance matrices are computed consistently.} for sink time $t_f$ as
\begin{equation}\begin{aligned}
    \Sigma_{\Delta_\tau}(t_f) 
    &= \sum_{\tau=\Delta_\tau}^{t_f-\Delta_\tau} R(t_f-\tau, \tau),
    \\
    J^\mathrm{eff}_{00,\Delta_\tau}(t_f) &= \Sigma_{\Delta_\tau}(t_f+1) - \Sigma_{\Delta_\tau}(t_f)
    \\ &= J_{00} + (\text{excited states}) ,
\end{aligned}
\label{eq:summation-Jeff}
\end{equation}
for each choice of $\Delta_\tau$, the summation cut.
$J^\mathrm{eff}$ is expected to plateau to $J_{00}$ as $t_f$ increases and excited states decay away.
Increasing $\Delta_\tau$ further removes contamination, and curve collapse is expected as $\Delta_\tau$ increases.

\section{Lanczos Method}
\label{sec:lanczos}

The previous section discussed how standard lattice analysis methods may be thought of as implementing the power iteration algorithm to resolve the ground eigenstate of the transfer matrix $T$.
The Lanczos algorithm improves upon power iteration~\cite{Parlett:1982,Golub:1989,Kuczynski:1992,Parlett:1995,Meurant:2006,Saad:2011,Golub:2013} by making use of the full set of Krylov vectors $\propto T^t \ket{\psi}$ obtained by iterative application of $T$, rather than discarding all but the last.
As explored in Ref.~\cite{Wagman:2024rid}, Lanczos defines a procedure to manipulate two-point correlators to extract eigenvalues of $T$.
Here, we extend that formalism to evaluate matrix elements in the basis of transfer-matrix eigenstates.

Specifically, the method proposed here is to evaluate
\begin{equation}
    J^{(m)}_{fi} = \braket{{y'_f}^{(m)} | J | y^{(m)}_i},
\end{equation}
where $\ket{{y'_f}^{(m)}} \approx \ket{f'}$ and $\ket{y^{(m)}_i} \approx \ket{i}$ are the initial- and final-state Ritz vectors after $m$ Lanczos iterations, the Lanczos algorithm's best approximation of the corresponding eigenstates.
The steps to do so, as worked through in the subsections below, are as follows:
\begin{enumerate}[noitemsep]
    \item Apply an oblique Lanczos recursion to compute the transfer matrix in bases of Lanczos vectors with appropriate quantum numbers. Diagonalize to obtain Ritz values and the change of basis between Ritz and Lanczos vectors. (\cref{sec:lanczos:oblique-lanczos})
    \item Compute the coefficients relating Lanczos and Krylov vectors. (\cref{sec:lanczos:krylov-polys})
    \item Compute the coefficients relating Ritz and Krylov vectors. (\cref{sec:lanczos:krylov-polys})
    \item Compute overlap factors to normalize the Ritz vectors. (\cref{sec:lanczos:overlaps})
    \item Repeat the above steps on initial- and final-state two-point functions to obtain Ritz vectors with initial- and final-state quantum numbers. 
    \item Project the three-point function onto the Ritz vectors to obtain matrix elements. (\cref{sec:lanczos:matrix-elements})
    \item Identify and discard spurious states. (\cref{sec:lattice})
\end{enumerate}
We also discuss how bounds may be used to characterize Lanczos convergence in \cref{sec:lanczos:bounds}.
Statistical noise introduces additional complications---especially, the final step above---as discussed in \cref{sec:lattice}.
A detailed summary of the steps that must be explicitly carried out to execute the algorithm in the noisy case is given in \cref{app:algo}.

We note immediately that the oblique formalism used here formally constructs an approximation of the transfer matrix of the form
\begin{equation}
    T \approx \sum_k \ket{\tilde{k}^R} \tilde{\lambda}_k \bra{\tilde{k}^L},
\end{equation}
with generally complex eigenvalues and distinct left and right eigenvectors.
Meanwhile, the true underlying transfer matrix is Hermitian, i.e.,
\begin{equation}
    T = \sum_k \ket{k} \lambda_k \bra{k},
\end{equation}
with real eigenvalues $\lambda_k$ and degenerate left and right eigenvectors.
Critically, physically sensible results must respect this underlying Hermiticity.
As discussed in \cref{sec:noiseless}, when the Hermiticity of the underlying transfer matrix is manifest in the data, the Lanczos approximation of $T$ is Hermitian as well.
However, statistical fluctuations obscure this underlying Hermiticity in noisy data.
As explored in \cref{sec:lattice}, this results in a Hermitian subspace and a set of unphysical noise-artifact states which must be discarded.

\subsection{The oblique Lanczos algorithm}
\label{sec:lanczos:oblique-lanczos}

This section serves primarily as a review of Ref.~\cite{Wagman:2024rid}, especially Sec.~C of its Supplemental Material. However, the notation and some of the definitions---notably, of the Ritz vectors---have been altered here to better accommodate the matrix element problem.

Oblique Lanczos is a generalization of the standard Lanczos algorithm which uses distinct bases of right and left Lanczos vectors, $\ket{v^R_i}$ and $\bra{v^L_i}$~\cite{Saad:1982,Parlett:1985,Nachtigal:1993}.
This generalization allows application to non-Hermitian operators and, in the lattice context, off-diagonal correlators with different initial and final interpolators.
As discussed in Ref.~\cite{Wagman:2024rid}, oblique Lanczos is also formally necessary to treat noisy correlator data even with diagonal correlators, but naive complexification of standard Lanczos provides an identical procedure when applied only to extracting the spectrum.
However, the matrix element problem requires treatment with the full oblique formalism.

We caution that the left and right vectors treated by oblique Lanczos should not be confused with initial and final states. These are distinct labels, and left and right spaces must be constructed for each of the initial and final state eigensystems.

We first present oblique Lanczos in full generality, then discuss the specific cases used in this work at the end of the subsection.
In this spirit, consider different right and left starting states, $\ket{\psi}$ and $\bra{\chi}$, and a non-Hermitian transfer matrix $T \neq T^\dagger$.
The oblique Lanczos process begins from the states
\begin{equation}
    \ket{v^R_1} = \frac{ \ket{\psi} }{ \sqrt{\braket{\chi|\psi}} }
    \quad \text{and} \quad 
    \bra{v^L_1} = \frac{ \bra{\chi} }{ \sqrt{\braket{\chi|\psi}} } 
\end{equation}
defined such that $\braket{v^L_1|v^R_1} = 1$, and after $m$ iterations constructs the right and left bases of Lanczos vectors, $\ket{v^R_j}$ and $\bra{v^L_j}$ with $j = 1,\ldots,m$.
The resulting right and left bases are mutually orthonormal by construction, i.e.,
\begin{equation}
    \braket{v^L_i | v^R_j} = \delta_{ij} \, ,
\end{equation}
but $\braket{v^R_i|v^R_j} \neq \braket{v^L_i|v^L_j} \neq \delta_{ij}$ in general.
In the process, oblique Lanczos necessarily also computes the elements $\alpha_j$, $\beta_j$, and $\gamma_j$ of the tridiagonal matrix
\begin{equation}
T^{(m)}_{ij} = \braket{v^L_i | T | v^R_j} = \begin{bmatrix}
\alpha_1 &  \beta_2 &          &          &        0 \\
\gamma_2 & \alpha_2 &  \beta_3 &          &          \\
         & \gamma_3 & \alpha_3 & \ddots   &          \\
         &          &   \ddots & \ddots   &  \beta_m \\
       0 &          &          & \gamma_m & \alpha_m \\
\end{bmatrix}_{ij} ~ .
\label{eq:lanczos:Tij-explicit}
\end{equation}
Beginning with $\alpha_1 = \braket{v^L_1|T|v^R_1}$ and defining $\beta_1=\gamma_1=0$ for notational convenience, the iteration proceeds via three steps:
\begin{equation}\begin{gathered}
\begin{aligned}
\raggedleft{\texttt{Step 1}}& \\
    \ket{r^R_{i+1}} &= (T - \alpha_i) \ket{v^R_i} - \beta_i \ket{v^R_{i-1}}, \\
    \ket{r^L_{i+1}} &= (T^\dagger - \alpha_i^*) \ket{v^L_i} - \gamma_i^* \ket{v^L_{i-1}},\\\\
\raggedleft{\texttt{Step 2}}& \\
\beta_{i+1} \gamma_{i+1} &= \braket{r^L_{i+1} | r^R_{i+1}}, \\\\
\raggedleft{\texttt{Step 3}}& \\
\ket{v^R_{i+1}} &= \frac{1}{\gamma_{i+1}} \ket{r^R_{i+1}}, \\
\ket{v^L_{i+1}} &= \frac{1}{\beta_{i+1}^*} \ket{r^L_{i+1}}, \\
\alpha_{i+1} &= \braket{v^L_{i+1} | T | v^R_{i+1}} ~.
\end{aligned}
\end{gathered}
\label{eq:oblique-lanczos}
\end{equation}
How precisely $\beta_j$ and $\gamma_j$ are determined from the product $\beta_{j} \gamma_{j}$ computed in step two is a matter of convention; any choice satisfying\footnote{Recall that repeated indices are not summed over in this work.} 
\begin{equation}
     \gamma_j = \frac{ \braket{r^L_j | r^R_j} }{ \beta_j }
\end{equation}
is correct.
It will be helpful to consider the symmetric convention
$\beta_j \equiv \gamma_j \equiv \sqrt{ \braket{r^L_j | r^R_j} }$
corresponding to naive complexification of the standard Lanczos algorithm.
We have in some cases observed improved numerical behavior with a different convention, ${\beta_j = \left| \sqrt{\braket{r^L_j | r^R_j}} \right|}$~\cite{Wagman:2024rid}.
All physical quantities computed with this formalism are invariant to the choice of oblique convention, although the Lanczos vectors are not.

The Lanczos approximation of the transfer matrix is
\begin{equation}
    T^{(m)} \equiv \sum_{i,j=1}^m \ket{v^R_i} \bra{v^L_i} T \ket{v^R_j}\bra{v^L_j} = \sum_{i,j=1}^m \ket{v^R_i} T^{(m)}_{ij} \bra{v^L_j} ~ .
    \label{eq:Tm-def}
\end{equation}
For any $t < m$ it exactly replicates the action of the transfer matrix on the starting vectors (see \cref{app:projectors}):
\begin{equation}
    T^t \ket{v^R_1} = [T^{(m)}]^t \ket{v^R_1}, 
     \quad \bra{v^L_1} T^t = \bra{v^L_1} [T^{(m)}]^t ~.
    \label{eq:T_eq_Tm}
\end{equation}
When applied to $T$ of rank $d$, Lanczos recovers the original operator exactly after $d$ iterations: $T^{(d)} = T$.
This motivates the identification of approximate physical eigenvalues and eigenvectors of $T$ with those of 
\begin{equation}
    T^{(m)} = \sum_{k=0}^{m-1} \ket{y^{R(m)}_k} \lambda^{(m)}_k \bra{y^{L(m)}_k} ~ ,
    \label{eq:Tm-ritz-decomp}
\end{equation}
where $\lambda^{(m)}_k$ are the Ritz values and $\ket{y^{R(m)}_k}$ and $\ket{y^{L(m)}_k}$ are the right and left Ritz vectors, already mentioned above.

We may relate the Ritz and Lanczos vectors by considering the eigendecomposition of the tridiagonal matrix
\begin{equation}
    T^{(m)}_{ij} = \sum_{k=0}^{m-1} \omega^{(m)}_{ik} \lambda^{(m)}_k (\omega^{-1})^{(m)}_{kj}
    \label{eq:Tm_ij-eigendecomp}
\end{equation}
where $\omega^{(m)}_{ik}$ is the $i$th component of the $k$th right eigenvector of $T^{(m)}_{ij}$ and $(\omega^{-1})^{(m)}_{kj}$ is the $j$th component of the $k$th left eigenvector.
Comparing Eqs.~\eqref{eq:Tm_ij-eigendecomp} and \eqref{eq:Tm-ritz-decomp}, we can write
\begin{equation}\begin{aligned}
    \ket{y^{R(m)}_k} &= \mathcal{N}^{(m)}_k \sum_{i=1}^m \ket{v^R_i} \omega^{(m)}_{ik}, \\
    \bra{y^{L(m)}_k} &= \frac{1}{\mathcal{N}^{(m)}_k} \sum_{j=1}^m (\omega^{-1})^{(m)}_{kj} \bra{v^L_j} ~ ,
\end{aligned}
\label{eq:ritz-vector-def}
\end{equation}
where $\mathcal{N}^{(m)}_k$ is an arbitrary constant included to set the normalization.
The Ritz values and right/left Ritz vectors are the best Lanczos approximation to the true eigenvalues and right/left eigenvectors, and recover them exactly in the limit $m = d$ where $d$ is the rank of $T$.
By construction, 
\begin{equation}
    \braket{y^{L(m)}_k | y^{R(m)}_{l}} = \delta_{k l},
\end{equation}
but  $\braket{y^{L(m)}_k | y^{L(m)}_{l}} \neq \braket{y^{R(m)}_k | y^{R(m)}_{l}} \neq \delta_{kl}$ in general.

Eigenvectors are defined only up to an overall (complex) constant, which must be set by convention.
We use unit-normalized right eigenvectors such that $\sum_i |\omega_{ik}|^2 = 1$ and set the phase by $\omega_{1k} = |\omega_{1k}|$.
The left eigenvector matrix $\omega^{-1}$ is fully defined by this convention via matrix inversion of $\omega$. 
The left eigenvectors are not, in general, unit-normalized if the right eigenvectors are.
This same observation applies for the Ritz vectors and in \cref{sec:lattice} is used to identify and understand unphysical states which arise due to violations of Hermiticity by statistical noise.

In lattice applications, we do not have direct access to vectors and operators in the Hilbert space of states, only correlator data of the form
\begin{equation}
    A_1(t)
    \equiv \braket{v^L_1 | T^t |v^R_1} 
    = \frac{ \braket{\chi | T^t |\psi} }{ \braket{\chi|\psi} }
    = \frac{ C(t) }{ C(0) } ~.
\end{equation}
where here $C(t)$ is off-diagonal for the general case.
However, as shown in Ref.~\cite{Wagman:2024rid}, $T^{(m)}_{ij}$ may be computed only in terms of these quantities using recursion relations, which iteratively construct generalized correlators evaluated between higher-order Lanczos vectors,
\begin{equation}\begin{aligned}
    A_j(t) &= \braket{v^L_j | T^t | v^R_j}, \\
    G_j(t) &= \braket{v^L_j | T^t | v^R_{j-1}}, \\
    B_j(t) &= \braket{v^L_{j-1} | T^t |v^R_j} ~ .
\end{aligned}\end{equation}
These recursions may be derived by inserting \cref{eq:oblique-lanczos} into the above expression, which gives
\begin{widetext}
\begin{equation}\begin{aligned}
A_{j+1}(t) 
= \braket{v^L_{j+1} | T^t | v^R_{j+1}} 
= \frac{1}{\beta_{j+1} \gamma_{j+1}} 
\bigg[&
\braket{v^L_j | T^{t+2} - 2 \alpha_j T^{t+1} + \alpha_j^2 T^t | v^R_j}
+ \beta_j \gamma_j \braket{v^L_{j-1} | T^t | v^R_{j-1}}
\\ &- \gamma_j \braket{v^L_{j-1} | T^{t+1} - \alpha_j T^t | v^R_j}
- \beta_j \braket{v^L_j | T^{t+1} - \alpha_j T^t | v^R_{j-1}} \bigg]
\\
= \frac{1}{\beta_{j+1} \gamma_{j+1}} \bigg[
       & A_j(t+2) - 2 \alpha_j A_j(t+1) 
    + \alpha_j^2 A_j(t) + \beta_j \gamma_j A_{j-1}(t)
    \\  &- \gamma_j B_j(t+1) + \alpha_j \gamma_j B_j(t)
    - \beta_j G_j(t+1) + \alpha_j \beta_j G_j(t)
    \bigg],
\end{aligned}
\label{eq:lanczos:recursion-1}
\end{equation}
\begin{equation}\begin{aligned}
G_{j+1}(t) = \bra{v^L_{j+1}} T^t \ket{v^R_j} 
= \frac{1}{\beta_{j+1}} \big[
    \braket{v^L_j | T^{t+1} - \alpha_j T^t | v^R_j} - \gamma_j \braket{v^L_{j-1} | T^t |v^R_j}
\big]
= \frac{1}{\beta_{j+1}}  \big[
    A_j(t+1) - \alpha_j A_j(t) - \gamma_j B_j(t)
\big],
\end{aligned}
\label{eq:lanczos:recursion-2}
\end{equation}
\begin{equation}\begin{aligned}
B_{j+1}(t) = \bra{v^L_j} T^t \ket{v^R_{j+1}} 
= \frac{1}{\gamma_{j+1}} \big[
    \braket{v^L_j | T^{t+1} - \alpha_j T^t | v^R_j} - \beta_j \braket{v^L_j | T^t |v^R_{j-1}}
\big]
= \frac{1}{\gamma_{j+1}} \big[
    A_j(t+1) - \alpha_j A_j(t) - \beta_j G_j(t)
\big],
\end{aligned}
\label{eq:lanczos:recursion-3}
\end{equation}
with $A_0(t) = G_1(t) = B_1(t) = 0$ defined for notational convenience.
Similarly, one may derive
\begin{equation}
\beta_{j+1} \gamma_{j+1} = \braket{r^L_{j+1} | r^R_{j+1}}
= \braket{v^L_j | T^2 | v^R_j} - \alpha_j^2 - \beta_j \gamma_j
= A_j(2) - \alpha_j^2 - \beta_j \gamma_j  .
\label{eq:lanczos:recursion-4}
\end{equation}
\end{widetext}
Using these expressions, starting from the (normalized) correlator data $A_1(t)$, each iteration proceeds by computing first $\gamma_{j+1}$ and $\beta_{j+1}$, then $G_{j+1}(t)$ and $B_{j+1}(t)$, then $A_{j+1}(t)$, and finally
\begin{equation}
    \alpha_{j+1} = \braket{v^L_{j+1} | T | v^R_{j+1}} = A_{j+1}(1) ~ .
\label{eq:lanczos:recursion-5}
\end{equation}
Each step incorporates two more elements of the original correlator, i.e., the $\alpha_j$, $\beta_j$, and $\gamma_j$ are functions of $A_1(t)$ from $t=0,\ldots,2j-1$.
Thus, the generalized correlators grow shorter with each iteration: $A_j(t)$, $G_j(t)$, and $B_j(t)$ are defined for $t=0,\ldots,N_t-2(j-1)$.
The recursion must terminate after all $N_t$ elements of the original correlator have been incorporated, producing $\alpha_j$, $\beta_j$, and $\gamma_j$ for $j=1, \ldots, N_t/2$ after $m=N_t/2$ steps.
Note that $G_j(1) = \gamma_j$ and $B_j(1) = \beta_j$ self-consistently.

The matrix element problem requires only a subcase of this formalism.
In particular, we use only diagonal two-point correlators with $\chi = \psi$ and hereafter formally assume 
\begin{equation}
\ket{v^R_1} = \ket{v^L_1} = \frac{ \ket{\psi} }{ \sqrt{\braket{\psi|\psi}} },
\end{equation}
such that
\begin{equation}
    \braket{v^L_1|T^t|v^R_1} = \braket{v^R_1|T^t|v^R_1} = \braket{v^L_1|T^t|v^L_1} 
    = \frac{ \braket{\psi|T^t|\psi} }{ \braket{\psi|\psi} } ~ .
\end{equation}
As already discussed at the top of this section, the underlying transfer matrix is Hermitian, i.e.~$T = T^\dagger$.
For any symmetric convention $|\beta_j| \equiv |\gamma_j|$, applying this to the formalism gives $\ket{v^R_j} = \ket{v^L_j}$ for all $j$, and the oblique Lanczos process reduces to standard Lanczos (identically, if $\beta_j \equiv \gamma_j$).
In this case, all left ($L$) and right ($R$) quantities become identical and the distinction may be dropped.\footnote{If $|\beta_j| \neq |\gamma_j|$, then $\ket{v^R_j} \neq \ket{v^L_j}$ even when $T = T^\dagger$ and $\ket{v^R_1} = \ket{v^L_1}$. However, the $L$ and $R$ versions of any physical quantity will still coincide, as they are necessarily convention-independent.}
However, when statistical noise obscures the Hermiticity of $T$, the distinction $\ket{v^R_j} \neq \ket{v^L_j}$ for $j>1$ remains important as discussed in \cref{sec:lattice}.

\subsection{Convergence \& bounds}
\label{sec:lanczos:bounds}

Explorations of the convergence properties of the Lanczos algorithm have produced several distinct classes of bounds on the approximation errors in its results.
This subsection reviews two, one of which can be used in practice to assess convergence and the other of which is useful to understand the convergence properties of the method.
Versus Ref.~\cite{Wagman:2024rid}, the definitions here have been adapted to include factors of $\mathcal{N}^{(m)}_k$ and accommodate complex $\gamma_j$ and $\beta_j$.

Formally restricting to Hermitian $T = T^\dagger$, the Lanczos formalism allows computation of a rigorous two-sided bound on the distance between a given Ritz value $\lambda^{(m)}_k$ and the nearest true eigenvalue $\lambda$~\cite{Paige:1971,Parlett:1979,Parlett:1995}.
Specifically, as derived for the oblique formalism in Ref.~\cite{Wagman:2024rid}, in the special case of Hermitian $T = T^\dagger$,
\begin{equation}
    \min_{\lambda \in \{\lambda_n\}} \left| \lambda^{(m)}_k - \lambda \right|^2 \leq \left| B^{R/L(m)}_k \right|,
    \label{eq:lanczos:bound}
\end{equation}
holds simultaneously for both
\begin{equation}
    B^{R(m)}_k \equiv 
     \frac{ R^{R(m)} }{ \braket{y^{R(m)}_k | y^{R(m)}_k} }
    ~, \;\;
    B^{L(m)}_k \equiv
    \frac{ R^{L(m)} }{ \braket{y^{L(m)}_k | y^{L(m)}_k} },
\label{eq:lanczos:residual-B}
\end{equation}
defined in terms of the residuals\footnote{Recall that the first index of $\omega^{(m)}_{ik}$ and second index of $(\omega^{-1})^{(m)}_{ki}$ correspond to Lanczos vectors and run from 1 to $m$.}
\begin{equation}
\begin{aligned}
    R^{R(m)}_k &= |\gamma_{m+1}|^2 |\omega^{(m)}_{mk}|^2 |\mathcal{N}^{(m)}_k|^2 \braket{v^R_{m+1}|v^R_{m+1}},
    \\
    R^{L(m)}_k &= |\beta_{m+1}|^2 |(\omega^{-1})^{(m)}_{km}|^2 \frac{1}{|\mathcal{N}^{(m)}_k|^2} \braket{v^L_{m+1}|v^L_{m+1}} ~ .
\end{aligned}
\label{eq:lanczos:residual-R}
\end{equation}
Regrouping terms admits a convenient simplification,
\begin{equation}
\begin{aligned}
    B^{R(m)}_k &= |\gamma_{m+1}|^2 
        |\omega^{(m)}_{mk}|^2 
        V^{R(m)}_k,
    \\
    B^{L(m)}_k &= |\beta_{m+1}|^2 
        |(\omega^{-1})^{(m)}_{km}|^2
        V^{L(m)}_k ~ ,
\end{aligned}
\end{equation}
where
\begin{equation}\begin{aligned}
    V^{R(m)}_k 
    &\equiv
    \braket{v^R_{m+1}|v^R_{m+1}}
        \frac{ |\mathcal{N}^{(m)}_k|^2 }{ \braket{y^{R(m)}_k | y^{R(m)}_k} }
    \\ &= \frac{ \braket{v^R_{m+1}|v^R_{m+1}} }
        { \sum_{ij} \omega^{(m)*}_{ik} \braket{v^R_i | v^R_j} \omega^{(m)}_{jk} },
    \\
    V^{L(m)}_k &\equiv
        \braket{v^L_{m+1}|v^L_{m+1}}
        \frac{ 1/|\mathcal{N}^{(m)}_k|^2 }{ \braket{y^{L(m)}_k | y^{L(m)}_k} }
    \\ &= \frac{ \braket{v^L_{m+1}|v^L_{m+1}} }
        { \sum_{ij} (\omega^{-1})^{(m)}_{ki} \braket{v^L_i | v^L_j} (\omega^{-1})^{(m)*}_{kj} }
    ~ .
\end{aligned}
\label{eq:lanczos:residual-V}
\end{equation}
The second equation in each of the above inserts the Ritz vector definition \cref{eq:ritz-vector-def}.
The cancellation of factors of $\mathcal{N}^{(m)}_k$ reflect independence of Ritz vector normalization convention.
The $\braket{v^R_i|v^R_j}$ and $\braket{v^L_i|v^L_j}$ factors may be computed as discussed in \cref{sec:lanczos:krylov-polys}, and factors of $\braket{y^{R/L(m)}_k | y^{R/L(m)}_k}$ and $\mathcal{N}^{(m)}_k$ as in \cref{sec:lanczos:ritz-rotators} and \cref{sec:lanczos:overlaps}.
These bounds are directly computable whenever Lanczos is applied and can therefore be used to monitor convergence in practice. 

Separately, Ritz values and vectors converge to true eigenvalues and eigenvectors with a rate governed by Kaniel-Paige-Saad (KPS) convergence theory~\cite{Kaniel:1966,Paige:1971,Saad:1980} even for infinite-dimensional systems.
In the infinite-statistics limit of the case of interest ($\ket{v^L_1} = \ket{v^R_1}$ and $T = T^\dagger$), differences between Ritz values and transfer-matrix eigenvalues satisfy the KPS bound\footnote{Note that $T_{m-n-1}(\Gamma_n)$ appears here and below in place of $T_{m-n}(\hat{\gamma}_n)$ in Ref.~\cite{Saad:1980}; the Chebyshev arguments are identical while the order differs by 1 because the largest eigenvalue is labeled $\lambda_0$ here as opposed to $\lambda_1$ in  Ref.~\cite{Saad:1980}. }
\begin{equation}\label{eq:KPS}
    0 \leq \lambda_n - \lambda_n^{(m)} \leq (\lambda_n - \lambda_{\infty}) \left[ \frac{ K_n^{(m)} \tan \phi_n}{ T_{m-n-1}(\Gamma_n)} \right]^2,
\end{equation}
where the $T_k(x)$ are Chebyshev polynomials of the first kind defined by $T_k(\cos x) = \cos(k x)$,
\begin{equation}
\begin{split}
    \Gamma_n &\equiv 1 + \frac{2(\lambda_n - \lambda_{n+1})}{\lambda_{n+1} - \lambda_{\infty}} = 2 e^{E_{n+1}-E_n} - 1,
    \end{split}
\end{equation}
$\lambda_\infty$ is the smallest eigenvalue of $T$, and
\begin{equation}
    K_n^{(m)} \equiv \prod_{l=0}^{n-1} \frac{\lambda_l^{(m)} - \lambda_{\infty}}{\lambda_l^{(m)} - \lambda_n}, \hspace{20pt} n > 0 \, .
\end{equation}
with $K_0^{(m)} \equiv 1$.
For the ground state and infinite-dimensional $T$ such that $\lambda_{\infty} = 0$, this simplifies to
\begin{equation}\label{eq:KPS0} \begin{split}
    \frac{\lambda_0 - \lambda_0^{(m)}}{\lambda_0} &\leq  \left[ \frac{\tan \arccos z_0}{T_{m-1}(2 e^{\delta} - 1)} \right]^2
    \end{split}
\end{equation}
where $z_n \equiv \braket{n|v_1}$ and $\delta \equiv E_1 - E_0$.
For large $k$, $T_{k}(x) \approx \frac{1}{2}(x + \sqrt{x^2 - 1})^{k}$, and this further simplifies to
\begin{equation}
    \frac{\lambda_0 - \lambda_0^{(m)}}{\lambda_0 } \lesssim  \frac{4 (1 - z_0^2)}{z_0^2} \times \begin{cases} e^{-2 (m-1) \delta }    &  \delta \gg 1 \\
    e^{-4(m-1)\sqrt{\delta}} &  \delta \ll 1 
    \end{cases} .
\end{equation}
As discussed in Ref.~\cite{Wagman:2024rid}, near the continuum limit where $\delta \ll \sqrt{\delta} \ll 1$ the $e^{-4 m \sqrt{\delta}} \sim e^{-2 t \sqrt{\delta}}$ convergence of Lanczos is exponentially faster than the $e^{-t \delta}$ convergence of the power-iteration method and standard effective mass.

An analogous KPS bound applies to the overlaps $Y_n^{(m)} \equiv \bigl< n | y_n^{(m)} \bigr>$ between Ritz vectors and transfer-matrix eigenvectors.
Defining the angle $\phi_n^{(m)} \equiv \arccos Y_n^{(m)}$ between these vectors, the KPS bound
 on $\tan \phi_n^{(m)} = \tan \arccos Y_n^{(m)}$
 is given by~\cite{Kaniel:1966,Paige:1971,Saad:1980}
\begin{equation}
  \tan \phi_n^{(m)} \leq \frac{K_n^{(m)}}{T_{m-n-1}(\Gamma_n)} \tan \arccos z_n.
\end{equation}
For the ground state this simplifies to
\begin{equation}
  \tan \arccos Y_n^{(m)} \leq \frac{1}{T_{m-1}(2 e^{\delta} - 1)} \tan \arccos z_0,
\end{equation}
which can be expanded similarly as
\begin{equation}
  \tan \arccos Y_n^{(m)} \lesssim  \frac{2 \sqrt{1 - z_0^2}}{z_0} \times \begin{cases} e^{-(m-1) \delta }    &  \delta \gg 1 \\
    e^{-2(m-1)\sqrt{\delta}} &  \delta \ll 1 
    \end{cases} .
\end{equation}
This demonstrates that $|Y_n^{(m)}|^2$ converges to 1, which indicates that $\ket{y_n^{(m)}}$ is identical to $\ket{n}$, with the same exponential rate that the Ritz values converge to transfer-matrix eigenvalues.

\subsection{Krylov polynomials}
\label{sec:lanczos:krylov-polys}

The right and left Lanczos vectors $\ket{v^R_j}$ and $\ket{v^L_j}$ are related to the right and left Krylov vectors
\begin{equation}
    \ket{k^R_t} \equiv T^t \ket{v^R_1} ,
        \qquad  \ket{k^L_t} \equiv (T^\dagger)^t \ket{v^L_1},
\end{equation}
by the Krylov coefficients $K^R_{jt}$ and $K^L_{jt}$ as
\begin{equation}
    \ket{v^R_j} = \sum_{t=0}^{j-1} K^R_{jt} \ket{k^R_t},
    \quad
    \ket{v^L_j} = \sum_{t=0}^{j-1} K^{L*}_{jt} \ket{k^L_t} ~ .
    \label{eq:krylov-coeff-def}
\end{equation}
Equivalently, these coefficients may be thought of as the coefficients of polynomials in $T$.
These polynomials are Hilbert-space operators $K^{R/L}_j$ which excite the Lanczos vectors from the starting vectors $\ket{v^R_1}$ and $\ket{v^L_1}$ as
\begin{equation}\begin{aligned}
    \ket{v^R_j} &= K^R_j \ket{v^R_1} = \sum_{t=0}^{j-1} K^R_{jt} T^t \ket{v^R_1}, \\
    \ket{v^L_j} &= K^{L\dagger}_j \ket{v^L_1} = \sum_{t=0}^{j-1} K^{L*}_{jt} (T^\dagger)^t \ket{v^L_1} ~.
\end{aligned}\end{equation}
It is convenient to consider these objects when relating quantities defined in terms of Lanczos vectors to correlation functions, including especially Ritz vectors.

The Krylov coefficients can be computed using a simple recursion.
Beginning with
\begin{equation}
\begin{gathered}
    K^R_{1t} = K^L_{1t} = \left[ 1, ~ 0 \cdots  \right]_t,
\\
\begin{aligned}
    K^R_{2t} &= \left[ -\frac{ \alpha_1 }{ \gamma_2}, ~ \frac{1}{\gamma_2}, ~ 0 \cdots  \right]_t, \\
    K^L_{2t} &= \left[ -\frac{ \alpha_1 }{ \beta_2 }, ~ \frac{1}{\beta_2}, ~ 0 \cdots \right]_t,
\end{aligned}
\end{gathered} 
\label{eq:krylov-coeff-start}
\end{equation}
the coefficients are obtained for each subsequent $j$ from
\begin{equation}\begin{aligned}
    K^R_{j+1,t} &= \frac{1}{\gamma_{j+1}} \left[ K^R_{j,t-1} - \alpha_j K^R_{jt} - \beta_j K^R_{j-1,t} \right], \\
    K^L_{j+1,t} &= \frac{1}{\beta_{j+1}} \left[ K^L_{j,t-1} - \alpha_j K^L_{jt} - \gamma_j K^L_{j-1,t} \right],
\end{aligned}
\label{eq:krylov-coeff-recursion}
\end{equation}
using $K^R_{j,-1} = K^L_{j,-1} = 0$ for notational convenience.
Note that $K^{R/L}_{jt} = 0$ for all $t > j-1$.

Once computed, $K^R_{jt}$ and $K^L_{jt}$ provide a convenient means to compute the Lanczos-vector matrix elements that appear in the eigenvalue bound \cref{eq:lanczos:residual-V}:
\begin{equation}
\begin{gathered}
\begin{aligned}
    \braket{v^R_i | v^R_j} 
    &= \sum_{s=0}^{i-1} \sum_{t=0}^{j-1} \braket{v^R_1 | (T^\dagger)^s K^{R*}_{is} K^R_{jt} T^t | v^R_1}
    \\ &= \sum_{st} K^{R*}_{is} \frac{C(s+t)}{\sqrt{C(0)}} K^R_{jt},
\end{aligned}
\\
\begin{aligned}
    \braket{v^L_i | v^L_j}
    &= \sum_{s=0}^{i-1} \sum_{t=0}^{j-1} \braket{v^L_1 | T^s K^L_{is} K^{L*}_{jt} (T^\dagger)^t | v^L_1}
    \\ &= \sum_{st} K^L_{is} \frac{C(s+t)}{\sqrt{C(0)}} K^{L*}_{jt},
\end{aligned}
\end{gathered}
\label{eq:lanczos:wv}
\end{equation}
restricting to Hermitian $T = T^\dagger$ in the second line of each equation.

With the symmetric convention $\beta_j \equiv \gamma_j$ the definitions \cref{eq:krylov-coeff-start,eq:krylov-coeff-recursion} are identical and $K^R_{jt} = K^L_{jt}$ always.\footnote{The right and left Lanczos vectors $\ket{v^R_j}$ and $\ket{v^L_j}$ are then excited by $K_j = \sum_t K_{jt} T^t$ and its conjugate $K^\dagger_j$, respectively.}
In this case, $\braket{v^R_i | v^R_j} = \braket{v^L_i | v^L_j}^*$ as computed by \cref{eq:lanczos:wv}, given real $C(t)$.
For other conventions, these quantities may differ nontrivially.

\subsection{Ritz rotators}
\label{sec:lanczos:ritz-rotators}

The tridiagonal matrix eigenvectors $\omega^{(m)}$ and right/left Krylov coefficients $K^{R/L}$ may be combined to compute the Ritz coefficients
\begin{equation}
\begin{aligned}
    P^{R(m)}_{kt} &\equiv \mathcal{N}^{(m)}_k \sum_{i=1}^m \omega^{(m)}_{ik} K^R_{it}, \\
    P^{L(m)}_{kt} &\equiv \frac{1}{\mathcal{N}^{(m)}_k} \sum_{i=1}^m (\omega^{-1})^{(m)}_{ki} K^L_{it},
\end{aligned}
\label{eq:ritz-coeffs}
\end{equation}
which directly relate the Ritz and Krylov vectors as
\begin{equation}
\begin{aligned}
    \ket{y^{R(m)}_k} &= \sum_{t=0}^{m-1} P^{R(m)}_{kt} \ket{k^R_t}, \\
    \ket{y^{L(m)}_k} &= \sum_{t=0}^{m-1} P^{L(m)*}_{kt} \ket{k^L_t} ~ .
\end{aligned}
\end{equation}
The Ritz coefficients are independent of oblique $\beta,\gamma$ convention.
They may equivalently be thought of as the coefficients of a polynomial in the transfer matrix $T$.
These operators are the right/left Ritz rotators $P^{R/L(m)}$, which excite the Ritz vectors from the starting ones as
\begin{equation}
\begin{aligned}
    \ket{y^{R(m)}_k} 
        &= P^{R(m)}_k \ket{v^R_1}
        \equiv \sum_{t=0}^{m-1} P^{R(m)}_{kt} T^t \ket{v^R_1} , \\
    \ket{y^{L(m)}_k} 
        &= P^{L(m)\dagger}_k \ket{v^L_1}
        \equiv \sum_{t=0}^{m-1} P^{L(m)*}_{kt} [T^\dagger]^t \ket{v^L_1} ~ .
\end{aligned}
\label{eq:ritz-rotators}
\end{equation}
These objects allow straightforward relation of quantities defined in terms of Ritz vectors with expressions in terms of correlation functions.
Note that the Ritz rotators $P^{R/L(m)}_k$ are not proper projection operators.
However, true Ritz projectors may be constructed using $T^{(m)}$ in place of $T$ with the same coefficients; see \cref{app:projectors}.

The normalization and phase of the Ritz vectors---as encoded by $\mathcal{N}^{(m)}_k$---is in principle a matter of convention, but in this application unit normalization
\begin{equation}
    \braket{y^{R(m)}_k|y^{R(m)}_k} = 1
\end{equation}
is required so that $\ket{y^{R(m)}_k} = \ket{y^{L(m)}_k}$ may hold for physical states; this cannot occur if their normalizations differ.
Furthermore, as discussed further in \cref{sec:lanczos:matrix-elements}, extractions of matrix elements from off-diagonal three-point functions depend on this convention; a physical choice is necessary.
Determining $\mathcal{N}^{(m)}_k$ is most straightforwardly accomplished by computing the overlap factors, as shown in the next section.

The Ritz coefficients $P^{R/L(m)}_{kt}$ afford an alternative means of computing the $\braket{y^{R(m)}_k | y^{R(m)}_k}$ and $\braket{y^{L(m)}_k | y^{L(m)}_k}$ factors in \cref{eq:lanczos:residual-V}.
Inserting \cref{eq:ritz-rotators} gives
\begin{equation}\begin{aligned}
    \braket{y^{R(m)}_k | y^{R(m)}_l } 
    &= \sum_{st} \braket{v^R_1 | (T^\dagger)^s P^{R(m)*}_{ks} P^{R(m)}_{lt} T^t | v^R_1}
    \\&= \sum_{st} P^{R(m)*}_{ks} \frac{C(s+t)}{C(0)} P^{R(m)}_{lt},
\\
    \braket{y^{L(m)}_k | y^{L(m)}_l } 
    &= \sum_{st} \braket{v^L_1 | T^s P^{L(m)}_{ks} P^{L(m)*}_{lt} (T^\dagger)^t | v^L_1}
    \\&= \sum_{st} P^{L(m)}_{ks}  \frac{C(s+t)}{C(0)} P^{L(m)*}_{lt} ,
\end{aligned}
\label{eq:ritz-rotator-norm}
\end{equation}
invoking $T = T^\dagger$ in the second equality of each.
This expression provides a useful consistency check.
In the noiseless case, when $P^{R/L(m)}_{kt}$ are properly normalized, \cref{eq:ritz-rotator-norm} should yield $\braket{y^{R(m)}_k | y^{R(m)}_l} = \braket{y^{L(m)}_k | y^{L(m)}_l} = \delta_{kl}$.
In the noisy case of \cref{sec:lattice}, similar holds for the subset of physical states.

\subsection{Overlap factors}
\label{sec:lanczos:overlaps}

The overlap factors may be obtained directly from the eigenvectors $\omega^{(m)}_{ik}$ of the tridiagonal matrix as
\begin{equation}
\begin{aligned}
    [Z^{R(m)}_k]^* = \braket{\psi | y^{R(m)}_k}
    &= \mathcal{N}^{(m)}_k \sum_{i=1}^m |\psi| \braket{v^L_1 | v^R_i} \omega^{(m)}_{ik}
    \\ &= \mathcal{N}^{(m)}_k |\psi| \omega^{(m)}_{1k},
    \\
    Z^{L(m)}_k = \braket{y^{L(m)}_k | \psi}
    &= \frac{1}{\mathcal{N}^{(m)}_k} \sum_{j=1}^m (\omega^{-1})^{(m)}_{kj} \braket{v^L_j | v^R_1} |\psi|
    \\ &= \frac{|\psi|}{\mathcal{N}^{(m)}_k} (\omega^{-1})^{(m)}_{k1}    ,
\end{aligned}
\label{eq:lanczos:Z}
\end{equation}
using the definition \cref{eq:ritz-vector-def}.
The specific choices of $L$ versus $R$ in these definitions are motivated further in \cref{sec:lattice}, where in the noisy case they provide useful intuition.
However, the two definitions coincide for all physical states with degenerate left and right eigenvectors, and other equally correct ones are possible.

The overlap factors provide a convenient means of determining $\mathcal{N}^{(m)}_k$ to enforce unit normalization for the Ritz vectors.
Specifically, note that $Z^{L(m)}_k = Z^{R(m)}_k$ only if $\ket{y^{R(m)}_k} = \ket{y^{L(m)}_k}$, which in turn requires compatible normalizations.
Demanding that this holds gives 
\begin{equation}
    |\mathcal{N}^{(m)}_k|^2 = \frac{ (\omega^{-1})^{(m)*}_{k1}  }{ \omega^{(m)}_{1k} }.
    \label{eq:lanczos:ritz-norms}
\end{equation}
When $T^{(m)}$ is Hermitian as in the noiseless case, $(\omega^{-1})^{(m)}_{ki} = \omega^{(m)*}_{ik}$ so that $|\mathcal{N}^{(m)}_k|^2 = 1$ automatically.
However, in the noisy case explored in \cref{sec:lattice}, it must be set manually; for unphysical noise-artifact states this will not be possible, as $(\omega^{-1})^{(m)*}_{k1} / \omega^{(m)}_{1k}$ will be complex in general when $\ket{y^{R(m)}_k} \cancel{\propto} \ket{y^{L(m)}_k}$.
The $\mathcal{N}^{(m)}_k$ similarly cannot be computed in this manner for off-diagonal correlators where $\ket{\psi} \neq \ket{\chi}$ and the $R/L$ overlaps differ nontrivially.

As mentioned previously, some convention is also required to set the phase of the Ritz vectors (and thus of $\mathcal{N}^{(m)}_k$).
The standard convention for the phase of the true eigenstates is typically set to give real overlap factors, so we adopt this for the Ritz vectors as well.
From \cref{eq:lanczos:Z} we see that the convention $\omega^{(m)}_{1k} = |\omega^{(m)}_{1k}|$ is equivalent to choosing $Z^{R(m)}_k$ real if $\mathcal{N}^{(m)}_k$ is.
In the noiseless case where $\omega^{(m)}$ is unitary, this convention is inherited by $Z^{L(m)}_k$.
However, in the noisy case, unphysical states with complex $Z^{L(m)}_k$ arise as discussed in \cref{sec:lattice}.

Properly normalized Ritz coefficients allow a different but equivalent computation,
\begin{equation}
\begin{aligned}
    Z^{L(m)}_k
    &= \sum_{t=0}^{m-1} P^{L(m)}_{kt} \braket{v^L_1|T^t| v^R_1} |\psi|
    = \sum_t P^{L(m)}_{kt} \frac{ C(t) }{ |\psi| },
\\
    [Z^{R(m)}_k]^*
    &= \sum_{t=0}^{m-1} |\psi| \braket{v^L_1|T^t| v^R_1} P^{R(m)}_{kt}
    = \sum_t P^{R(m)}_{kt} \frac{ C(t) }{ |\psi| } ,
\end{aligned}
\label{eq:lanczos:Z-with-P}
\end{equation}
inserting the Ritz rotators \cref{eq:ritz-rotators} into the definitions of \cref{eq:lanczos:Z}.
This allows derivation of several nontrivial identities and can be useful for cross-checks.
For example, in the noiseless case, complex values indicate $\omega^{(m)}_{1k} = |\omega^{(m)}_{1k}|$ has not been enforced correctly; in the noisy case, this may signal the appearance of unphysical states as discussed below in \cref{sec:lattice}.

\subsection{Matrix elements}
\label{sec:lanczos:matrix-elements}

With the right/left Ritz coefficients $P^{R/L(m)}_{kt}$ computed and normalized, we can derive an expression to directly compute matrix elements from three-point functions.
Using \cref{eq:ritz-rotators}, the derivation proceeds as 
\begin{equation}\begin{aligned}
    J_{fi}^{(m)} 
    &= \braket{{y'}^{L (m)}_f | J | y^{R(m)}_i}
    \\ &= \sum_{\sigma,\tau=0}^{m-1} {P'}^{L(m)}_{f\sigma} \braket{v^{\prime L}_1| T^\sigma J T^\tau |v^R_1} P^{R(m)}_{i \tau}
    \\ &= \sum_{\sigma \tau} {P'}^{L(m)}_{f \sigma} 
    \frac{ \braket{\psi'| T^\sigma J T^\tau |\psi} }{ |\psi'| |\psi| }
    P^{R(m)}_{i \tau} 
    \\ &= \sum_{\sigma \tau} {P'}^{L(m)}_{f \sigma} 
    \frac{ C^\mathrm{3pt}(\sigma, \tau) }{ \sqrt{ C'(0) C(0) } }
    P^{R(m)}_{i \tau},
\end{aligned}
\label{eq:lanczos:mx-elts}
\end{equation}
where the primed ${P'}^{L(m)}_{f \sigma}$ are computed from the final-state two-point function $C'(t)$, and the unprimed $P^{R(m)}_{i \tau}$ from the initial-state $C(t)$.
This expression is the main result of this work.
Note that for any state of physical interest \cref{eq:lanczos:mx-elts} will reduce to \cref{eq:intro:mx_elt}---this is all states in the noiseless case where the data is manifestly Hermitian, but only a subset in the presence of noise as discussed in \cref{sec:lattice}.
The generalization to matrix elements of products of currents or other temporally nonlocal operators follows immediately by replacing $J$ in Eq.~\eqref{eq:lanczos:mx-elts} with the corresponding composite or nonlocal operator.

The extracted value is in general sensitive to the choice of Ritz vector normalization as $J_{fi}^{(m)} \propto \mathcal{N}^{(m)}_i / \mathcal{N}^{'(m)}_f$.
In the special case of diagonal three-point correlators where $\psi' = \psi$, these factors cancel and the value is convention-independent.
However, convention dependence remains in the off-diagonal case, emphasizing the importance of enforcing unit normalization as noted above to obtain physically interpretable results.

It is natural to ask why \cref{eq:lanczos:mx-elts} should be preferred over $\braket{{y'}^{R (m)}_f | J | y^{R(m)}_i}$ or $\braket{{y'}^{L (m)}_f | J | y^{L(m)}_i}$.
The definition employed is privileged in that, for the other two options, the equivalents of the last equality in \cref{eq:lanczos:mx-elts} must invoke $T=T^\dagger$.
However, the distinction is irrelevant in practice: for physical states the right and left Ritz vectors are degenerate, in which case all three definitions produce identical results.

The Lanczos matrix-element extraction has stricter data requirements than standard analysis methods.
Specifically, evaluating $J^{(m)}_{fi}$ requires $C^\mathrm{3pt}(\sigma, \tau)$ for all $0 \leq \sigma \leq m-1$ and $0 \leq \tau \leq m-1$, which includes data for all sink times $ 0 \leq t_f \leq 2(m-1)$ for $m$ Lanczos iterations.
Formally this is always the case, but some simple redefinitions allow application to more general datasets.
Taking the initial state to be $T^{t_0} \ket{\psi}$ rather than $\ket{\psi}$ defines Lanczos applied to the trimmed correlators $C(t + 2t_0)$ and $C^\mathrm{3pt}(\sigma+t_0, \tau+t_0)$, with $t,\sigma,\tau \geq 0$.
Separately, defining Lanczos from an iteration in the operator $T^n$ with integer $n>1$ rather than $T$ defines a method applicable to sparsely (but regularly) evaluated correlators, $C(nt)$ and $C^\mathrm{3pt}(n\sigma, n\tau)$.
In combination, they allow a Lanczos analysis incorporating three-point function data evaluated with regular spacing $n$ in $t_f,\tau$ starting from any nonzero $t_0$.

These stricter requirements are no concern for quark-line disconnected and gluonic operators, where three-point data is naturally available for all $0 \leq \tau < N_t$ and $0 \leq t_f < N_t$, or wherever $C(t_f)$ is available.
However, this clashes with the typical strategy used when computing connected contributions with sequential source methods.
The typical mode of inverting through the sink requires a separate calculation for each $t_f$ desired, each obtaining all $\tau$ at fixed $t_f$.
Thus, to avoid computation, $C^\mathrm{3pt}$ is often computed sparsely in $t_f$, typically avoiding small $t_f$ and skipping points in the evaluated range.
For standard analysis methods, this strategy gives better confidence in control over excited-state effects given a fixed budget.
However, for a Lanczos analysis, this restricts the number of iterations that may be evaluated. Moreover, the requirement of regular spacing in $t_f$ means Lanczos may not be able to fully incorporate the entirety of existing connected three-point function datasets.
While inconvenient, these data requirements are not necessarily a disadvantage.
The small-$t_f$ points typically discarded have good SNR, and Lanczos may usefully incorporate them without the possibility of worsening ESC.

At most, the estimate \cref{eq:lanczos:mx-elts} incorporates only $1/4$ of the computable lattice three-point function, corresponding to slightly less than half of the useful data where operator ordering $\sim \psi' J \overline{\psi}$ is satisfied.
The useful data correspond to all $t_f, \tau$ in $0, \ldots, N_t-1$ satisfying $t_f \geq \tau$, but the sums in \cref{eq:lanczos:mx-elts} over $\tau$ and $\sigma = t_f-\tau$ run only to $N_t/2-1$ at maximal $m=N_t/2$, excluding all $\sigma, \tau \geq N_t/2$.
While it would be desirable to incorporate all data available in the noisy case, the formalism does not allow it; however, we note that all excluded points lie in the region $t_f \geq N_t/2$, where thermal effects are significant or dominant.
Remarkably, as seen in the $N_t/2$-dimensional example of \cref{sec:noiseless}, the subset of three-point data incorporated is sufficient to solve for all matrix elements exactly in a finite-dimensional setting.

\section{Manifest Hermiticity \& Noiseless Example}
\label{sec:noiseless}

For our problems of interest, the transfer matrix $T$ is Hermitian, with real eigenvalues and degenerate left and right eigenvectors.
As emphasized throughout \cref{sec:lanczos}, physical interpretability requires that this also holds for the Lanczos approximation of the transfer matrix, at least for states of physical interest.
As explored in this section, in the absence of statistical noise Lanczos produces a fully Hermitian eigensystem.
We demonstrate that not only do Lanczos matrix-element estimates converge, they do so much more rapidly than estimates with previously available approaches.

It is straightforward to see that Lanczos respects Hermiticity when it is manifest in the correlator data.
In this case, $T = T^\dagger$ may be applied in the formalism of \cref{sec:lanczos} without introducing any inconsistencies.
Taking the symmetric convention\footnote{As discussed in \cref{sec:lanczos}, all physical quantities are independent of the choice of oblique convention. The reductions discussed in this paragraph apply for any symmetric convention $|\beta_j| \equiv |\gamma_j|$, which produce identical results with manifestly Hermitian data.}
$\beta_j \equiv \gamma_j$, oblique Lanczos reduces identically to standard Lanczos, which produces degenerate right/left Lanczos vectors $\ket{v^R_j} = \ket{v^L_j} = \ket{v_j}$ by construction.
It follows immediately that
\begin{equation}
    T^{(m)}_{ij} = \sum_{ij} \ket{v_i} \braket{v_i | T | v_j} \bra{v_j}
\end{equation}
is Hermitian, thus the Ritz values $\lambda^{(m)}_k$ are real and the right/left Ritz vectors are degenerate ${\ket{y^{R(m)}_k} = \ket{y^{L(m)}_k} \equiv \ket{y^{(m)}_k}}$.
All left and right quantities coincide, and the $L/R$ distinction may be dropped.

\begin{figure}
    \includegraphics[width=\linewidth]{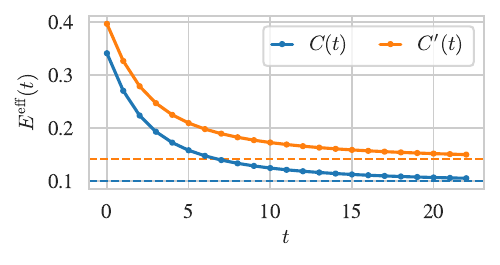}
    \includegraphics[width=\linewidth]{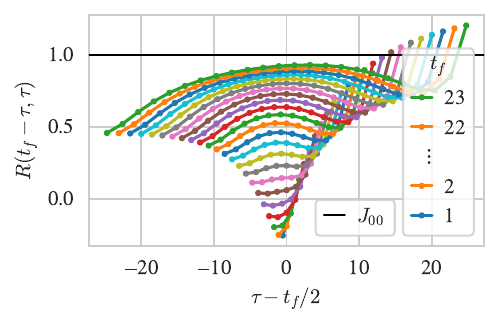}
    \caption{
        Noiseless mock-data example as defined in \cref{eq:lanczos:example-params}.
        At top, the effective energies (\cref{eq:meff}) for the initial- and final-state two-point correlators, $C(t)$ and $C'(t)$. Severe excited-state contamination is visible in the slow decay towards the ground-state energies.
        At bottom, the standard ratio (\cref{eq:std-ratio}) of three- and two-point functions to isolate the ground-state matrix element, shown for $0 \leq \tau \leq t_f$ for each $t_f$ (different colors).
        Excited-state contamination decreases in $t_f$, as seen in the approach of the value towards the true ground-state matrix element $J_{00}$ (black line), and increases in $|\tau - t_f/2|$, as seen in the curvature at fixed $t_f$.
        The curl upwards at right is an indication of severe excited-state contamination.
    }
    \label{fig:noiseless:data}
\end{figure}

To verify these statements and demonstrate the Lanczos method, we apply it to a finite-dimensional, exactly Hermitian mock-data example.
The simulated problem is the most general one that can be treated with the procedure defined in \cref{sec:lanczos}: an off-diagonal three-point function and corresponding pair of diagonal initial- and final-state two-point functions.
The two-point functions are defined as
\begin{equation}
\begin{gathered}
\begin{aligned}
    E_i &= 0.1 (i+1) ,   
        \qquad &  E'_f &= \left[ E_f^2 + 0.1^2 \right]^{1/2}, \\
    Z_i &= \frac{1}{\sqrt{2 E_i}} ,
        \qquad &  Z'_f &= \frac{1}{\sqrt{2 E'_f}}, \\
    C(t) &= \sum_{i=0}^{N_t/2-1} Z_i^2 e^{-E_i t}, 
        \; & C'(t) &= \sum_{f=0}^{N_t/2-1} Z_f^{\prime 2} e^{-E'_f t},
\end{aligned}
\end{gathered}
\end{equation}
while the three-point function is defined as 
\begin{equation}
\begin{gathered}
\begin{aligned}
    J_{fi} &= \sqrt{\frac{4 E'_0 E_0}{4 E'_f E_i}} \tilde{J}_{fi}, \\
    \tilde{J}_{00} &= J_{00} = 1, \\
    \tilde{J}_{fi} &\sim \mathcal{N}(0,1), \;\;\;\; (f+i) > 0,
\end{aligned}
\\
\begin{aligned}
    C^\mathrm{3pt}(\sigma, \tau) &= \sum_{i,f=0}^{N_t/2-1} Z'_f Z_i J_{fi} e^{-E'_f \sigma -E_i \tau},
\end{aligned}
\end{gathered}
\label{eq:lanczos:example-params}
\end{equation}
where $\tilde{J}_{fi} \sim \mathcal{N}(0,1)$ indicates that those values have been drawn from a unit-width normal distribution centered at zero.\footnote{The precise values of $J_{fi}$ used are provided in a supplemental data file.}
Initial- and final-state effective energies and the standard ratio \cref{eq:std-ratio} are shown in \cref{fig:noiseless:data}.
The energies $E_i$ and $E'_f$ are chosen to resemble an off-forward two-point function, with the final-state spectrum a boosted version of the initial one.
The overlap factors are flat up to the single-particle relativistic normalization of states, simulating the case of severe excited-state contamination.\footnote{In fact, if left untruncated, this choice of overlaps is unphysically severe: the infinite sum $C(0) = \sum_k |Z_k|^2 \sim \sum_k \frac{1}{k}$ does not converge, whereas $C(0)$ is always finite in practice.}
The excited-state and transition matrix elements have mixed signs, with the only structure in their magnitudes from the imposed single-particle relativistic normalization.
The value of the ground-state matrix element is fixed to 1; this is much larger than the typical magnitude of $J_{fi}$ with $f+i > 0$, so that $C^\mathrm{3pt}$ is ground-state dominated.

\begin{figure}
    \includegraphics[width=\linewidth]{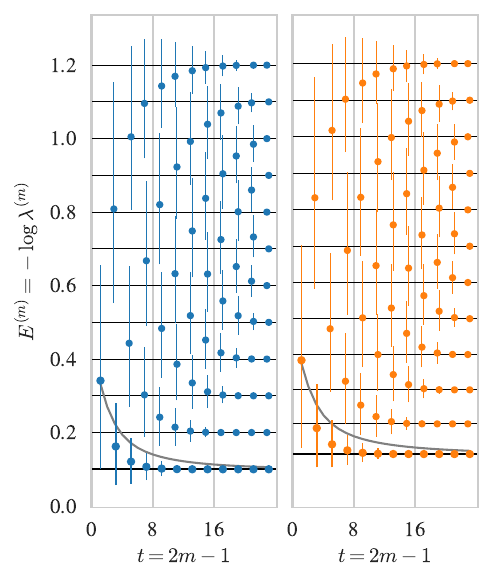}
    \caption{
    Independent Lanczos extractions of the initial- (blue, at left) and final-state (orange, at right) spectra from $C(t)$ and $C'(t)$, respectively, for the noiseless mock-data example \cref{eq:lanczos:example-params}.
    Each point corresponds to a Ritz value (i.e.~an approximate transfer-matrix eigenvalue).
    The associated uncertainties are not statistical, and instead correspond to the (asymmetric, after mapping to $E$) extent of the window of values of $\lambda^{(m)}$ allowed by the bound \cref{eq:lanczos:bound}.
    Bolded points correspond to the largest eigenvalues at each number of Lanczos steps $m$, identified as the ground state.
    Points are offset slightly in $t$ so that their uncertainties may be distinguished.
    The solid gray lines correspond to the effective ground-state energies \cref{eq:meff}, reproduced from \cref{fig:noiseless:data} but plotted as $E^\mathrm{eff}(t-1)$ to align with Lanczos at $t=m=1$.
    The x-axis $t=2m-1$ corresponds to the maximum $t$ included in the analysis after $m$ steps.
    Black horizontal lines correspond to the true energies $E_i$ and $E'_f$.
    At the final step $m=N_t/2=12$, the $N_t/2$-state spectra are solved exactly.
    }
    \label{fig:noiseless:spectrum}
\end{figure}

We proceed following the steps laid out at the top of \cref{sec:lanczos} separately to each of the initial- and final-state correlators, $C(t)$ and $C'(t)$.
It is straightforward to numerically verify the claims above.
Note that all statements made in this section should be taken to apply for exact arithmetic.\footnote{The Lanczos algorithm is notoriously susceptible to numerical instabilities due to round-off error at finite precision. \Cref{app:precision} discusses where precisely high-precision arithmetic is required.}
With any $|\beta_j| = |\gamma_j|$, the tridiagonal matrices $T^{(m)}_{ij}$ are real and symmetric, and thus have real Ritz values $\lambda^{(m)}_k$ and unitary eigenvector matrices  $\omega^{(m)}$.
The right/left Krylov coefficients are real and identical, $K^R_{jt} = K^L_{jt}$, such that $K^R_j = K^{L \dagger}_j$, implying $\ket{v^R_j} = \ket{v^L_j}$.
Consistently, evaluating \cref{eq:lanczos:wv} confirms $\braket{v^R_i|v^R_j} = \braket{v^L_i|v^L_j} = \delta_{ij}$.
The right and left Ritz coefficients $P^{R/L(m)}$ are also identical, such that the Ritz rotators are Hermitian as necessary for $\ket{y^{R(m)}_j} = \ket{y^{L(m)}_j}$.
Similar statements apply for primed final-state quantities.
The $L/R$ distinction is thus dropped in the discussion of results that follows.

\Cref{fig:noiseless:spectrum} shows the spectra of Ritz values extracted for different numbers of Lanczos iterations $m$.
One additional Ritz value is produced after each iteration, and the spectra are recovered increasingly accurately as $m$ increases.
This accuracy is reflected in the decreasing size of the eigenvalue bounds \cref{eq:lanczos:bound}, represented by the error bars in \cref{fig:noiseless:spectrum}; \cref{eq:lanczos:residual-V} simplifies when $\ket{v^R_j} = \ket{v^L_j}$, so these may be computed without further effort.
Because each spectrum has only $N_t/2$ states, Lanczos recovers all energies exactly at the maximal $m=N_t/2$ iterations.

\begin{figure}
    \includegraphics[width=\linewidth]{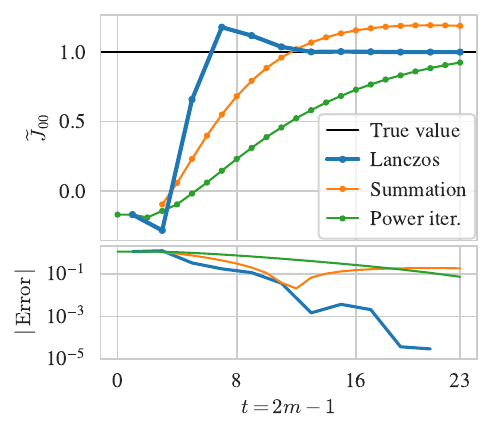}
    \caption{
        Ground-state matrix element $J_{00}$ extracted using various methods (top) and their deviations from the true value (bottom), for the noiseless mock-data example \cref{eq:lanczos:example-params}.
        The curves for Power iteration and Summation are effective matrix elements as defined in \cref{eq:power-iter-Jeff} and \cref{eq:summation-Jeff}; no fits are involved.
        The Lanczos estimate $J^{(m)}_{00}$ is computed as defined in \cref{eq:lanczos:mx-elts} using Ritz rotators constructed from the initial- and final-state two-point correlators, $C(t)$ and $C'(t)$.
        At the final step $m=N_t/2=12$, Lanczos recovers the true ground state-matrix element exactly for this 12-state example, i.e.~$J^{(12)}_{00} = J_{00}$, hence the error for this point is not shown in the log-scaled bottom panel.
    }
    \label{fig:noiseless:gsme}
\end{figure}

After obtaining the initial- and final-state Ritz coefficients ${P'}^{(m)}_{ft}$ and $P^{(m)}_{it}$, the matrix elements may be computed using \cref{eq:lanczos:mx-elts}, which reduces to
\begin{equation}
    J^{(m)}_{fi} = \sum_{\sigma \tau} (P')^{(m)*}_{f \sigma} \frac{C^\mathrm{3pt}(\sigma, \tau)}{\sqrt{C'(0) C(0)}} P^{(m)}_{i \tau}
\end{equation}
in the noiseless case, recovering the form of \cref{eq:intro:mx_elt} for all states $f,i$.
The resulting estimates of the ground-state matrix element $J_{00}$ are shown in \cref{fig:noiseless:gsme}, alongside effective matrix elements computed with power iteration (\cref{eq:power-iter-Jeff}) and the summation method (\cref{eq:summation-Jeff}).
The Lanczos estimate converges rapidly to the true value, reproducing it exactly at maximal $m=N_t/2$ where Lanczos solves the system.
The advantages in comparison to the other methods are apparent, neither of which converge near to the true value before the full Euclidean time range available is exhausted.
As a more qualitative advantage, the Lanczos estimate does not approach the true value smoothly, advertising that results are unstable until convergence is achieved.
The benefit is made apparent by considering the summation curve, which appears to be asymptoting---but to an incorrect value.
This disceptibility is investigated more directly in \cref{sec:attack}.

\begin{figure}
    \includegraphics[width=\linewidth]{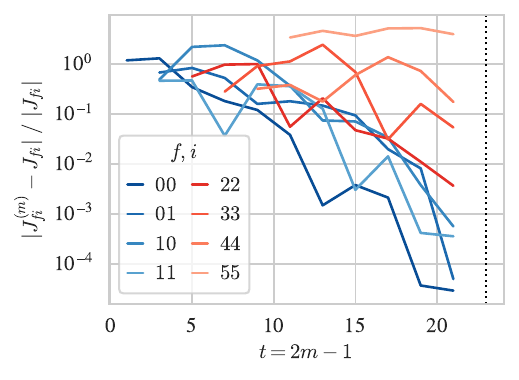}
    \caption{
        For the noiseless example~\cref{eq:lanczos:example-params}, absolute fractional deviations of Lanczos matrix element estimates $J^{(m)}_{fi}$ from the true values $J_{fi}$ for the two lowest-lying states (blue) and increasingly high-energy diagonal states (red).
        Higher states not included in this plot have more severe deviations.
        At the final step $m=N_t/2=12$ (dashed vertical line), Lanczos recovers the full set of matrix elements exactly for this 12-state example, i.e.~$J^{(12)}_{fi} = J_{fi}$, which cannot be shown.
    }
    \label{fig:noiseless:all-me}
\end{figure}

At maximal $m = N_t/2$, applied to this $N_t/2$-dimensional example, Lanczos extracts not only the true $J_{00}$ exactly but \emph{all} $(N_t/2)^2$ matrix elements: ${J^{(N_t/2)}_{fi} = J_{fi}}$.
As illustrated in \cref{fig:noiseless:all-me}, the convergence is more rapid for lower-lying states.
Comparing with \cref{fig:noiseless:spectrum}, this may be associated to a combination of the faster convergence of lower-lying eigenvalues, as expected given the relationship between eigenvalue and eigenvector convergence discussed in \cref{sec:lanczos:bounds}, and misidentification of higher-lying states with true ones.

\begin{figure}
    \includegraphics[width=\linewidth]{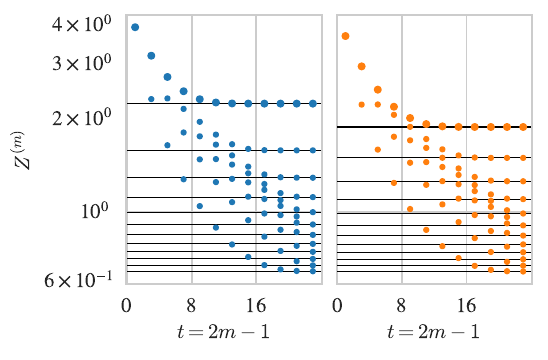}
    \caption{
        Independent Lanczos extractions of the initial- (blue, at left) and final-state (orange, at right) overlap factors $Z^{(m)}_k$ for the noiseless mock-data example \cref{eq:lanczos:example-params}, using \cref{eq:lanczos:Z-with-P} from $C(t)$ and $C'(t)$, respectively.
        For this noiseless example, $Z_k^{R(m)} = Z_k^{L(m)} \equiv Z^{(m)}_k$.
        Bolded points correspond to the ground state, identified as the state with the largest eigenvalue.
        Black horizontal lines correspond to the true overlaps $Z_i$ and $Z'_f$.
        The x-axis $t=2m-1$ corresponds to the maximum $t$ included in the analysis after $m$ steps.
        At the final step $m=N_t/2=12$, Lanczos recovers the true overlap factors exactly for these 12-state examples, i.e.~$Z^{(12)}_k = Z_k$ and $Z^{\prime (12)}_k = Z'_k$.
    }
    \label{fig:noiseless:Zs}
\end{figure}

Finally, \Cref{fig:noiseless:Zs} shows the initial- and final-state overlap factors computed using \cref{eq:lanczos:Z}.
As with the other quantities estimated, the $Z^{(m)}_k$ rapidly converge to their true values, with lower-lying states converging more quickly, and recovers the true $Z_k$ exactly at $m = N_t/2 = 12$.

\section{Noise \& the Strange Scalar Current}
\label{sec:lattice}

The previous section explored application of the Lanczos matrix-element procedure to a noiseless example.
However, as emphasized throughout the preceding discussion, important differences arise when introducing statistical noise.
Without noise, the underlying Hermiticity of the transfer matrix is manifest in the correlator data, but noise obscures this Hermiticity, resulting in unphysical states which must be identified and discarded.
This section explores these issues and techniques to treat them in an application to noisy lattice data.
The discussion emphasizes modifications versus the noiseless case; a detailed summary of the complete algorithm for the noisy case is given in \cref{app:algo}.

We use bootstrap median estimators to reduce uncertainties associated with spurious states arising from noise. Their uncertainties are estimated using nested bootstrap resampling as introduced in Ref.~\cite{Wagman:2024rid} and discussed in more detail in Ref.~\cite{Hackett:2024nbe}; a strategy involving fits to more standard outlier-robust estimators without nested bootstrap methods achieves similar uncertainties on final results is discussed in \cref{app:lattice-std}.

\subsection{Problem statement \& data}

We apply the Lanczos method to extract the forward matrix element of the strange scalar current
\begin{equation}
    J(\tau) = \sum_\vec{y} \overline{s}(\vec{y},\tau) \, s(\vec{y},\tau)
\end{equation}
in the nucleon and its first few excited states.
We employ a single ensemble of $N_\mathrm{cfg} = 1381$ configurations generated by the JLab/LANL/MIT/WM groups~\cite{ensembles}, using the tadpole-improved L\"uscher-Weisz gauge action~\cite{Luscher:1984xn} and $N_f = 2+1$ flavors of clover fermions~\cite{Sheikholeslami:1985ij} defined with stout smeared~\cite{Morningstar:2003gk} links on a $48^3 \times 96$ lattice volume. Action parameters are tuned such that $a \approx 0.091~\mathrm{fm}$ and $M_\pi \approx 170~\mathrm{MeV}$~\cite{Park:2021ypf,Yoon:2016jzj,Mondal:2020ela}. 
The data for the example are a diagonal three-point function with zero external momentum (i.e. $\vec{p} = \vec{p}' = 0$) and the single corresponding nucleon two-point function projected to zero momentum, all generated in the course of the studies published in Refs.~\cite{Hackett:2023nkr,Hackett:2023rif}.
Details are largely as in those references, but reproduced here for completeness.
Ref.~\cite{Wagman:2024rid} used data generated independently on configurations from the same ensemble.

\begin{figure}
    \includegraphics[width=\linewidth]{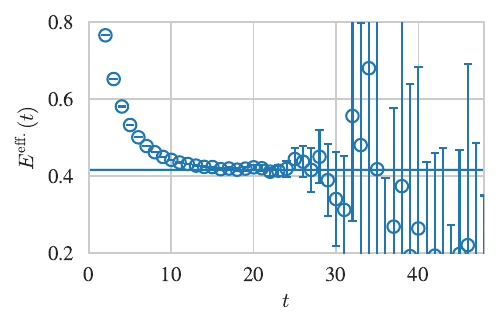}
    \includegraphics[width=\linewidth]{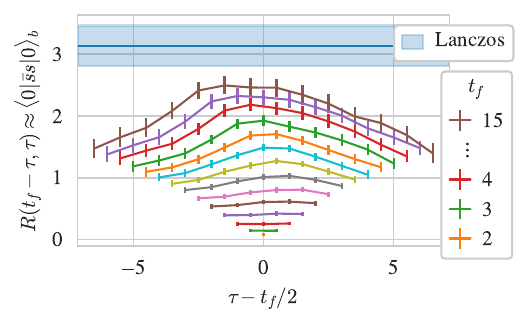}
    \caption{
        Data for the nucleon strange scalar current example.
        At top, the effective energies (\cref{eq:meff}) for the zero-momentum ($\vec{p}=0$) nucleon two-point correlator.
        The horizontal line is the fitted value $M_N = 0.4169(18)$ taken from the analysis of a superset of this data in Refs.~\cite{Hackett:2023nkr,Hackett:2023rif}; its uncertainties are not visible on the scale of the plot.
        At bottom, the standard ratio (\cref{eq:std-ratio}) constructed from the same two-point function and the forward three-point function with zero external momenta (i.e.~$\vec{p}=\vec{p'}=\vec{q}=0$) for the $\bar{s} s$ operator insertion.
        Data are available up to $t_f=96$, but signal-to-noise rapidly degrades for $t_f > 15$.
        The blue band is the final Lanczos estimate ($m=48$) from \cref{fig:lattice:gsme}, discussed below.
        Uncertainties of all estimators are computed using the nested bootstrap median approach described in the text; see \cref{fig:lattice-std:data} for analogous results without nested bootstrap.
        Uncertainties on $E^\mathrm{eff}(t)$ are computed for $\lambda^\mathrm{eff}_0(t) \equiv C(t+1)/C(t)$ then propagated linearly to avoid missingness induced by negative logarithm arguments on noisier points.
    }
    \label{fig:lattice:data}
\end{figure}

Each nucleon two-point function measurement is computed as
\begin{equation}
    C(t; x_0) = \sum_\vec{x} \Tr \left[ \Gamma \braket{ \chi(\vec{x}, t+t_0) \overline{\chi}(x_0) } \right],
\end{equation}
where $x_0 = (\vec{x}_0, t_0)$ is the source position and with the trace over implicit Dirac indices, including those of the spin projector
\begin{equation}
    \Gamma = P_+ (1 + \gamma_x \gamma_y) \text{   with   } P_+ = \frac{1}{2} (1+\gamma_t) ~ .
\end{equation}
The interpolator employed is
\begin{equation}
    \chi(x) = \epsilon^{abc} [u_b^S(x)^T C \gamma_5 d_c^S(x)] u_a^S(x),
\end{equation}
where $C$ is the charge conjugation matrix, and $u^S(x)$ and $d^S(x)$ are up- and down-quark fields smeared using gauge-invariant Gaussian smearing to radius 4.5 with the smearing kernel defined using spatially stout-smeared link fields~\cite{Morningstar:2003gk}.
This is evaluated at 1024 different source positions $x_0$ on each configuration, arranged in two interleaved $4^3 \times 8$ grids with an overall random offset.
Averaging over all source positions yields the per-configuration measurements of $C(t)$ used in this analysis.
\Cref{fig:lattice:data} shows the effective mass computed from it.

Crucially, we invoke the underlying Hermiticity of $T$ to discard the measured imaginary part of the two-point correlator $C(t)$, which is real in expectation.
While the procedure is well-defined for complex correlators, the clear separability of physical from noise-artifact states discussed below appears to arise as a result of manually enforcing Hermiticity at this level.

The three-point function is computed as
\begin{equation}
    C^\mathrm{3pt}(\sigma, \tau; x_0) = \sum_\vec{x}
    \Tr \left[ \Gamma
        \braket{ \chi(\vec{x}, t_f+t_0) J(\tau+t_0) \overline{\chi}(x_0)}
    \right] ,
\end{equation}
where $t_f = \sigma+\tau$.
Integrating over quark fields results in a quark-line disconnected diagram.
The strange quark loops are evaluated stochastically~\cite{Hutchinson1990} using one shot of $Z_4$ noise per configuration, computing the spin-color trace exactly and diluting in spacetime using hierarchical probing~\cite{Stathopoulos:2013aci,Gambhir:2017the} with a basis of 512 Hadamard vectors.
These are convolved with the grids of two-point functions and vacuum subtracted to produce the three-point function.
As with the two-point correlator, we discard the measured imaginary part.
\Cref{fig:lattice:data} shows the standard ratio computed from these data.

Note that the methodology assumes unit-normalized eigenvectors and the corresponding definitions
\begin{equation}\begin{aligned}
    C(t) &= \sum_k |Z_k|^2 e^{-E_k t}, \\
    C^\mathrm{3pt}(\sigma, \tau) &= \sum_{fi} Z^{\prime *}_f Z_i J_{fi} e^{-E'_f \sigma - E_i \tau} ~ .
\end{aligned}\end{equation}
This absorbs kinematic factors and relativistic normalizations into the definitions of $Z$ and $J$.
In general, the matrix elements and overlaps extracted must be rescaled by appropriate factors to isolate the quantities of interest; this is no different than when using methods based on correlator ratios.
However, in this example, all such normalization and kinematic factors cancel other than a factor of $\sqrt{2}$ absorbed into the overlap factors (for single-particle states).
The matrix elements extracted here thus correspond directly to the physically normalized ones, up to adjustments required if any of the resolved states is a multi-particle one.
However, we emphasize that the quantities extracted are bare.
Accounting for renormalization and operator mixing to obtain a physical quantity would requires secondary calculations unrelated to the subject of this work.
In this case, treating mixing is particularly important: the strange scalar current mixes with the light one, whose matrix element is much larger.

\subsection{Lanczos with noise}

Applying Lanczos to noisy correlator data yields spurious noise-artifact states which must be discarded to obtain physically meaningful results.
Ref.~\cite{Wagman:2024rid} introduced one strategy to identify them based on the Cullum-Willoughby (CW) test~\cite{Cullum:1981,Cullum:1985}; we employ it here as well.
In addition, consideration of the Lanczos approximation of the transfer matrix eigensystem provides a different and complementary view of this issue.

After $m$ iterations, it can be shown (see \cref{app:corr-decomp}) from \cref{eq:T_eq_Tm} that Lanczos reproduces the incorporated correlator data (i.e., for $t \leq 2m-1$) exactly:\footnote{We thank Anthony Grebe for this important insight.}
\begin{equation}
\begin{aligned}
    C(t) &= \braket{\psi | [T^{(m)}]^t |\psi} 
    \\ &= \sum_k \braket{\psi | y^{R(m)}_k} (\lambda^{(m)}_k)^t \braket{y^{R(m)}_k | \psi}
    \\ &= \sum_k Z^{R(m)*}_k Z^{L(m)}_k  e^{-E^{(m)}_k t} ,
\end{aligned}
\label{eq:lattice:corr-decomp}
\end{equation}
where $E^{(m)}_k = -\log \lambda^{(m)}_k$.
For noisy $C(t)$, this requires contributions from states with complex eigenvalues that oscillate in $t$, as well as generally complex overlap products $Z^{R(m)*}_k Z^{L(m)}_k$, which may only occur with distinct left and right Ritz vectors.
Due to the enforced reality of $C(t)$, these states necessarily contribute in pairs with complex-conjugate Ritz values and overlap products.
However, we find that we are able to identify a Hermitian subspace of states $H$ with real Ritz values and degenerate left and right Ritz vectors
\begin{equation}
    \ket{y^{R(m)}_k} = \ket{y^{L(m)}_k} = \ket{y^{(m)}_k} \quad \forall ~ k \in H ,
\end{equation}
such that, for that part of the approximation of $T$,
\begin{equation}
    T^{(m)}_H \equiv \sum_{k \in H} \ket{y^{(m)}_k} \lambda^{(m)}_k \bra{y^{(m)}_k}
\end{equation}
is manifestly Hermitian.
These states are physically interpretable, while the others are clearly associated with (or at least contaminated by) noise.
This separation may provide a mechanistic explanation for why Lanczos estimators do not exhibit exponential SNR degradation, as discussed in the conclusion.
Besides insight, in practice this observation also provides a hyperparameter-free prescription for state filtering as detailed below.
While additional filtering with the CW test remains necessary, the requirement for tuning is reduced.

We employ bootstrap resampling to study the effects of statistical fluctuations and to estimate uncertainties in the next subsection.
In particular, we employ a nested bootstrap scheme to compute uncertainties of outlier-robust median estimators.
We first construct $B=200$ ``outer'' bootstrap ensembles by randomly drawing $N_\mathrm{cfg} = 1381$ configurations with replacement from the original ensemble.
We then repeat this procedure within each outer ensemble, to construct a set of $B$ ``inner'' bootstrap ensembles for each.
Two- and three-point correlators are then averaged within each inner ensemble, producing a total of $B \times B$ measurements of each.
On each such pair of two- and three-point correlators, we independently apply the steps laid out in \cref{sec:lanczos} to compute all the various quantities therein.
Unlike in the noiseless example, high-precision arithmetic is necessary in only a few places, none of which are computationally expensive; see \Cref{app:precision} for details.
Spurious state filtering is applied independently for each inner bootstrap as detailed in the remainder of this subsection, along with general observations.
Median estimators and their uncertainties are computed as described in the next subsection.

Running the oblique Lanczos recursion of \cref{sec:lanczos:oblique-lanczos} produces in $N_t/2=48$ iterations the elements $\alpha_j$, $\beta_j$, $\gamma_j$ of the tridiagonal matrices $T^{(m)}$.
All $\alpha_j$ and products $\beta_j \gamma_j$ are real because $C(t)$ is, but may be negative; which $\beta_j \gamma_j < 0$ varies per bootstrap.
With the symmetric convention $\beta_j \equiv \gamma_j \equiv \sqrt{\beta_j \gamma_j}$, negative fluctuations produce pure imaginary $\beta_j$ and $\gamma_j$.

Diagonalizing $T^{(m)}$ for each $m$ yields Ritz values $\lambda^{(m)}_{k}$ and eigenvector matrices $\omega^{(m)}$.
The majority of Ritz values extracted are complex.\footnote{Diagnosed as $|\mathrm{Im} \, \lambda| / |\lambda| > 10^{-8}$ numerically. This convention is used for all similar statements in this section.}
The corresponding states may be excluded from the Hermitian subset $H$ immediately.
The number of real and complex Ritz values at fixed $m$ varies per bootstrap draw; for all $m \geq 3$ Lanczos iterations, there are a minimum of 3 real Ritz values in each.

\begin{figure*}
    \includegraphics[width=\linewidth]{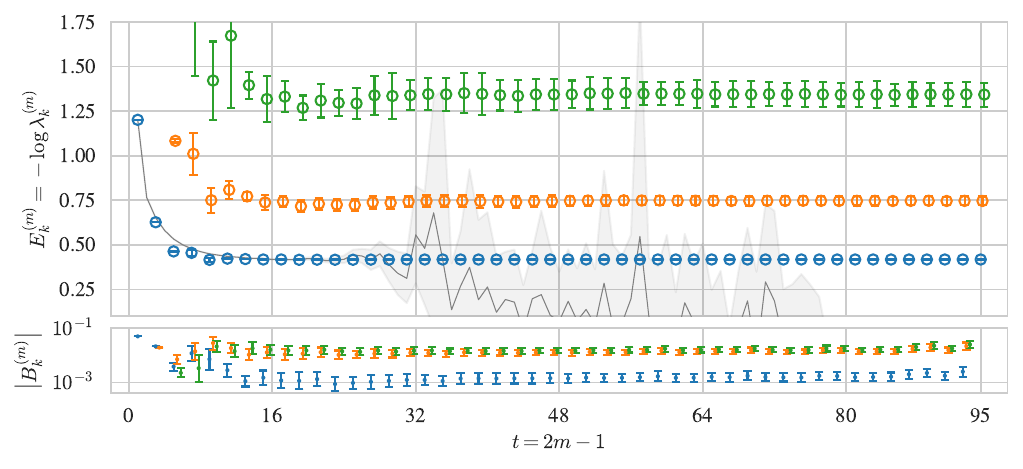}
    \caption{
        Spectrum extracted by Lanczos from the zero-momentum nucleon two-point correlator (top) and noisy estimates of the bounding values $\left| B_k^{(m)} \right|$ (\cref{eq:lanczos:residual-B}).
        The gray band reproduces the effective energy from \cref{fig:lattice:data} but plotted as $E^\mathrm{eff}(t-1)$ to align with Lanczos at $t=m=1$.
        The central values and uncertainties on the Ritz values are computed using bootstrap median estimators with nested bootstrap uncertainties as described in the main text; see \cref{fig:lattice-std:spectrum} for comparison of analogous results without nested bootstrap.
        Note that presenting results with $t=2m-1$ in this way means that both Lanczos and effective energies are plotted as a function of the largest time $t$ for which $C(t)$ is involved in constructing the estimator. In particular, Lanczos results plotted at $t$ use correlators with times $[0,\ldots,t]$ while effective energies use correlators with $[t-1,t]$.
    }
    \label{fig:lattice:spectrum}
\end{figure*}

Unit normalization cannot be simultaneously enforced for states outside the Hermitian subspace while maintaining our definitions, providing a useful means of identifying them---the ``norm trick''.
Defined as they appear in the eigendecomposition, the conventions of the left eigenvectors are fully determined by those of the right.
Attempting to compute $|\mathcal{N}^{(m)}_k|^2$ using \cref{eq:lanczos:ritz-norms} for such states thus yields complex-valued $(\omega^{-1})^{(m)*}_{k1} / \omega^{(m)}_{1k}$, with the apparent contradiction because $\braket{\psi|y^R}$ and $\braket{\psi|y^L}$ cannot be made equal if $\ket{y^L} \neq \ket{y^R}$.
We thus identify the Hermitian subspace $H$ as those states $k$ for which $(\omega^{-1})^{(m)*}_{k1} / \omega^{(m)}_{1k}$ is real and positive (and with real $\lambda^{(m)}_k$).
Manual normalization is necessary for states in $H$ after the first $m$ where $\beta_j \gamma_j < 0$, for which $|\mathcal{N}^{(m)}_k|^2 \neq 1$ in general.\footnote{This corresponds to the iteration where the standard Lanczos process would terminate or refresh, and oblique Lanczos is required to proceed. Before this, oblique and standard Lanczos coincide.}
We note that the norm trick is not applicable to off-diagonal correlators where $\ket{\psi} \neq \ket{\chi}$, and more generally that the Hermitian subspace logic presented here applies specifically to the case of diagonal correlators.

We diagnose the remaining spurious states using the CW test~\cite{Cullum:1981,Cullum:1985} as in Ref.~\cite{Wagman:2024rid}.
To do so, we define $\tilde{T}^{(m)}$ as $T^{(m)}$ with the first row and first column removed and diagonalize it to obtain the $m-1$ CW values $\tilde{\lambda}^{(m)}_l$.
Ritz values of spurious states will have a matching CW value $\tilde{\lambda}^{(m)}_l$; non-spurious states will not.
Thus, we keep all states in $H$ which satisfy
\begin{equation}
    \Delta^{\mathrm{CW}(m)}_k > \epsilon^{\mathrm{CW}(m)},
\end{equation}
where $\epsilon^{\mathrm{CW}(m)}$ is some threshold value, and
\begin{equation}
    \Delta^{\mathrm{CW}(m)}_k = \min_{\tilde{\lambda} \in \{\tilde{\lambda}^{(m)}_l\}_\mathbb{R}} | \lambda^{(m)}_k - \tilde{\lambda} |,
\end{equation}
restricting the minimum to only the real CW values; when there are none, we accept all states.
As noted in Ref.~\cite{Wagman:2024rid}, results are sensitive to the cut $\epsilon^{\mathrm{CW}(m)}$ and thus the procedure to choose it introduces the primary source of hyperparameter dependence in these methods (studied further in \cref{sec:lattice:xchecks}).
In this work, we use a simple heuristic choice versus the more extensive analysis employed in Ref.~\cite{Wagman:2024rid}, taking 
\begin{equation}
    \epsilon^{\mathrm{CW}(m)} = 
    \frac{
        \max_k [\Delta^{\mathrm{CW}(m)}_k] - \min_k [\Delta^{\mathrm{CW}(m)}_k]
    }{
        a |H| + b
    },
\label{eq:lattice:cw-cut-heuristic}
\end{equation}
where $|H|$ is the number of states in $H$.
We adopt the relatively permissive $a = 10$ and $b = 1$ and rely on outlier-robust estimators to compensate for mistuning, as discussed below.
The surviving subset are identified as the physical ones.

Filtering to the Hermitian subspace and with the CW test are highly redundant.
The CW test removes 100\% of complex eigenvalues.
Of the real eigenvalues, demanding normalizability---i.e., filtering to the Hermitian subspace with the ``norm trick''---removes $O(1\%)$ of states admitted by CW alone, while CW removes $O(10\%)$ of states admitted by normalizability alone.
Of the surviving states, none have $\lambda^{(m)}_k > 1$ (corresponding to negative $E^{(m)}_k$, which may be a thermal state).
Due to statistical fluctuations, $O(2\%)$ have $\lambda^{(m)}_k < 0$ (corresponding to imaginary $E^{(m)}_k$); notably, the CW test removes $O(75\%)$ of such states admitted by normalizability.
Note that these statistics depend on the choice of CW cut.

\subsection{Results}

With state filtering complete, we may proceed to computing observables and estimating their statistical uncertainties.
We note that for all states that survive filtering, the left and right Ritz projector coefficients are equal up to round-off error, as expected for states from the Hermitian subspace.
It follows that the $L$ and $R$ definitions of all observables will coincide for these states, so we may drop the distinction for the results in this section.
For matrix elements in particular, \cref{eq:intro:mx_elt} is recovered for all states $f,i$.

Different numbers of states survive filtering in each different (inner) bootstrap ensemble.
To avoid dealing with the complications of error quantification with data missingness, we present results for only three states, which are present in $\approx$ 99.99\% of inner ensembles.
However, we note that $\approx 80\%$ of ensembles have at least four states,\footnote{This is sufficient to calculate some quantities for this state with some reasonable but ad-hoc definitions; see \cref{app:3es}.} and $\approx 7\%$ have at least five, with the precise fraction depending on $m$.

Uncertainty quantification requires associating states between different bootstrap ensembles.
There is no unique or correct prescription for doing so, so this represents another primary source of hyperparameter dependence.
In this analysis, we make the simple choice of associating the states by sorting on their Ritz values $\lambda^{(m)}_k$ and taking the ground state as the one with the largest $\lambda^{(m)}_k$, the first excited state the one with the second largest $\lambda^{(m)}_k$, etc.
Inspection of bootstrap distributions indicates frequent misassociations by this procedure.
Rather than tuning our filtering and association schemes, we compensate by using the nested bootstrap median approach of Ref.~\cite{Wagman:2024rid}.
To do so, after filtering and sorting states, we take the median over the inner bootstraps associated with an outer bootstrap to define the estimator for that outer bootstrap.\footnote{For $\approx 10\%$ of outer ensembles, values for the second excited state ($k=2$) are not present in 100\% of inner ensembles, but are always present in at least 95\% of inner ensembles. We thus define the median as over only the values present for up to 5\% missingness, and undefined otherwise. }
Central values and uncertainties are then obtained as the mean and standard deviation over (outer) bootstraps as usual.

We now present the results of this analysis, beginning with quantities computed from $C(t)$ only.
\Cref{fig:lattice:spectrum} shows the energies of the three lowest-lying states as extracted by Lanczos.
The results are similar to those seen in Ref.~\cite{Wagman:2024rid}: Lanczos energy estimates exhibit no exponential decay in SNR, in contrast to the effective energy, which is meaningfully resolved only up to $t \approx 30$.
However, it is important to note that this absence of SNR decay is associated with the onset of large correlations between results with large $m$, as discussed further below.
The ground state is resolved with excellent signal. Noise increases moving up the spectrum.
\Cref{app:3es} shows results for the nearly-resolved third excited state.

\begin{figure}
    \includegraphics[width=\linewidth]{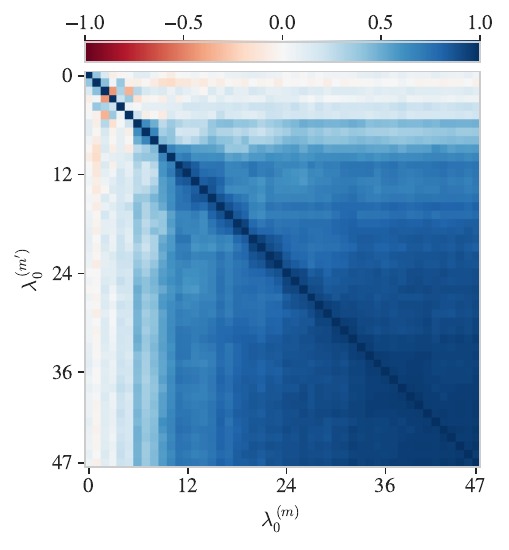}
     \includegraphics[width=\linewidth]{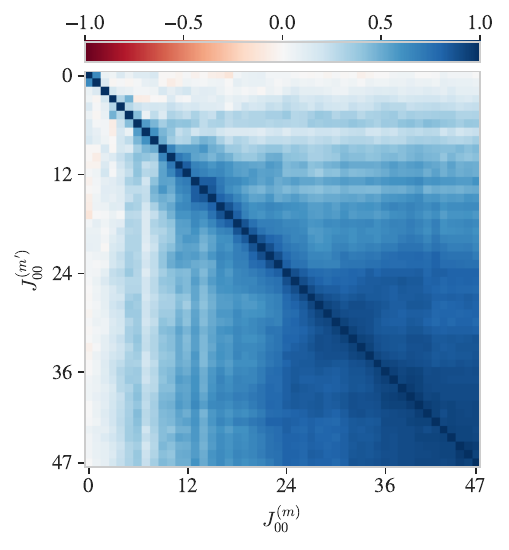}
    \caption{
        For the ground state, correlations between bootstrap median Ritz values (top) and matrix element estimators (bottom) computed using nested bootstrap resampling.
    }
    \label{fig:lattice:corrs}
\end{figure}

Bootstrap median Ritz value results for $m \gtrsim 12$ are highly correlated with one another as shown in \cref{fig:lattice:corrs}.
It is therefore reasonable to quote the results from a single large $m$ as final results.
Using the maximal $m = N_t/2= 48$ for example gives $E_k = [0.4179(21), 0.747(25), 1.343(67)]$ for $k=[0,1,2]$.
These can be compared with the results of constant fits to \texttt{gvar}-style~\cite{peter_lepage_2020_4290884} outlier-robust estimators instead of nested median uncertainties as described in \cref{app:lattice-std}, which give $E_k = [0.4175(17), 0.736(34), 1.296(83)]$ for $k=[0,1,2]$. 
An analysis of a superset of this data using standard methods in Refs.~\cite{Hackett:2023nkr,Hackett:2023rif} found $E_0 = 0.4169(18)$.

\begin{figure*}
    \includegraphics[width=\linewidth]{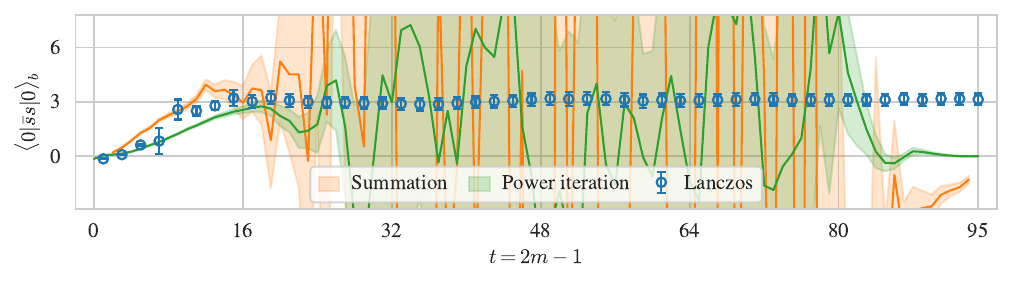}
    \caption{
        Ground-state bare nucleon matrix element of the strange scalar current with zero momentum transfer estimated using various methods.
        The curves for power iteration and summation are effective matrix elements as defined in \cref{eq:power-iter-Jeff} and \cref{eq:summation-Jeff}; no fits are involved.
        The summation curve is computed with $\Delta_\tau=1$, corresponding to the ratio data shown in \cref{fig:lattice:data}.
        Lanczos results show bootstrap median estimators with nested bootstrap uncertainties; see \cref{fig:lattice-std:gsme} for comparison of analogous results without nested bootstrap.
        Note that presenting results with $t=2m-1$ in this way means that both Lanczos and effective energies are plotted as a function of the largest time $t=\sigma+\tau$ for which $C^{\rm 3pt}(\sigma,\tau)$ is involved in constructing the estimator. In particular, Lanczos results plotted at $t$ use correlators with times $[0,\ldots,t]$ while summation and power-iteration matrix elements use correlators with $[t-1,t]$.
    }
    \label{fig:lattice:gsme}
\end{figure*}

Stability of results at large $m$ should not be interpreted as indicating ground-state (or excited-state) dominance has been achieved: this behavior is necessary, but not sufficient.
As discussed in the introduction, oblique Lanczos can exhibit a phenomenon known as stagnation, where not-yet-converged results remain stable under additional iterations.
In application to correlator analyses, this may manifest as partially-converged results ``freezing in'' after reaching the noise-dominated part of the correlator, which is dominated by finite-sample effects arising from large phase fluctuations~\cite{Wagman:2016bam}.
The large-time correlations visible in \cref{fig:lattice:corrs} and asymptotic constancy of Lanczos uncertainties together suggest that such stagnation behavior occurs in analyses of nucleon correlators at finite statistics.
In contrast, if Lanczos analyses could extract meaningful signals from points in the noise region, one would expect asymptotically vanishing SNR corresponding to improvement of signal precision with all ranges of correlator times analyzed; this is not what is observed.

\begin{figure*}
    \includegraphics[width=\linewidth]{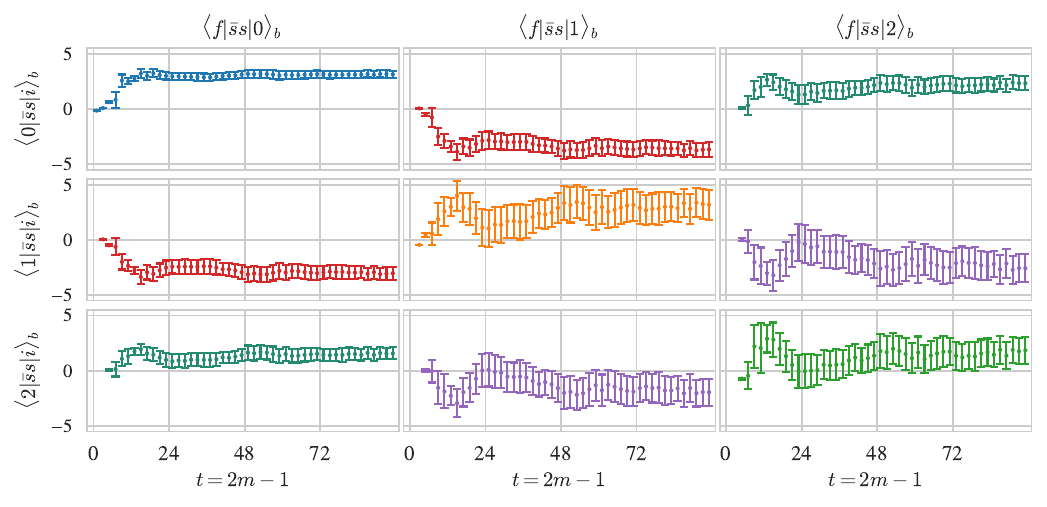}
    \caption{
        Lanczos extractions of bare forward matrix elements $\braket{f'|J|i}_b$ of the strange scalar current for three low-lying states in the nucleon spectrum.
        Note that eigenvectors are unit-normalized, not relativistically normalized. 
        Rows are indexed as the final state $f$, while columns are indexed as the initial state $i$.
        $\braket{0|\bar{s}s|0}$ is reproduced from \cref{fig:lattice:gsme}.
        Lanczos results show bootstrap median estimators with nested bootstrap uncertainties; see \cref{fig:lattice-std:MEs} for comparison of analogous results without nested bootstrap.
    }
    \label{fig:lattice:MEs}
\end{figure*}

With the statistical precision available, Lanczos does not resolve several known intermediate states in the spectrum.
With $M_\pi \approx 0.078$ and $M_N \approx 0.42$, the $N(1)\pi(-1)$ and $N\pi\pi$ multi-particle states both lie near $E \approx 0.6$, between the ground and first excited state found by Lanzcos.
This is to be expected: at finite precision, Lanczos is known to miss eigenvectors (here, states) with small overlap with the initial vector (here, $\ket{\psi}$)~\cite{Parlett:1990,Kuijlaars:2000}, and these states are known to have very small overlaps with the single-hadron interpolating operators used here~\cite{Bar:2016uoj,Gupta:2017dwj,Jang:2019vkm,RQCD:2019jai,Gupta:2021ahb,Jang:2023zts,Gupta:2023cvo,Grebe:2023tfx,Barca:2024sub,Gupta:2024krt}.
Their absence in the results points immediately to several topics requiring further study: the dynamics that determine which states are extracted by Lanczos, and how badly such missed intermediate states contaminate Lanczos matrix-element estimates.

With the analysis of the two-point correlator data understood, we move on to matrix-element estimates incorporating the three-point correlator.
\Cref{fig:lattice:gsme} shows the Lanczos estimate $J^{(m)}_{00}$ of $\braket{0|\bar{s}s|0})_b$, the bare forward matrix element of the strange scalar current in the nucleon, as compared to effective matrix elements defined with summation and power iteration.
As immediately apparent, Lanczos provides clear signals across the full range of $m$, with no exponential SNR decay.
However, bootstrap median estimators for $J^{(m)}_{00}$ show a qualitatively similar pattern of large correlations at large $m$ as for $\lambda^{(m)}_0$ as seen in \cref{fig:lattice:corrs}, although large correlations do not appear until somewhat larger $m$ in the matrix element case.
The summation and power-iteration estimates are less noisy than Lanczos for small $t$ but break down after $t \gtrsim 20$.

\Cref{fig:lattice:gsme} shows several indications that Lanczos provides better control over excited-state effects than either other method, as expected from the analyses of \cref{sec:noiseless} and \cref{sec:attack}.
The value of Lanczos estimates stabilizes within error after $m \approx 8$, corresponding to $t \approx 15$, where both the other estimators still show clear indications of large excited-state effects.
Before losing signal, the cleaner power iteration estimator may be read as suggesting an asymptote at an incompatible, smaller value than the one found by Lanczos.
The analyses in the noiseless case suggest that this behavior is most likely deceptive and that power iteration remains contaminated by excited states.
The summation estimator loses signal before achieving any convincing plateau, but suggests a value compatible with Lanczos or slightly greater.
It is interesting to note that this ordering of values---power iteration, Lanczos, then summation---is the same as observed in the example of \cref{sec:noiseless}.
The Lanczos estimate at maximal $m=48$ is shown in the ratio plot of \cref{fig:lattice:data}; the substantial extrapolation from the data is another indication of large excited-state effects.

Unlike summation and power iteration, Lanzcos allows direct and explicit computation of estimates for transition and excited-state matrix elements.
\Cref{fig:lattice:MEs} shows the results for all combinations of the three states fully resolved; \cref{tab:lattice:MEs} lists the values at maximal $m=48$.
While noisier than the ground-state matrix element, useful signals are available at all $m$ for all excited-state and transition matrix elements.
Matrix elements involving the ground state are less noisy, but otherwise noise for estimates involving either excited state is similar.
The data in \cref{fig:lattice:MEs} and listed in \cref{tab:lattice:MEs} all satisfy the expected symmetry $\braket{f|\bar{s}s|i} = \braket{i|\bar{s}s|f}$ within error.

\begin{table}
    {
    \newcommand{\p}{\phantom{+}}
    \begin{equation*}
        \braket{f | \bar{s} s | i}_b =
        \begin{bmatrix}
            \p 3.14(33) &   -3.67(64)  & \p 2.34(63)  \\
              -3.03(59) & \p 3.2(1.4)  &   -2.6(1.3) \\
            \p 1.57(54) &   -2.0(1.2)  & \p 1.8(1.2) 
        \end{bmatrix}_{fi} 
    \end{equation*}
    }
    \caption{
        Results for Lanczos estimates $J^{(m)}_{fi}$ for $m=N_t=2=48$ for bare matrix elements $\braket{f|\bar{s}s|i}_b$ of the strange scalar current. $f$ indexes rows and $i$ indexes columns. Values are as shown in corresponding panels of \cref{fig:lattice:MEs} and computed as described there.
        Results of constant fits to \texttt{gvar}-style~\cite{peter_lepage_2020_4290884} outlier-robust estimators without nested median uncertainties described in \cref{app:lattice-std} as shown in \cref{tab:lattice-std:MEs} and for the ground state give $3.03(24)$.
    }
    \label{tab:lattice:MEs}
\end{table}

\begin{figure}
    \includegraphics[width=\linewidth]{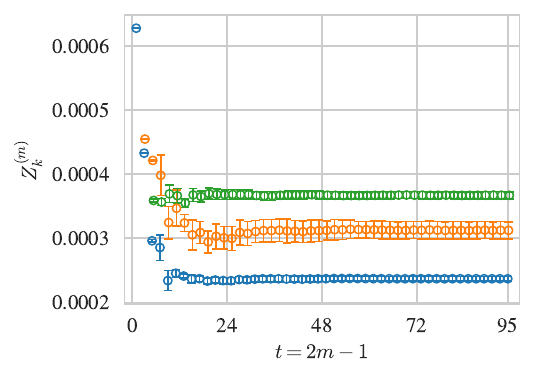}
    \caption{
        Overlap factors $Z^{(m)}$ extracted by Lanczos from the zero-momentum nucleon two-point correlator.
        Blue, orange, and green markers correspond to the ground, first excited, and second excited state respectively.
        Note that eigenvectors are unit-normalized, not relativistically normalized.
        Lanczos results show bootstrap median estimators with nested bootstrap uncertainties; see \cref{fig:lattice-std:Zs} for comparison of analogous results without nested bootstrap.
    }
    \label{fig:lattice:Zs}
\end{figure}

Finally, \Cref{fig:lattice:Zs} shows the overlap factors for the three states fully resolved.
Similar correlations between bootstrap median results for $Z_k^{(m)}$ are found as for $\lambda_k^{(m)}$.
Results using $m = N_t/2 = 48$ give 
$Z_k = [2.369(37), 3.12(13), 3.672(54)] \times 10^{-4}$.
For comparison, $\sqrt{C(0)} = \sqrt{\sum_k |Z_k|^2} = 6.2802(13) \times 10^{-4}$.
Results of constant fits to \texttt{gvar}-style~\cite{peter_lepage_2020_4290884} outlier-robust estimators without nested median uncertainties described in \cref{app:lattice-std} give $Z_k = [2.360(41), 3.03(18), 3.670(66)] \times 10^{-4}$ for $k=[0,1,2]$.
Fits of a three-state model to the same correlator data find compatible values.
These are not required for the matrix element calculation and do not correspond to any quantity of physical interest in this calculation.
However, in other settings, overlap factors are extracted to determine quantities like decay constants and quark masses.
These results suggest that Lanczos can provide an advantage in these calculations as well.

\subsection{Cross-checks}
\label{sec:lattice:xchecks}

Here, we provide some cross-checks of the results of the previous section which assess stability under different variations.
The parameters whose specification uniquely defines Lanczos results are the minimum and maximum $t$ entering the analysis and the choice of $\varepsilon_{\rm CW}$ used for spurious-state filtering.
As discussed below, the results demonstrate useful insensitivity to analysis hyperparameter choices, and provide further guidance on the interpretation of our results.

\begin{figure}
    \includegraphics[width=\linewidth]{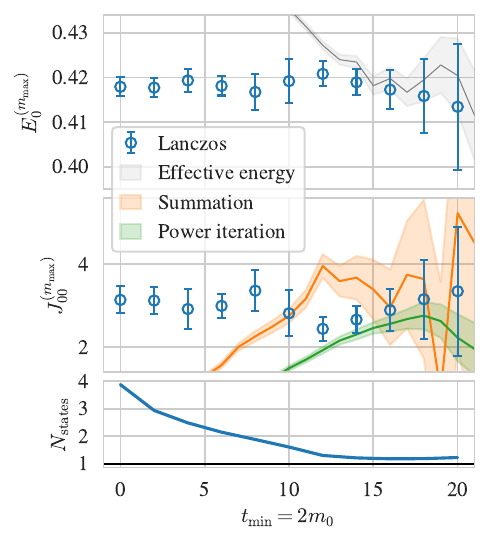}
    \caption{
        Ground-state energies (top) and bare nucleon matrix elements (middle) as a function of the minimum time $t_{\rm min}$ incorporated. For Lanczos estimators, $t_\mathrm{max} = 95$, while for effective estimators $t_\mathrm{max} = t_\mathrm{min}+1$.
        The bottom panel shows the number of non-spurious states after Hermitian subspace and Cullum-Willoughby filtering.
    }
    \label{fig:lattice:tmin}
\end{figure}

As discussed in \cref{sec:lanczos:matrix-elements}, the minimum $t$ incorporated in the Lanczos analysis can be varied by redefining the initial vector as $\ket{\psi} \rightarrow T^{m_0} \ket{\psi}$.
Following through the formalism,\footnote{Note that this also involves formally assuming $T^{m_0} \ket{\psi} \equiv [T^\dagger]^{m_0} \ket{\psi}$.} this prescribes applying the same steps to $C(t)$ with the first $2 m_0$ points removed and to $C^\mathrm{3pt}(\sigma,\tau)$ with the first $m_0$ points removed in both $\sigma$ and $\tau$.
\Cref{fig:lattice:tmin} shows the effects of varying $t_\mathrm{min} = 2 m_0$ on estimators of the ground-state energy $E^{(m_\mathrm{max})}_0$ and ground-state matrix-element $J^{(m_\mathrm{max})}_{00}$ evaluated at the final iteration $m_\mathrm{max} = N_t/2 - m_0$.
Several features are worth noting.

The ground-state energy $E_0$ is fully stable within errors for all $t_\mathrm{min}$ considered.
The same is true for $J_{00}$ at earlier $t_\mathrm{min}$, although it fluctuates outside errors beyond $t_\mathrm{min} = 10$.
Stability against removing early-time points contradicts two conflicting pieces of intuition.
The usual ``power iteration'' logic applied as for multi-state fits suggests that removing early, highly contaminated data will \emph{decrease} excited-state effects resulting from incomplete modeling of exited states.
Meanwhile, Lanczos logic suggests that removing vectors from the Krylov space will \emph{increase} excited-state effects due to reduced convergence/information.
Neither effect is observed; instead, varying $t_\mathrm{min}$ provides a consistent estimate, suggesting that $t_\mathrm{min} = 0$ can be taken without concern (apart from the possibility of contact terms, see e.g. Ref.~\cite{Zhang:2025hyo}).

Although the $t_\mathrm{min}$ dependence is not smooth, noise generally increases progressively as better-resolved early-time points are trimmed away. 
Generally, Lanczos results have larger uncertainties than power-iteration results, which can be understood from the fact that power-iteration results only involve $C(t)$ with $t \in [t_{\rm min}, t_{\rm min}+1]$, while Lanczos results involve noisier $C(t)$ with larger $t$.
Fewer non-spurious states are resolved as $t_{\rm min}$ is increased, and the relatively large fluctuations in Lanczos results around $t_{\rm min} \sim 12$ are coincident with the point where the first excited state becomes poorly resolved.

\begin{figure}
    \includegraphics[width=\linewidth]{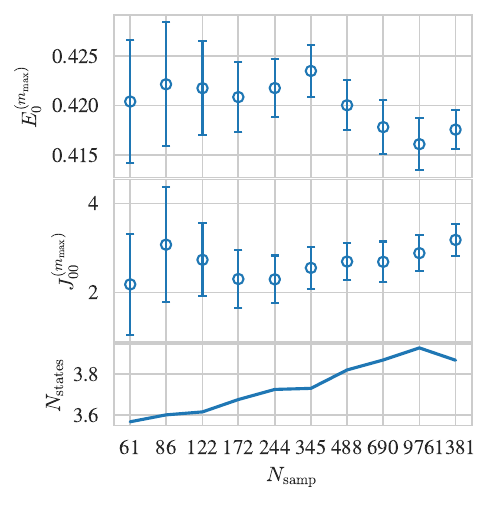}
    \caption{
        Ground-state energies (top) and bare nucleon matrix element (middle) estimators as in \cref{fig:lattice:gsme} with maximum $t=95$ and $t_{\rm min} = 0$ plotted as a function of the size of the statistical ensemble of gauge-field configurations used to compute correlation functions.  The bottom panel shows the number of non-spurious states.
    }
    \label{fig:lattice:stats}
\end{figure}

The dependence of Lanczos ground-state energy and matrix element estimators on the size of the statistical ensemble of gauge-field configuration is shown in \cref{fig:lattice:stats}.
We study this dependence by first randomizing the order of the 1381 configurations included, then taking the first $N_\mathrm{samp}$; results for larger $N_\mathrm{samp}$ thus include a superset of those for smaller $N_\mathrm{samp}$.
The expected $1/\sqrt{N_{\rm samps}}$ decrease of uncertainties can be observed.
There are also hints of a decrease in $E_0^{(m_{\rm max})}$ and increase in $J_{00}^{(m_{\rm max})}$ with increasing $N_{\rm samps}$ that could point towards finite-sample bias,
as would be expected from a result which converges until stagnating at the noise-dominated part of the correlator.
The number of non-spurious states is only very weakly dependent on $N_{\rm samps}$, varying between 3.6 and 3.8 as statistics are varied by more than an order of magnitude.

\begin{figure}
    \includegraphics[width=\linewidth]{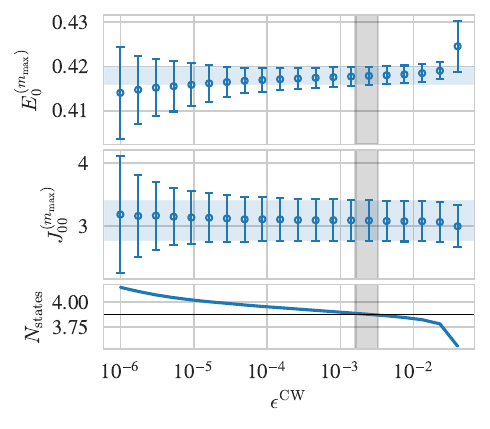}
    \caption{
        Ground-state energies (top) and bare nucleon matrix elements (middle) estimators as in \cref{fig:lattice:gsme} with maximum $t=95$ and $t_{\rm min} = 0$ plotted as a function of the threshold entering Cullum-Willoughby test for spurious eigenvalues. The bottom panel shows the number of non-spurious states.
        The vertical gray band in each panel indicates the values of $\epsilon^\mathrm{CW}$ chosen adaptively.
        The horizontal blue bands indicate estimates of each observable made with the adaptive choice.
        The horizontal black line in the bottom panel is the average number of states retained with the adaptive choice.
    }
    \label{fig:lattice:CW-scan}
\end{figure}

The main results of this work and those of Ref.~\cite{Wagman:2024rid} considered adaptive choices of CW cut $\epsilon^\mathrm{CW}$ using different schemes.
However, one can also simply consider choosing some fixed cut, taken commonly across all bootstraps.
In \Cref{fig:lattice:CW-scan}, we consider sensitivity of the estimators $E^{(m_\mathrm{max})}_0$ and $J^{(m_\mathrm{max})}_{00}$ to this choice.
We observe very weak sensitivity to $\epsilon^\mathrm{CW}$ across several orders of magnitude.
At low values, $\epsilon^\mathrm{CW} \sim 10^{-5}$, the errors begin to increase; 
this corresponds to too-small cuts admitting more and more spurious eigenvalues.
After several decades of stability, the result rapidly destabilizes just before  $\epsilon^\mathrm{CW} \sim 0.1$; this corresponds to an overly aggressive cut which removes all useful information.
The adaptively chosen cut lies within the stable region; the fixed choice of $\epsilon^\mathrm{CW}$ produces values and uncertainties closely compatible with the adaptive choice.
The number of states labeled non-spurious varies smoothly with $\ln \epsilon^\mathrm{CW}$ over many orders of magnitude but drops sharply at the last point.

Typical analyses involve statistical fits, either to three- and two-point correlation functions or directly to the ratios entering power-iteration estimators.
In contrast to the summation and power-iteration estimators shown in \cref{fig:lattice:gsme}, fit results have uncertainties that do not increase as the maximum $t_f$ analyzed is increased.
In this way, fit results are qualitatively more analogous to Lanczos than power-iteration estimators.
Quantitative comparison of Lanczos and fit results is complicated by the large space of fit possibilities and number of hyperparameters needed to define a fitting methodology.
Many choices of two-state fits achieve results consistent with Lanczos with somewhat larger uncertainties, e.g.~$3.00(59)$ and $2.79(43)$ with $t_{f,{\rm min}} = 15$ and $t_{f,{\rm min}} =17$, respectively, and $\tau_{\rm min} = 3$ and $t_{f,{\rm max}} = 20$ in both cases.
On the other hand, one-state fits restricted to larger imaginary times give somewhat smaller matrix element values, e.g.~$2.35(12)$ and $2.49(17)$ with $t_{f,{\rm min}}=16$ / $\tau_{\rm min} = 5$ and $t_{f,\rm min} =17$ / $\tau_{\rm min} = 6$, respectively.
All four of these examples give acceptable fits, with $\chi^2/\text{dof}$ of 0.38, 0.34, 0.47, and 0.32, respectively.
There are multiple reasonable choices for how to combine these and other acceptable fits including model selection based e.g. on the Akaike information criterion (AIC)~\cite{AkaikeAIC} and weighted averaging based of Bayesian model averaging~\cite{Jay:2020jkz} or other criteria~\cite{BMW:2014pzb,Rinaldi:2019thf,Neil:2022joj}.
Using the same model selection and averaging procedure detailed in Ref.~\cite{Shanahan:2020zxr} with hyperparameter $\text{tol}_{AIC} = -2$, linear shrinkage~\cite{Ledoit:2004} parameter 0.1, conjugate-gradient optimization, and other default hyperparameter choices specified in that work, a weighted average over $t_{f, \rm min}$ and $\tau_{\rm min}$ gives $2.19(14)$, while with $\text{tol}_{AIC} = 0$, an analogous weighted average gives $2.52(43)$. A weighted average over two-state fits only gives $3.3(1.2)$. 
Different procedures for combining one- and two-state fits therefore lead to qualitatively different answers for e.g.~whether Lanczos results are more or less precise than fit results.

These results suggest that the outputs of a Lanczos analysis depend less on analyzer choices than a fitting analysis.
For fixed data, fits are sensitive to the number of states included in the fit model and other choices regarding covariance matrix regularization and numerical optimization needed to define a statistical inference scheme, while Lanczos results depend on the scheme used for spurious eigenvalue filtering and the choice of iteration presented as final results.
In this noisy data example, the results shown in \cref{fig:lattice:gsme,fig:lattice:tmin,fig:lattice:CW-scan} indicate that both the central values and uncertainties of Lanczos ground-state matrix element results have  weaker sensitivity to $t_{\rm min}$, $t_{\rm max}$, and the order of magnitude of $\epsilon^{\rm CW}$ than the sensitivity of fits to $t_{\rm min}$ and the number of states included in the truncated spectral expansion used as a fit model.

\section{Adversarial Testing}
\label{sec:attack}

Absent a framework of rigorous bounds as is available for energy levels, it is worthwhile to develop more qualitative intuition about the practical reliability of Lanczos extractions.
In this section, we construct adversarial attacks to test both Lanczos and previous methods for ground-state matrix element estimation.
Specifically, in a noiseless finite-dimensional setting, we attempt to construct pathological examples which lead the different methods to report deceptive results which confidently suggest an incorrect answer.
We find that Lanczos appears to be qualitatively more robust than the other methods considered.

To construct the attacks, we fix the value of the ground state matrix element to the ``true'' value $J_{00} = \tilde{J}_{00} = 1$ and attempt to construct examples where the methods report the ``fake'' value $\breve{J}_{00} = 0.5$, and restrict all parameters varied to physically reasonable values to avoid unrealistic fine-tuning.
For simplicity, we consider a diagonal example where $\psi = \psi'$.
We take $N_t=32$ and $N_t/2 = 16$ states with energies and overlaps fixed to
\begin{equation}
    E_k = 0.1 (k+1), \quad Z_k = \frac{1}{\sqrt{2 E_k}},
\end{equation}
for both the initial- and final-state spectrum.
These choices are as employed for the initial-state spectrum in the example of \cref{sec:noiseless}; as observed there, this provides an example with severe excited state contamination.

In the attacks, we vary only the matrix elements, adversarially optimizing them based on criteria described below.
For this diagonal example, we enforce a symmetric matrix element $J_{ij} = J_{ji}$.
This also serves to make the attack more difficult by preventing fine-tuned near-cancellations between contributions with similar energies and opposite signs.
We also put in that $J_{ij}$ scales with energy as the single-particle normalization of states by defining
\begin{equation}
    J_{ij} = \frac{2 E_0}{\sqrt{4 E_i E_j}} \tilde{J}_{ij},
\end{equation}
as in \cref{sec:noiseless} and optimizing the variables $\tilde{J}_{ij}$ (up to symmetrization and fixing $J_{ij}=1$).
For all of the methods considered, matrix-element estimates are linear in the three-point function and thus linear in $\tilde{J}_{ij}$.
Fixing $Z_k$ and $E_k$ and using the $\chi^2$ functions defined in the subsections below provides a quadratic optimization problem that may be minimized analytically.

\subsection{Attack on the summation method}
\label{sec:attack:summation}

\begin{figure}
    \includegraphics[width=\linewidth]{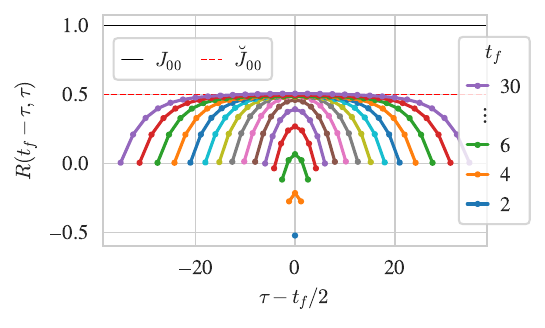}
    \caption{
        Standard ratio \cref{eq:std-ratio} for the example constructed to deceive the summation method.
        The ratio appears to approach 0.5 (dashed red line), but the true ground-state matrix element $J_{00}=1$ (black line).
    }
    \label{fig:attack:summation-ratio}
\end{figure}

\begin{figure}
    \includegraphics[width=\linewidth]{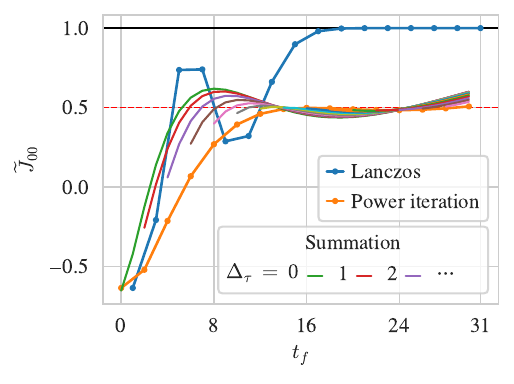}
    \caption{
        Ground-state matrix element $J_{00}$ extracted using various methods for the example constructed to deceive the summation method.
        The curves for Power iteration and Summation are effective matrix elements as defined in \cref{eq:power-iter-Jeff} and \cref{eq:summation-Jeff}; no fits are involved.
        Different summation curves correspond to different choices of the summation cut $\Delta_\tau$.
        The black horizontal line indicates the true ground-state matrix element $J_{00}=1$.
        The dashed red line indicates the faked value.
        The Lanczos estimate $J^{(m)}_{00}$ is computed as defined in \cref{eq:lanczos:mx-elts}.
        At the final step $m=N_t/2=16$, Lanczos recovers the true ground state-matrix element exactly for this 16-state example, i.e.~$J^{(12)}_{00} = J_{00}$.
    }
    \label{fig:attack:summation-meffs}
\end{figure}

As defined in \cref{sec:review}, the summation method may be used to define an effective matrix element $J^\mathrm{eff}_{00,\Delta_\tau}(t)$.
Their interpretation is similar to effective energies: we expect $J^\mathrm{eff}_{00,\Delta_\tau}(t)$ to asymptote to the true value $J_{00} = 1$ as $t$ increases and excited states decay away. Increasing the summation cut $\Delta_\tau$ is also expected to reduce contamination. 
With this usage in mind, we construct the minimization objective $\chi^2 = \chi^2_\mathrm{SM} + \chi^2_\sigma$.
The first term 
\begin{equation}
    \chi^2_\mathrm{SM} = \sum_{\Delta_\tau=2}^{N_t/2-1} \sum_{t=2\Delta_\tau + 2}^{N_t-1} \left[
        J^\mathrm{eff}_{00,\Delta_\tau}(t) - \breve{J}_{00}
    \right]^2 ~ ,
\end{equation}
where $J^\mathrm{eff}_{00,\Delta_\tau}(t)$ is a function of the optimized $\tilde{J}_{ij}$,
attempts to induce a deceptive ``pseudo-plateau'' at $\breve{J}_{00} = 0.5$, with estimates for early $t$ and small $\Delta_\tau$ unconstrained. Note that $J^\mathrm{eff}_{00,\Delta_\tau}(t)$ is only defined for $t \geq 2 \Delta_\tau$ and the maximal $\Delta_\tau = N_t/2-1$.
The second term 
\begin{equation}
    \chi^2_\sigma = \sum_{i \leq j} \frac{(\tilde{J}_{ij})^2}{ \sigma^2 },
    \label{eq:attack:prior}
\end{equation}
serves to keep the values of $\tilde{J}_{ij}$ reasonably physical; we take $\sigma = 10$ to enforce $\sim O(1)$ matrix elements, up to scaling with energy.
This term is also necessary to regulate the otherwise underconstrained fit.

Optimizing yields the example shown in \cref{fig:attack:summation-ratio,fig:attack:summation-meffs}.
While we have directly attacked the summation method, we also examine the ratio \cref{eq:std-ratio} and the power iteration effective matrix element defined in \cref{sec:review}, as might be done for cross-checks in an analysis.
The ratio, \cref{fig:attack:summation-ratio}, is exactly as expected if the ground state matrix element were $\breve{J}_{00} = 0.5$; its behavior is visually indistinguishable from a well-behaved decay of excited states as $t_f$ increases.
\cref{fig:attack:summation-meffs} shows effective matrix elements for both power iteration and summation for all possible $\Delta_\tau$; all appear to asymptote near $\breve{J}_{00}$.
While some noticeable curvature remains for the summation curves, it is subtle enough to be concealed by even a small amount of noise.
Taking these points together, a naive analysis of this example with these methods would likely conclude with high confidence that $J_{00} = 0.5$, a factor of 2 off from the true value.

Analyzing the examples found by an adversarial attack can provide insight into what mechanisms may cause a method to fail.
Inspection of the fitted matrix%
{
\newcommand{\p}{\phantom{+}}
\begin{equation*}
\tilde{J}^{(*)}_{fi} =
\begin{pmatrix}
   \p1.000 &  -4.137   & \p9.047 &  -4.979 &  -3.467 &         \\
    -4.137 &  \p17.344 &  -9.85  &  -6.139 &  -1.884 &         \\
   \p9.047 &  -9.850   &  -3.635 &  -1.297 & \p0.683 & \cdots  \\
    -4.979 &  -6.139   &  -1.297 & \p0.547 & \p1.317 &         \\
    -3.467 &  -1.884   & \p0.683 & \p1.317 & \p0.518 &         \\
           &           &  \vdots &         &         & \ddots
\end{pmatrix}
\end{equation*}
}%
reveals that the pathological behavior may be attributed to a small cluster of low-lying states with larger-magnitude matrix elements than the ground state.
Such a scenario may easily arise in nature if the ground-state matrix element happens to be small.
This situation resembles closely the situation with $N\pi$ and $N\pi\pi$ contamination speculated to cause problems in lattice calculations of axial form factors~\cite{Gupta:2017dwj,RQCD:2019jai,Jang:2023zts,Gupta:2023cvo,Barca:2024sub}.

Also shown in \cref{fig:attack:summation-meffs} is the Lanczos estimate $J^{(m)}_{00}$ for the same example.
After an initial period of violent reconfiguration with no pseudo-plateau, the estimate quickly converges to the true value.
This convergence occurs long before the maximal $m=N_t/2$ where the system is solved exactly.
This represents a qualitative improvement in treatment of this example, and suggests immediately that Lanczos is more robust against such pathological scenarios.

\subsection{Attack on Lanczos}
\label{sec:attack:lanczos}

Applying the same adversarial strategy against the Lanczos method allows its improved robustness to be assessed more directly.
We use the optimization objective $\chi^2 = \chi^2_{LM} + \chi^2_\sigma$ where $\chi^2_\sigma$ is as in \cref{eq:attack:prior} and
\begin{equation}
    \chi^2_{LM} = \sum_{m \in M} \left[ J^{(m)}_{00} - \breve{J}_{00} \right]^2,
\end{equation}
with $M$ some set of $m$ to target.
Note that high-precision arithmetic is especially important in the inversion involved in computing the solution to the optimization.

\begin{figure}
    \includegraphics[width=\linewidth]{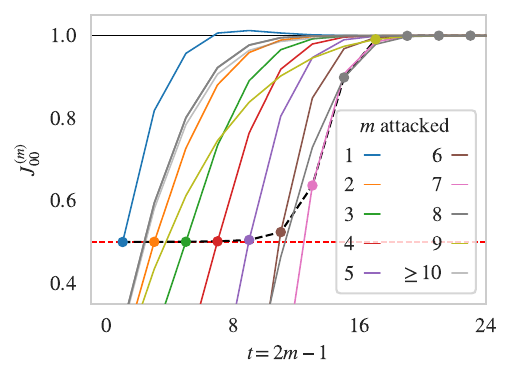}
    \caption{
        Ground-state matrix elements $J^{(m)}_{00}$ extracted using Lanczos for different examples constructed to deceive Lanczos (different color curves).
        The black horizontal line indicates the true $J_{00}=1$.
        Each example attempts to shift $J^{(m)}_{00}$ to the fake value 0.5 (dashed red line) at only a single point (indicated by the same-color marker), with regulator $\sigma=10$.
        The dashed black line through the targeted points suggests the increasingly difficulty of shifting points at greater $m$.
    }
    \label{fig:attack:lanczos}
\end{figure}

We were unable to produce a similarly pathological example as in the previous subsection.
Minimal attacks on $J^{(m)}_{00}$ for single values of $m$ provide a clear picture of the difficulty.
Note this is less difficult than attempting to shift multiple points.
\Cref{fig:attack:lanczos} shows the results of a set of experiments with the regulator $\sigma = 10$ as in the previous example.
In \cref{fig:attack:lanczos}, each curve corresponds to a different example, each attempting to shift $J^{(m)}_{00}$ to the fake value $\breve{J}_{00}=0.5$ at the indicated value of $m$.
Attacks on single estimates at small $m$ are successful. However, starting at $m \sim 5$, the values begin visibly drifting from $\breve{J}_{00}$. By $m \sim 10$, Lanczos converges to the true value and the attacks fail completely.

\begin{figure}
    \includegraphics[width=\linewidth]{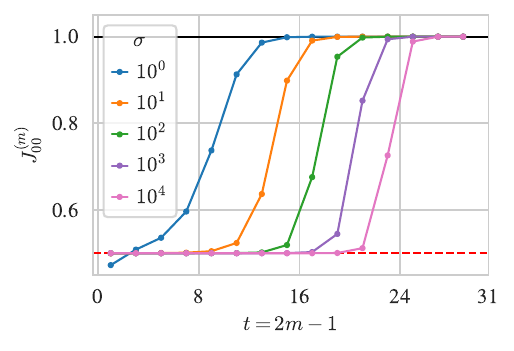}
    \caption{
        Efficacy of attacking the Lanczos estimate of the ground-state matrix element $J^{(m)}_{00}$ at different values of $m$, given different regulators $\sigma$.
        Larger values of $\sigma$ allow increasingly large and unnatural values of the true matrix elements.
        The black horizontal line indicates the true $J_{00}=1$; the dashed red line indicates the fake value 0.5.
        The orange $\sigma=10^1$ curve corresponds to the dashed black line in \cref{fig:attack:lanczos}.
    }
    \label{fig:attack:lanczos-sigmas}
\end{figure}

Increasing $\sigma$ allows more extreme values of $J_{ji}$ and thus greater freedom for fine-tuning.
\cref{fig:attack:lanczos-sigmas} shows the results of further experiments varying $\sigma$.
The data shown are now only the attacked values of $m$; the curve for $\sigma=10$ corresponds to the dashed black line in \cref{fig:attack:lanczos}.
As expected, we find that allowing more unnatural values allows deception of Lanczos at later $m$.
Increasing $\sigma$ to $10^4$ is sufficient to push convergence to nearly the maximal $m$ where Lanczos solves the finite-dimensional system exactly.
However, large hierarchies are unlikely to arise in QCD matrix elements.

We conclude this exercise by noting that it is neither robust nor exhaustive, and it is important not to overinterpret its specific results, which depend on the precise details of our problem setup and strategy.
These results should not, for example, be taken to mean that a Lanczos matrix-element extraction will always converge within 10 iterations so long as QCD matrix elements have natural values.
However, they provide strong suggestive evidence that the Lanczos method is qualitatively more robust than the methods presently in common use.

\section{Conclusions}
\label{sec:conclusions}

The Lanczos formalism is a promising new approach to analyzing lattice correlation functions.
This work demonstrates that the successes of Ref.~\cite{Wagman:2024rid} in spectroscopy extend to the task of extracting matrix elements as well.
Lanczos matrix element estimators provide useful signals for not only ground-state matrix elements but low-lying excited states as well, analogous to multi-state fits but without involving statistical fitting.
Testing in the noiseless case provides strong suggestive evidence that Lanczos estimates provide qualitatively better treatment of excited-state contamination than presently preferred methods.
In practice, the Lanczos method is also simpler to apply:
matrix elements are obtained from three-point functions by simply applying change-of-basis matrices computed from two-point functions.
This requires no statistical modeling or numerical optimization and has few analysis hyperparameters to vary.
Lanczos methods may therefore 
permit more reliable determinations of observables whose uncertainties are dominated by excited-state systematics as well as 
enable applications previously out of reach of the lattice toolkit; it is imperative to deploy them immediately so that their full capabilities may be assessed.

As discussed in Refs.~\cite{Wagman:2024rid,Ostmeyer:2024qgu,Chakraborty:2024scw,Abbott:2025yhm}, the unfiltered Ritz values obtained by the Lanczos algorithm are identical to the polynomial roots obtained by Prony's method~\cite{Prony}.
LQCD applications of Prony's method used a ``sliding window'' approach with a fixed iteration count $\leq 4$ and variable starting time~\cite{Lin:2007iq,Fleming:2009wb,Beane:2009kya,Fischer:2020bgv}, noting that the appearance of unphysical solutions obstructs straightforward applications with more iterations~\cite{Fleming:2009wb}.
Through the Lanczos correspondence, unphysical solutions can be identified as eigenvalues associated with spurious states, and the Hermitan subspace and CW test can be used to remove them.
These mathematically well-understood spurious-state filtering techniques are a primary advantage of the Lanczos perspective. Other advantages are the ability to quantify convergence rates using KPS theory and the two-sided bounds on finite-iteration approximation errors provided by the residual bound.
A further important advantage is provided by the results of this work: matrix-element estimators with analogous convergence guarantees.
We are not aware of a direct construction of analogous matrix-element estimators using Prony's method, and we note that direct application of Prony's method to one or the other time argument of a three-point function is inequivalent to the Ritz vector matrix element approach introduced here.

Importantly, many possibilities remain for improvements and extensions.
As noted in \cref{sec:lattice}, the primary sources of analysis hyperparameter dependence in the Lanczos-based method are involved in filtering spurious states and associating states between bootstrap ensembles.
Better approaches to these tasks will help improve both the reliability and precision of Lanczos spectroscopy and matrix element results.

The methodology presented here applies straightforwardly to lattice four-point functions or higher-point functions.
Such cases may be treated simply by considering the higher-point functions as three-point functions of a composite operator involving powers of transfer matrices, e.g.~$J_1 T^{\delta} J_2$ with $\delta$ the operator-operator separation.

As discussed in \cref{sec:lanczos:matrix-elements}, the Lanczos method requires three-point functions evaluated with regular spacing in sink and operator times.
This means that while existing disconnected three-point function datasets can be analyzed with Lanczos immediately, the standard strategy of generating data at only some sink times when using sequential source methods means that Lanczos will be awkward to apply to existing connected three-point datasets.
If the method proves as effective as our results suggest, data generation strategies should be adjusted to take advantage.

As noted in \cref{sec:lattice}, the approximate eigenstates resolved by Lanczos can be separated into states admitting a physical interpretation and states which are clearly noise artifacts.
The Lanczos transfer matrix approximation acts as a Hermitian operator on the physical subspace but acts with complex eigenvalues and distinct left- and right-eigenvectors on the noise artifact subspace.
This provides an exact representation of a noisy correlation function as a sum of purely decaying exponentials plus terms which oscillate to capture the effects of noise. 
The ability to distinguish spurious and non-spurious states---through identification of the Hermitian subspace and the Cullum-Willoughby test---then provides a mechanism for isolating and removing unphysical noise effects.
This provides a mechanistic explanation, complementary to the formal projection operation description discussed in Ref.~\cite{Wagman:2024rid}, for the noise properties of Lanczos estimators with large iteration counts.
The apparent convergence of Lanczos results for physical states even in the presence of statistical noise may be a manifestation of the so-called ``Lanczos phenomenon''~\cite{Cullum:1981,Cullum:1985,Druskin:1991,Parlett:1995}: the accurate convergence of an identifiable subset of Lanczos results in the face of numerical errors that might be expected to spoil the results entirely.

\begin{acknowledgments}
We thank Anthony Grebe, Dimitra Pefkou, Ryan Abbott, George Fleming, Rajan Gupta, William Jay, Fernando Romero-L\'opez, and Ruth Van de Water for stimulating discussions and helpful comments. We also thank Dimitra Pefkou for assistance preparing data for the lattice example and Phiala Shanahan for collaborating on its generation.
This manuscript has been authored by FermiForward Discovery Group, LLC under Contract No.\ 89243024CSC000002 with the U.S.\ Department of Energy, Office of Science, Office of High Energy Physics.
This research used resources of the National Energy Research Scientific Computing Center (NERSC), a U.S. Department of Energy Office of Science User Facility operated under Contract No.~DE-AC02-05CH11231.
This research used facilities of the USQCD Collaboration, which are funded by the Office of Science of the U.S. Department of Energy.
The authors thank Robert Edwards, Rajan Gupta, Balint Jo{\'o}, Kostas Orginos, and the NPLQCD collaboration for generating the ensemble used in this study.
The Chroma~\cite{Edwards:2004sx}, QUDA~\cite{Clark:2009wm,Babich:2011np,Clark:2016rdz}, QDP-JIT~\cite{6877336}, and Chromaform~\cite{chromaform} software libraries were used to generate the data in this work, as well as code adapted from LALIBE~\cite{lalibe} including the hierarchical probing implementation by Andreas Stathopoulos~\cite{Stathopoulos:2013aci}. Numerical analysis was performed using NumPy~\cite{harris2020array}, SciPy~\cite{2020SciPy-NMeth}, pandas~\cite{jeff_reback_2020_3715232,mckinney-proc-scipy-2010}, lsqfit~\cite{peter_lepage_2020_4037174}, gvar~\cite{peter_lepage_2020_4290884}, and mpmath~\cite{mpmath}.
Figures were produced using matplotlib~\cite{Hunter:2007} and seaborn~\cite{Waskom2021}.
\end{acknowledgments}

\appendix

\section{Algorithm summary}
\label{app:algo}

In this Appendix, we detail the precise sequence of operations used to compute the results presented in the application to noisy lattice data of \cref{sec:lattice}.
However, we emphasize that this does not represent the unique or best possible implementation of a Lanczos analysis.
For example, as discussed in \cref{sec:lanczos}, different choices of oblique Lanczos convention are possible and may provide better or worse sensitivity to numerical precision issues.
Separately, different implementations of the CW test are possible; for example, Ref.~\cite{Wagman:2024rid} provides a different procedure to choose the CW cut.
Uncertainty quantification in the presence of outliers is not uniquely defined, and different approaches than the nested bootstrap median estimators employed in the main text and the outlier robust estimators of \cref{app:lattice-std} are possible.
These methods are new and their usage is still being explored; the version presented herein should not be considered the ultimate implementation.
With that understood, we may proceed with the description of the algorithm.

The algorithm as defined here is executed separately for each (inner) bootstrap draw and for each fixed number of Lanczos iterations $m$.
To compute the input data for a given bootstrap draw and $m$:
\begin{itemize}[leftmargin=*]
    \item Take the ensemble average over all configurations in the bootstrap draw to obtain the three-point correlator $C^\mathrm{3pt}(t)$ and its corresponding initial-state and final-state two-point correlators, $C(t)$ and $C'(t)$, respectively.
    \item Take the real part of each two-point correlator.
    \item Trim to take the first $2m$ points of each two-point correlator $C(t)$ and $C'(t)$, and the leading $m \times m$ part of $C^\mathrm{3pt}(\sigma, \tau)$ (i.e., restrict to $t \in 0,\ldots,2m-1$ and $\sigma, \tau \in 0,\ldots,m-1$.
\end{itemize}

With the input data computed, the algorithm proceeds by first analyzing each of $C(t)$ and $C'(t)$ separately.
We consider first an analysis of the initial-state correlator $C(t)$. The steps are as follows:
\begin{itemize}[leftmargin=*]
    \item Evaluate the Lanczos recursion defined by \cref{eq:lanczos:recursion-1,eq:lanczos:recursion-2,eq:lanczos:recursion-3,eq:lanczos:recursion-4,eq:lanczos:recursion-5} to obtain $\alpha_i$, $\beta_i$, and $\gamma_i$ for all $i \in 1, \ldots, m$, using the oblique convention $\beta_j \equiv \gamma_j \equiv \sqrt{\braket{r^L_j|r^R_j}}$.
    \item Use $\alpha_i$, $\beta_i$, and $\gamma_i$ to construct the tridiagonal matrix $T^{(m)}_{ij}$ as defined in \cref{eq:lanczos:Tij-explicit}.
    \item Diagonalize $T^{(m)}_{ij}$ per \cref{eq:lanczos:Tij-explicit} to obtain the (unfiltered) Ritz values $\lambda^{(m)}_k$ for $k \in 0, \ldots, m-1$ and the eigenvector matrix $\omega^{(m)}_{jk}$. 
    \item Obtain $(\omega^{-1})^{(m)}$ by inverting $\omega^{(m)}$.
    \item Evaluate the auxiliary recursion defined by \cref{eq:krylov-coeff-start,eq:krylov-coeff-recursion} to obtain the right/left ($R/L$) Krylov coefficient matrices, $K^{R/L}_{jt}$, defined for all $j \in 1,\ldots,m$ and $t \in [0,m-1]$.
    \item Compute the Ritz vector norms $|\mathcal{N}^{(m)}_k|^2 = \frac{ (\omega^{-1})^{(m)*}_{k1} }{ \omega^{(m)}_{1k} }$; note that the value computed will be complex, negative, or zero for non-Hermitian states (see below).
    \item Combine $K^{R/L}$, $\mathcal{N}^{(m)}_k$, and $\omega$ to obtain the Ritz coefficients $P^{R/L}_{kt}$ per \cref{eq:ritz-coeffs}.
\end{itemize}

Spurious state filtering requires additional steps, which defines a filtered ``non-spurious'' subset $\{k\}_{\overline{\mathcal{S}}}$ of the indices $k \in [0,m-1]$ to retain.
The methods prescribed in this work are Hermitian subspace filtering and the CW test.
To apply the CW test, some precomputation is necessary:
\begin{itemize}[leftmargin=*]
    \item Construct the matrix $\tilde{T}^{(m)}_{ij}$ by knocking out the first row and column of $T^{(m)}_{ij}$, i.e., by taking $\tilde{T}^{(m)}_{ij} \equiv T^{(m)}_{i+1,j+1}$ for all $i,j \in [1,m-1]$.
    \item Diagonalize $\tilde{T}^{(m)}_{ij}$ to obtain $\tilde{\lambda}^{(m)}_l$, defined for all $l \in [0,m-2]$.
    \item Construct for each $k$ the CW distance $\Delta^\mathrm{CW(m)}_k = \min_l |\lambda_k - \tilde{\lambda}_l|$.
    \item Choose a CW cut $\epsilon^\mathrm{CW(m)}$ using \cref{eq:lattice:cw-cut-heuristic} with $a=10$ and $b=1$.
\end{itemize}
The full filtering prescription is then to retain any $k$ for which all of the following conditions hold:
\begin{itemize}[leftmargin=*]
    \item Hermitian subspace (1): $\lambda^{(m)}_k$ is real within numerical precision as diagnosed by e.g.~$|\mathrm{Im}[\lambda_k] / \lambda^{(m)}_k | < 10^{-8}$.
    \item Hermitian subspace (2; ``norm trick''): $\frac{ (\omega^{-1})^{(m)*}_{k1} }{ \omega^{(m)}_{1k} }$ is similarly real within numerical precision, and its real part is positive.
    \item CW test: $\Delta^\mathrm{CW(m)}_k > \epsilon^\mathrm{CW(m)}$.
\end{itemize}
The subset of indices $\{k\}_{\overline{\mathcal{S}}}$ surviving filtering are then identified with different states by simply sorting on $\lambda_k$.
The surviving $k$ with the largest $\lambda_k$ is identified as the ground state, the next-largest with the first excited state, etc.
The energy estimators for this bootstrap draw and iteration $m$ are defined as $E^{(m)}_k = - \log \lambda^{(m)}_k$.

In the general case when $C^\mathrm{3pt}$ is evaluated with distinct initial and final interpolators, the steps above must be repeated independently for the final-state correlator $C'(t)$ to obtain primed quantities, most importantly the (left) Ritz coefficients ${P'}^{L(m)}_{k't}$.
Spurious state filtering similarly defines a subset of the final-state indices to retain, $\{k'\}_{\overline{\mathcal{S}}}$, appropriately sorted to identify each $k'$ with physical states.
This selection is entirely independent of the initial-state subset $\{k\}_{\overline{\mathcal{S}}}$.
A different number of initial- and final-state indices may survive filtering.
In the subcase of a symmetric $C^\mathrm{3pt}$ with identical initial and final interpolators, the corresponding initial- and final-state correlators coincide, i.e.~$C(t) = C'(t)$, and thus so do all other unprimed and primed quantities---only a single two-point analysis is necessary.

Once initial and final state Ritz coefficients are evaluated---specifically, ${P'}^{L(m)}_{k't}$ and $P^{R(m)}_{kt}$---then matrix elements can be computed by simply evaluating \cref{eq:lanczos:mx-elts}, i.e.,
\begin{equation}
    J^{(m)}_{k'k} = \sum_{\sigma\tau}  {P'}^{L(m)}_{k' \sigma} 
    \frac{C^\mathrm{3pt}(\sigma, \tau)}{\sqrt{C'(0) C(0)}}
    P^{R(m)}_{k \tau} ~ .
\end{equation}
The result is a separate estimator for each of the matrix elements $\braket{k'|J|k}$ for all non-spurious states $\{k\}_{\overline{\mathcal{S}}}$ and $\{k'\}_{\overline{\mathcal{S}}}$.

To estimate uncertainties via bootstrapping, the algorithm must be executed over many different bootstrap draws.
The resulting set of bootstrapped estimates of $E^{(m)}_k$, ${E'}^{(m)}_{k'}$, and $J^{(m)}_{k'k}$ are then combined to produce estimates of the values and uncertainties.
Their central values and uncertainties as presented in \cref{sec:lattice} are computed using the nested bootstrap median approach of Ref.~\cite{Wagman:2024rid} discussed therein.
\Cref{app:lattice-std} instead uses the  standard outlier-robust estimators defined therein---i.e., the median along with the confidence-interval construction defined in \cref{eq:lattice-std:outlier-robust-errs}.
This entire procedure may similarly be repeated for different $m$ to obtain a sequence of estimates at different numbers of Lanczos iterations.
We note that in practice, computing results for all different $m \in [1, N_t/2]$ simultaneously allows substantial reduction of repeated computation; the procedure is defined here independently for each $m$ only for clarity.

\section{Where to use high-precision arithmetic}
\label{app:precision}

Applying the Lanczos methods described here sometimes requires high-precision arithmetic to avoid numerical instabilities.
This Appendix discusses where this is necessary to obtain the results presented above.
We implement this using the \texttt{mpmath} Python library for multiple-precision arithmetic~\cite{mpmath}.
We find 100 decimal digits of precision is sufficient to produce the results of this paper, but have made no effort to determine the minimum required.
We otherwise work in double precision.

In practice, higher-than-double precision is required primarily for the noiseless examples in \cref{sec:noiseless} and \ref{sec:attack}.
In particular, it is necessary in:
\begin{itemize}
    \item The sums over states when constructing the example two- and three-point functions;
    \item The recursions to construct the tridiagonal matrix coefficents $\alpha_j, \beta_j, \gamma_j$ from the two-point correlator;
    \item The recursions to compute the Krylov coefficients $K^{R/L}$;
    \item Computing the eigenvalues and eigenvectors of $T^{(m)}$;
    \item Matrix multiplications to construct the Ritz coefficients $P^{(m)}$, compute observables like $Z^{(m)}$ and $J^{(m)}$.
\end{itemize}
This amounts to everything except for the inversion of the eigenvector matrix $\omega^{(m)}$, which may be carried out in double precision.

In the lattice example of \cref{sec:lattice}, we find that high-precision arithmetic is only important for the initial recursion to construct $\alpha_j, \beta_j, \gamma_j$, which is relatively inexpensive.
Crucially, the tasks which dominate the computationally cost may be carried out in only double precision: computing the eigenvalues/vectors of $T^{(m)}$, inverting the eigenvector matrix $\omega^{(m)}$, and the various matrix multiplications.
As implemented for this work, running the full procedure 200 times for each bootstrap ensemble to produce the results of \cref{sec:lattice} takes $\approx 2$ minutes on a c.~2019 Intel MacBook Pro.
We caution that high-precision arithmetic may become more necessary for larger lattices and/or different parameters.

\section{Ritz projectors}
\label{app:projectors}

Although the right Ritz rotator
\begin{equation}
    P^{R(m)}_k = \sum_{t=0}^{m-1} P^{R(m)}_{kt} T^t,
\end{equation}
allows construction of the right Ritz vectors as
\begin{equation}
    P^{R(m)}_k \ket{v^R_1} = \ket{y^{R(m)}_k} ~,
\end{equation}
it is not a projection operator of the form $\ket{y^{R(m)}_k} \bra{y^{L(m)}_k}$.
This is straightforward to see: it is a finite polynomial in $T$, which has support outside the Krylov subspace spanned by the Ritz vectors.
However, as we show in this appendix, the equivalent operator constructed with
\begin{equation}
    T^{(m)} = \sum_{i,j=1}^{m} \ket{v^R_i} T^{(m)}_{ij} \bra{v^L_j}
\end{equation}
in place of $T$ is an unnormalized projector, i.e.,
\begin{equation}
    \mathcal{P}^{R(m)}_k 
    = \sum_{t=0}^{m-1} P^{R(m)}_{kt} [T^{(m)}]^t
    = \frac{ \ket{y^{R(m)}_k} \bra{y^{L(m)}_k} }{ \braket{y^{L(m)}_k | v^R_1} } ~.
    \label{eq:projectors:ritz-projector-def}
\end{equation}
Similar arguments apply for the left Ritz rotator $P^{L(m)}_k$.

To prove \cref{eq:projectors:ritz-projector-def}, first note that by the definition of matrix exponentiation,
\begin{equation}\begin{aligned}
    \mathcal{P}^{R(m)}_k 
    &= \sum_t P^{(m)}_{kt} [T^{(m)}]^t
    \\&= \sum_l \ket{y^{R(m)}_l} \sum_t P^{R(m)}_{kt} [\lambda_l^{(m)}]^t \bra{y^{L(m)}_l}
    \\&\equiv \sum_l \ket{y^{R(m)}_l} q^{(m)}_{kl} \bra{y^{L(m)}_l} ~ .
    \label{eq:projectors:q-def}
\end{aligned}\end{equation}
What remains is to show that the symbol $q^{(m)}_{kl}$ defined in the last line is diagonal in $k,l$ and normalized as claimed.

To proceed, note that by construction,
\begin{equation}
    [T - T^{(m)}] \ket{v^R_j} = \delta_{jm} \gamma_{m+1} \ket{v^R_{m+1}},
    \label{eq:projectors:T-Tm_vR}
\end{equation}
for all $j \leq m$, from which follows
\begin{equation}
    T^t \ket{v^R_1} = [T^{(m)}]^t \ket{v^R_1},
    \label{eq:projectors:Tt_vR1_eq_Tmt_vR1}
\end{equation}
for all $t < m$.
To see this, note that we can write
\begin{equation}
    T^t \ket{v^R_1} = \sum_{j=1}^{t+1} \ket{v^R_j} c_{jt}
    \label{eq:projectors:T_v1_sum}
\end{equation}
for some coefficients $c_{jt}$.
In principle these coefficients---related to the matrix inverse of the right Krylov coefficients $K^R_{jt}$---could be computed, but their values are irrelevant for the proof.
Each application of $T$ populates one higher Lanczos vector in the sum, i.e.,
\begin{equation}
    T T^t \ket{v^R_1} 
    = \sum_{j=1}^{t+1} T \ket{v^R_j} c_{jt} 
    = \sum_{j=1}^{t+2} \ket{v^R_j} c_{j(t+1)}.
    \label{eq:projectors:T-on-sum}
\end{equation}
The action of $T^{(m)}$ on this sum is identical to that of $T$ as long as $t \leq m-2$:
\begin{equation} \begin{aligned}
    T^{(m)} \sum_{j=1}^{t+1} \ket{v^R_j} c_{jt}
    &= \sum_{j=1}^{t+1} \bigg(
        T \ket{v^R_j} - \delta_{jm} \gamma_{m+1} \ket{v^R_{m+1}} 
    \bigg) c_{jt}
    \\ &= \sum_{j=1}^{t+1} T \ket{v^R_j} c_{jt}
    = T^{t+1} \ket{v^R_1} ,
\end{aligned}\end{equation}
where in the first equality we use \cref{eq:projectors:T-Tm_vR},
in the second uses that $\delta_{jm}=0$ for all $j \leq t+1 \leq m-1$ in the sum,
and the third uses \cref{eq:projectors:T-on-sum}.
Together with $T\ket{v^R_1} = T^{(m)}\ket{v^L_1}$, \cref{eq:projectors:Tt_vR1_eq_Tmt_vR1} follows by induction.

We can see from \cref{eq:projectors:Tt_vR1_eq_Tmt_vR1} that $P^{R(m)}_k$ and $\mathcal{P}^{R(m)}_k$ have identical action on $\ket{v^R_1}$, i.e.,
\begin{equation}\begin{aligned}
    \mathcal{P}^{R(m)}_k \ket{v^R_1}
    &= \sum_{t=0}^{m-1} P^{R(m)}_{kt} [T^{(m)}]^t \ket{v^R_1}
    \\&= \sum_{t=0}^{m-1} P^{R(m)}_{kt} T^t \ket{v^R_1}
    \\&= P^{R(m)}_k \ket{v^R_1} = \ket{y^{R(m)}_k},
\end{aligned}\end{equation}
because the sum over $t$ runs only to $m-1$.
Inserting \cref{eq:projectors:q-def}, we see that
\begin{equation}\begin{aligned}
    \mathcal{P}^{R(m)}_k \ket{v^R_1} 
    &= \sum_l \ket{y^{R(m)}_l} q^{(m)}_{kl} \braket{y^{L(m)}_l | v^R_1} = \ket{y^{R(m)}_k}
    \\ &\equiv \sum_l \alpha_{kl} \ket{y^{R(m)}_l} = \ket{y^{R(m)}_k}  ~ .
\end{aligned}
\end{equation}
Because the right Ritz vectors are linearly independent, it must be that
\begin{equation}
    \alpha_{kl} \equiv q^{(m)}_{kl} \braket{y^{L(m)}_l | v^R_1} = \delta_{kl} ~ ;
\end{equation}
no superposition of $\ket{y^{R(m)}_l}$ with $l \neq k$ has extent along $\ket{y^{R(m)}_k}$, so the term with $l=k$ must saturate the sum.
Because the factor $\braket{y^{L(m)}_l | v^R_1}$ has no dependence on $k$, it must be that $q^{(m)}_{kl} \propto \delta_{kl}$.
Separately, the factor $\braket{y^{L(m)}_l | v^R_1} = Z^{L(m)}_l / |\psi|$ (see \cref{sec:lanczos:overlaps}) and is generically nonzero for all $l$.
Given these constraints, it can only be that
\begin{equation}
    q_{kl}^{(m)} = \sum_t P^{R(m)}_{kt} [\lambda_l^{(m)}]^t 
    = \frac{\delta_{kl} }{ \braket{y^{L(m)}_l | v^R_1} },
\end{equation}
and \cref{eq:projectors:ritz-projector-def} holds as claimed.

\section{Correlator decomposition}
\label{app:corr-decomp}

This Appendix proves \cref{eq:lattice:corr-decomp} from the main text, i.e.~that Lanczos quantities provide an exact decomposition of any correlator $C(t)$ of the form
\begin{equation}
    C(t) = \sum_k Z^{R(m)*}_k Z^{L(m)}_k  (\lambda^{(m)}_k)^t ~ ,
\label{eq:corr-decomp:corr-decomp}
\end{equation}
which holds for all $t \leq 2m-1$, i.e., the full extent of the correlator incorporated after $m$ steps.
While the discussion in the main text primarily addressed diagonal correlators, the proof is equally straightforward for the more general case of an off-diagonal correlator,
\begin{equation}
    C(t) = \braket{\chi | T^t | \psi} \, ,
\end{equation}
recovering the diagonal case when $\ket{\chi} = \ket{\psi}$.

Under exponentiation,
\begin{equation}
    [T^{(m)}]^t = \sum_k \ket{y^{R(m)}_k} (\lambda^{(m)})^t \bra{y^{L(m)}_k} ,
\end{equation}
because $\braket{y^{R(m)}_k | y^{L(m)}_l} = \delta_{kl}$ by construction, so
\begin{equation}\begin{aligned}
    \braket{\chi | [T^{(m)}]^t | \psi} 
    &= \sum_k \braket{\chi | y^{R(m)}_k} (\lambda^{(m)})^t \braket{y^{L(m)}_k | \psi}
    \\
    &= \sum_k Z^{R(m)*}_k (\lambda^{(m)})^t Z^{L(m)}_k,
\end{aligned}\end{equation}
where $Z^{R(m)}_k = \braket{y^{R(m)}_k | \chi}$ and $Z^{L(m)}_k = \braket{y^{L(m)}_k | \psi}$.
Recall also that 
\begin{equation}
    \bra{v^L_1} \equiv \frac{\bra{\chi}}{\sqrt{\braket{\chi|\psi}}}
    \text{  and   }
    \ket{v^R_1} \equiv \frac{\ket{\psi}}{\sqrt{\braket{\chi|\psi}}} ~ ,
\end{equation}
i.e.~that $\ket{\psi}$ and $\ket{\chi}$ are simply related to $\ket{v^{R/L}_1}$ by a known constant $\braket{\chi|\psi} = C(0)$.
These results immediately establish that $\braket{v^L_1 | [T^{(m)}]^t | v^R_1}$ is proportional to the desired decomposition.
The nontrivial part of proving \cref{eq:corr-decomp:corr-decomp} is to show that 
\begin{equation}
    \braket{v^L_1 | T^t | v^R_1} = \braket{v^L_1 | [T^{(m)}]^t | v^R_1},
\label{eq:corr-decomp:vv-equality}
\end{equation}
for all $t \leq 2m-1$, to which we turn next.

\begin{figure*}[p]
    \includegraphics[width=\linewidth]{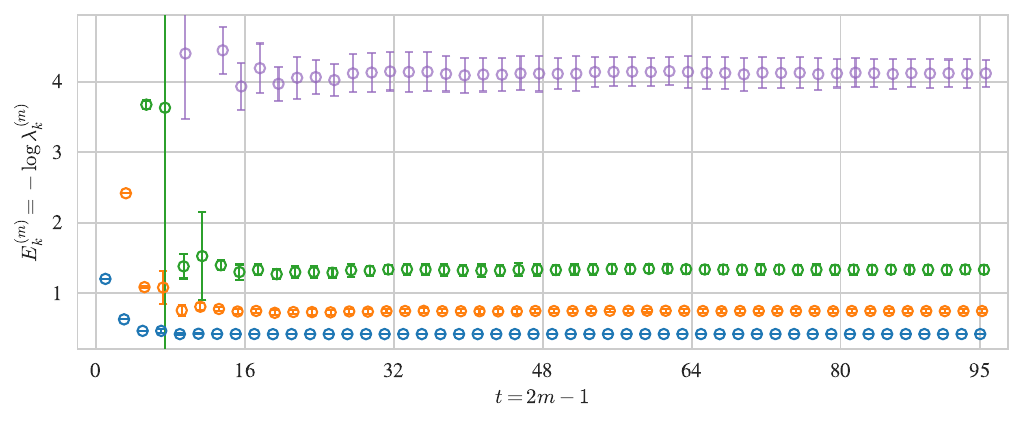}
    \caption{
        Spectrum extracted by Lanczos from the zero-momentum nucleon two-point correlator including the partially-resolved third excited state (cf.~\cref{fig:lattice:spectrum}).
        Blue, orange, green, and pale purple markers correspond to the ground state and first three excited states.
        For the third excited state, outer bootstrap estimates are computed as medians over whatever inner bootstrap values are present, allowing up to 32\% missingness and undefined otherwise.
        Uncertainties on the third excited state then correspond to $\approx 68\%$ confidence intervals computed after symmetrically setting missing outer bootstrap values to $\pm \infty$.
        Uncertainties are computed as in \cref{fig:lattice:spectrum} otherwise.
    }
    \label{fig:lattice:spectrum-4state}
\end{figure*}

\begin{figure*}[p]
    \includegraphics[width=\linewidth]{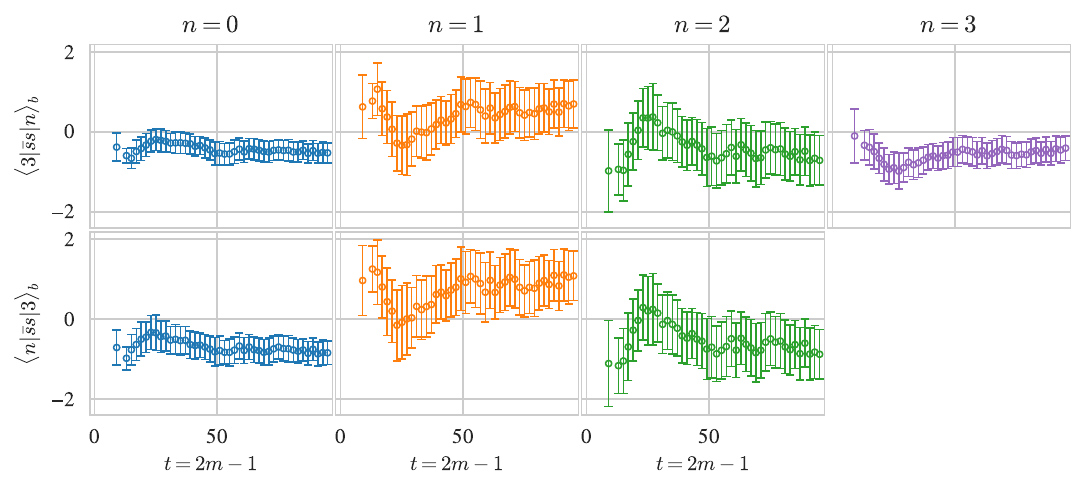}
    \caption{
        Lanczos extractions of forward matrix elements of the strange scalar current involving the partially resolved third excited state in the nucleon spectrum (cf.~\cref{fig:lattice:MEs}).
        Note that eigenvectors are unit-normalized, not relativistically normalized. 
        For the third excited state, outer bootstrap estimates are computed as medians over whatever inner bootstrap values are present, allowing up to 32\% missingness and undefined otherwise.
        Uncertainties on the third excited state then correspond to $\approx 68\%$ confidence intervals computed after symmetrically setting missing outer bootstrap values to $\pm \infty$.
        Uncertainties are computed as in \cref{fig:lattice:spectrum} otherwise.
    }
    \label{fig:lattice:MEs-4state}
\end{figure*}

The previous Appendix proved \cref{eq:projectors:Tt_vR1_eq_Tmt_vR1}, i.e.~that
\begin{equation}
    T^t \ket{v^R_1} = [T^{(m)}]^t \ket{v^R_1},
\label{eq:corr-decomp:T-vR-eq}
\end{equation}
for all $t \leq m-1$, and identical arguments give
\begin{equation}
    \bra{v^L_1} T^t  = \bra{v^L_1} [T^{(m)}]^t,
\label{eq:corr-decomp:vL-T-eq}
\end{equation}
as well.
We see immediately that \cref{eq:corr-decomp:vv-equality} holds for all $t \leq 2m-2$, because $T^t$ can be factored as $T^t = T^{s_0} T^{t_0}$ with both $s_0 \leq m-1$ and $t_0 \leq m-1$.
In more detail, the derivation proceeds as
\begin{equation}\begin{aligned}
    \braket{v^L_1 | T^t | v^R_1}
    &= \braket{v^L_1 | T^{s_0} T^{t_0} | v^R_1}
    \\
    &= \braket{v^L_1 | [T^{(m)}]^{s_0} [T^{(m)}]^{t_0} | v^R_1}
    \\
    &= \braket{v^L_1 | [T^{(m)}]^{t} | v^R_1},
\end{aligned}\end{equation}
where $t = s_0+t_0$ in the first equality,
we choose some $s_0,t_0 \leq m-1$ and apply \cref{eq:corr-decomp:T-vR-eq,eq:corr-decomp:vL-T-eq} in the second,
and combine terms in the third.

All that remains is to show that \cref{eq:corr-decomp:vv-equality} holds in the edge case $t = 2m-1$.
Given the arguments of the previous paragraph, 
it is sufficient to show that
\begin{equation}
    \braket{v^L_1 | T^{m-1} ~ [T - T^{(m)}] ~ T^{m-1} | v^R_1} = 0.
\end{equation}
First, note that we can rewrite the LHS as
\begin{equation}
    \sum_{j,l=1}^{m} c^{L}_{(m-1)j} \braket{v^L_j | [T - T^{(m)}] |v^R_l} c^R_{l(m-1)},
\end{equation}
where $c^R_{jt}$ are the same coefficients as in \cref{eq:projectors:T_v1_sum} and $c^L_{tj}$ are their left equivalents.
Following similar reasoning as used to prove \cref{eq:projectors:Tt_vR1_eq_Tmt_vR1} gives
\begin{equation}
\begin{aligned}
    [T-T^{(m)}] \sum_{j=1}^m \ket{v^R_j} c^R_{jt}
    &= \sum_{j=1}^m \delta_{jm} \gamma_{m+1} \ket{v^R_{m+1}} c^R_{jt}
    \\
    &= \ket{v^R_{m+1}} \gamma_{m+1} c^R_{mt} ~ ,
\end{aligned}
\end{equation}
evaluating the sum with the delta function in the last equality.
We thus find, as desired,
\begin{equation}\begin{aligned}
    &\sum_{j,l=1}^{m} c^{L}_{(m-1)j} \braket{v^L_j | [T-T^{(m)}] |v^R_l} c^R_{l(m-1)}
    \\
    = &\sum_{j=1}^m c^{L}_{(m-1)j} \braket{v^L_j | v^R_{m+1}} \gamma_{m+1} c^R_{m(m-1)} = 0 ,
\end{aligned}\end{equation}
where the final equality is simply because $\braket{v^L_j | v^R_{m+1}} = \delta_{j(m+1)}$ and the sum runs only to $j=m$.
This completes the proof that \cref{eq:corr-decomp:corr-decomp} holds for all $t \leq 2m-1$.

\section{Results for third excited state}
\label{app:3es}

The analysis presented in the main text fully resolves three states, meaning specifically that at least three states survive filtering in each bootstrap ensemble. 
However, values are available for a fourth state in $\approx 80\%$ of inner bootstraps.
To construct median estimators, we relax the definition of the median over inner bootstraps to allow up to 32\% (i.e.~$1\sigma$) of values to be missing, taking the median over values present.
This produces estimates for the third excited state in $\approx 90\%$ of outer bootstrap ensembles ($m$-dependently).
While insufficient to compute like-in-kind estimates to compare with those for the other three states, it is interesting to look at the results using some reasonable assumptions.
Note that the number of states resolved and present in each bootstrap will vary for different schemes to filter states and associate them between bootstraps.

It is not generally possible to do rigorous statistics with missing data if the mechanism which causes missingness is not understood.
This is the case here.
However, we may make a reasonable choice: we assume all missing data are outliers and also equally likely to be high- or low-valued.
In this case, the median and $1\sigma$ confidence interval definitions of values and errors as used in \cref{app:lattice-std} remain well-defined, as long as measurements are available for $\gtrsim 68\%$ of outer bootstraps.
Note that these assumptions are inequivalent to and more conservative than the assumption that missingness is uncorrelated with value. This prescribes computing whatever estimators only on the non-missing subset of data, which will compress the width of the uncertainties.

Under these assumptions, we may compute various observables involving the third excited state as well.
\Cref{fig:lattice:spectrum-4state} shows the spectrum including its energy.
It is not clear that this state is physical, as its mass is near the expected second layer of doublers~\cite[Ch.~5]{Gattringer:2010zz}.
\Cref{fig:lattice:MEs-4state} shows matrix elements involving the third excited state.
These are not substantially noisier than those for the lower three states in \cref{fig:lattice:MEs}, but we emphasize that this comparison is not between quantities defined equivalently.

\section{Analysis with standard estimators}
\label{app:lattice-std}

\begin{figure}
    \includegraphics[width=\linewidth]{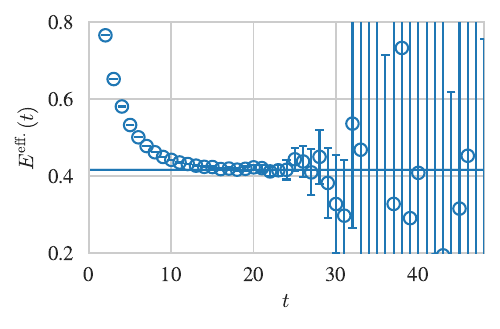}
    \includegraphics[width=\linewidth]{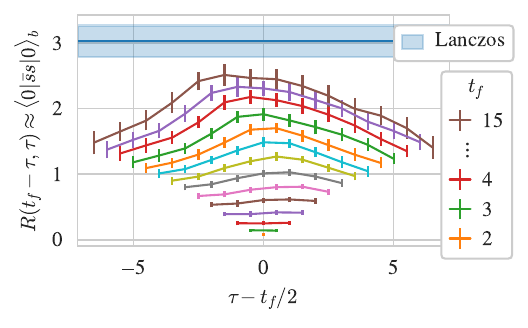}
    \caption{
        Data for the nucleon strange scalar current example as in \cref{fig:lattice:data} but using the outlier-robust uncertainty estimators described in this appendix instead of nested bootstrap uncertainties. 
        The ``Lanczos'' band is the same fit to Lanczos estimates as shown in \cref{fig:lattice-std:gsme}.
    }
    \label{fig:lattice-std:data}
\end{figure}

\begin{figure*}[p]
    \includegraphics[width=\linewidth]{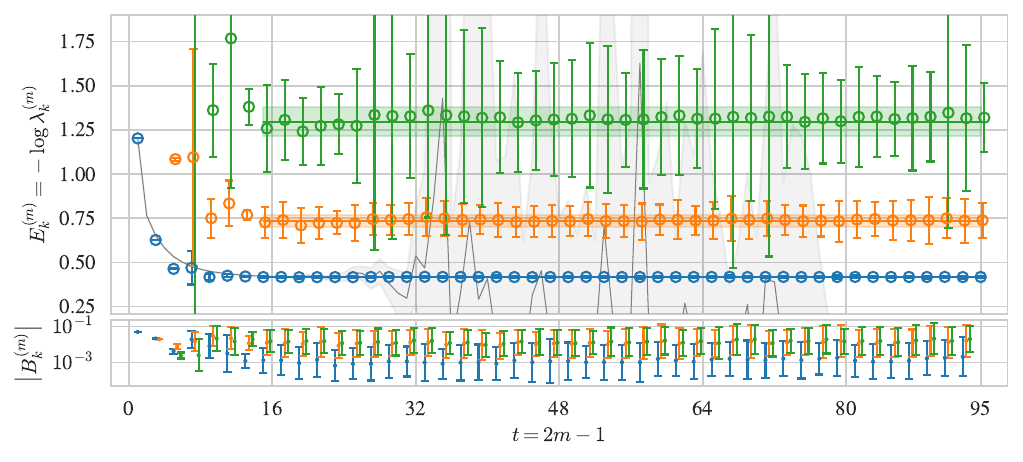}
    \caption{
        Spectrum extracted by Lanczos from the zero-momentum nucleon two-point correlator (top) and noisy estimates of the bounding values $\left| B_k^{(m)} \right|$ (\cref{eq:lanczos:residual-B}) as in \cref{fig:lattice:spectrum} but using the outlier-robust uncertainty estimators described in this appendix instead of nested bootstrap uncertainties.
        Blue, orange, and green markers correspond to the ground, first excited, and second excited state respectively.
        The blue, orange, and green bands indicate fits of a constant to the values they cover, finding $E_k = [0.4175(17), 0.736(34), 1.296(83)]$ for $k=[0,1,2]$.
        Uncertainties on the fitted values are linearly propagated from the covariance matrix of the $\lambda^{(m)}$.
        The fact that the central values fluctuate less than their error bars arises from the use of outlier-robust estimators without nested bootstrap uncertainties.
        Uncertainties on energies are propagated linearly through $E_k^{(m)} = -\log \lambda_k^{(m)}$ to avoid issues with artificial missingness due to negative arguments of logarithms for noisier points. 
    }
    \label{fig:lattice-std:spectrum}
\end{figure*}

\begin{figure*}
    \includegraphics[width=\linewidth]{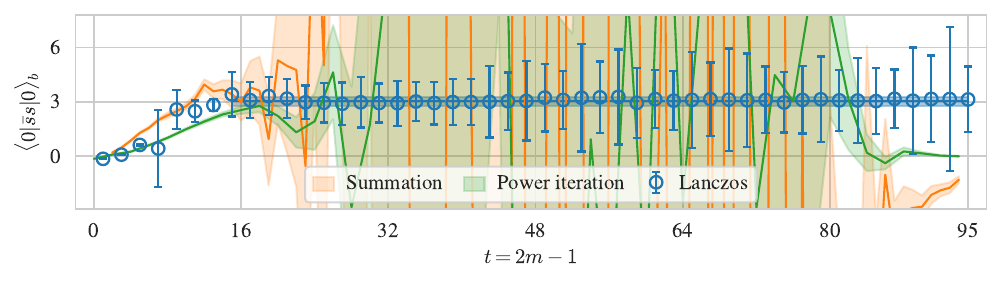}
    \caption{
        Ground-state bare nucleon matrix element of the strange scalar current with zero momentum transfer as in \cref{fig:lattice:gsme} but using the outlier-robust uncertainty estimators described in this appendix instead of nested bootstrap uncertainties.
        The blue band corresponds to a fit of a constant to the Lanczos estimates it covers, finding $J_{00} = 3.03(24)$.
        Uncertainties on the fit to Lanczos values, are linearly propagated from the covariance matrix of the $J^{(m)}_{00}$.
    }
    \label{fig:lattice-std:gsme}
\end{figure*}

\begin{figure*}
    \includegraphics[width=\linewidth]{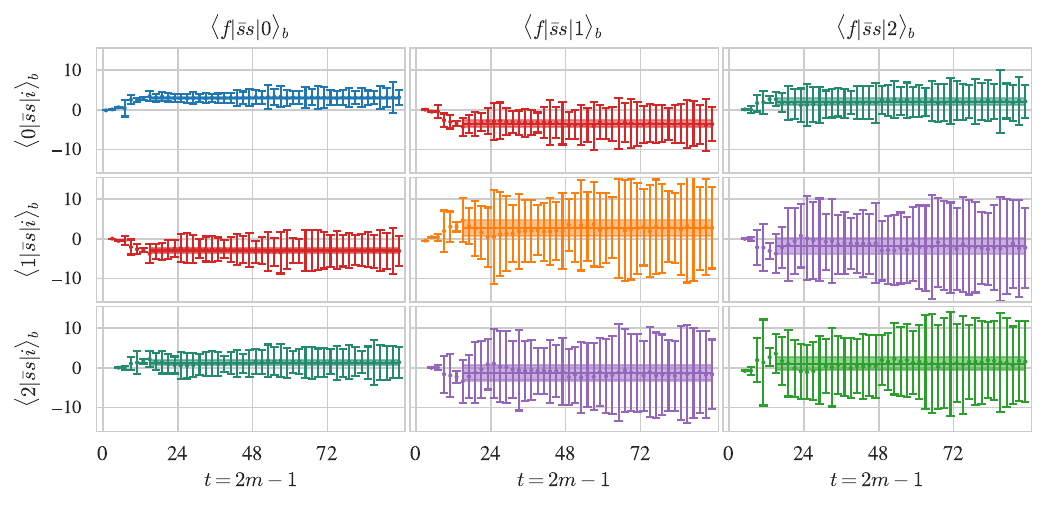}
    \caption{
        Lanczos extractions of bare forward matrix elements $\braket{f'|J|i}_b$ of the strange scalar current for three low-lying states in the nucleon spectrum as in \cref{fig:lattice:MEs} but using the outlier-robust uncertainty estimators described in this appendix instead of nested bootstrap uncertainties.
        The bands corresponds to fits of a constant to the $J^{(m)}_{fi}$ that they cover, with values listed in \cref{tab:lattice-std:MEs}.
        Fit uncertainties are linearly propagated from the data covariance matrix.
    }
    \label{fig:lattice-std:MEs}
\end{figure*}

\begin{figure}
    \includegraphics[width=\linewidth]{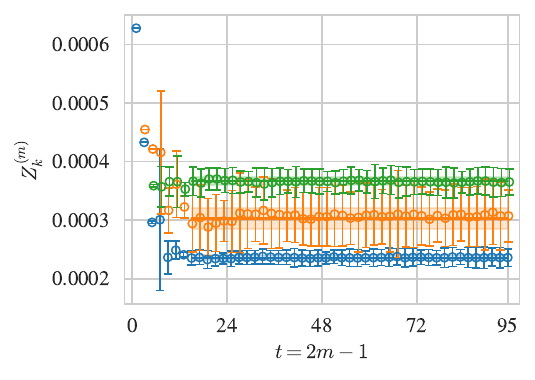}
    \caption{
        Overlap factors $Z^{(m)}$ extracted by Lanczos from the zero-momentum nucleon two-point correlator
        as in \cref{fig:lattice:Zs}
        but using the outlier-robust uncertainty estimators described in this appendix instead of nested bootstrap uncertainties.
        Blue, orange, and green markers correspond to the ground, first excited, and second excited state respectively.
        The blue, orange, and green bands indicate fits of a constant to the values they cover, finding $Z_k = [2.360(41), 3.03(18), 3.670(66)] \times 10^{-4}$ for $k=[0,1,2]$.
        Uncertainties on the fitted values are linearly propagated from the covariance matrix of the $Z^{(m)}$.
    }
    \label{fig:lattice-std:Zs}
\end{figure}

Results in the main text use bootstrap median estimators with nested bootstrap uncertainties.
For effective masses and other PI estimators, this leads to negligible differences compared to values and uncertainties computed with the sample-mean or with standard outlier-robust median/confidence interval estimators.
Conversely, for Lanczos results this leads to qualitatively different uncertainties and correlations for large $m$.
To illustrate these differences, this appendix shows alternative versions of all figures without nested bootstrap uncertainty estimates, instead computed using standard \texttt{gvar}-style outlier-robust estimators as described below.
In particular, 
\begin{itemize}
    \item \cref{fig:lattice-std:data} shows effective mass and matrix elements results analogous to \cref{fig:lattice:data}. The ``Lanczos'' band in the second subplot corresponds to a fit of Lanczos matrix element estimates in \cref{fig:lattice-std:gsme}.
    \item \cref{fig:lattice-std:spectrum} shows Lanczos energy estimators analogous to \cref{fig:lattice:spectrum}.
    \item \cref{fig:lattice-std:gsme} shows ground-state matrix element results analogous to \cref{fig:lattice:gsme}.
    \item \cref{fig:lattice-std:MEs} shows ground- and excited-state matrix element results analogous to \cref{fig:lattice:MEs}.
    \item \cref{tab:lattice-std:MEs} tabulates matrix element results analogous to \cref{tab:lattice:MEs}.
    \item \cref{fig:lattice-std:Zs} shows overlap factor results analogous to \cref{fig:lattice:Zs}.
\end{itemize}

Throughout this appendix, we use the de-facto lattice community standard for outlier-robust estimators as implemented in the \texttt{gvar} software package~\cite{peter_lepage_2020_4290884}: we take the median rather than the mean, an estimator based on the width of the $1\sigma$ confidence interval (CI) rather than the standard deviation,\footnote{
Specifically, we use
$\sigma = \max(\Delta^+, \Delta^-)$ where 
\begin{equation}\label{eq:lattice-std:outlier-robust-errs}
\begin{aligned}
    \hat{y} &= \mathrm{Median}_b[y_{b}] \\
    \Delta^- &= \hat{y} - \mathrm{Percentile}_b[y_{b}, 100 s] \\
    \Delta^+ &= \mathrm{Percentile}_b[y_{b}, 100 (1-s)] - \hat{y}
\end{aligned}\end{equation}
with $b$ indexing bootstrap samples and $s = \int_{-\infty}^{-1} \frac{1}{\sqrt{2 \pi}} e^{-x^2/2} \approx 0.16$
is the CDF of the unit normal distribution evaluated at $-1$.} and the usual Pearson correlation matrix to construct covariance matrices.
We observe that the median over bootstraps appears to have a regulating effect versus simply carrying out the analysis on the central value correlator (i.e.~the mean of $C(t)$ over bootstraps).
However, we note immediately that, across all results presented here, this gives estimates that fluctuate less than their errors and measured correlations would suggest.
This reflects an inconsistency between these definitions for the estimator (here, the median) and its uncertainties (here, the CI construction).
This is resolved by the nested bootstrap procedure introduced in Ref.~\cite{Wagman:2024rid} and employed in the main text, which provides a self-consistent approach to employing median estimators.
Versus this more principled treatment of the median,
the CI construction employed here drastically overestimates uncertainties but underestimates correlations.

\begin{table}
    {
    \newcommand{\p}{\phantom{+}}
    \begin{equation*}
        \braket{f | \bar{s} s | i}_b =
        \begin{bmatrix}
            \p 3.03(24) &   -3.51(84) & \p 2.06(81)  \\
              -2.93(72) & \p 2.7(2.0) &   -1.8(2.0)  \\
            \p 1.34(67) &   -1.2(2.0) & \p 1.1(1.7) 
        \end{bmatrix}_{fi} 
    \end{equation*}
    }
    \caption{
        Results of fits of a constant to Lanczos estimates $J^{(m)}_{fi}$ for all $m \geq 8$ of bare matrix elements $\braket{f|\bar{s}s|i}_b$ of the strange scalar current. $f$ indexes rows and $i$ indexes columns. Values are as shown in corresponding panels of \cref{fig:lattice-std:MEs} and computed as described there.
    }
    \label{tab:lattice-std:MEs}
\end{table}

When employing the more traditional approach to estimator construction and uncertainty quantification used in this appendix, Lanczos estimates at different $m$ carry independent information that can be combined to obtain a more precise estimate.
To demonstrate, for all estimates in this section, we fit a constant to all $m \geq 8$, with the range chosen to exclude the less-noisy points at early $m$.
For simplicity, uncertainties on fitted values are estimated by linear propagation from the data covariance matrix.
The reduction in uncertainty versus the average uncertainty of the data gives a notion of the amount of independent measurements in the $41$ points included in each fit.
For the energies $E_k$, the fitted values are $\approx 9,4,4$ times more precise than the data for $k=0,1,2$, respectively.
Reduction by a factor $\sqrt{41} \approx 6.4$ corresponds to the expected reduction for 41 statistically independent points, but
fluctuations about this value are expected due to noise.
Systematic deviations from this value arising from correlations between data points are not observed.

For the ground-state matrix elements, the fit of the Lanczos estimate is $\approx 7$ times more precise than the estimates with particular $m$ (on average).
The fit of the Lanczos estimate is {$\approx 4$-$5$} times more precise than the data for all excited matrix elements.
Fits to the overlap factors $Z_k$ are $\approx 2, 8, 3$ times more precise than the data for $k=0,1,2$, respectively.

It is important to note that this reduction is a consequence of the already-noted overestimation of uncertainties and underestimation of correlations by the CI construction.
As observed in Refs.~\cite{Wagman:2024rid,Hackett:2024nbe} and the main text, proper treatment of the median estimator with nested bootstrapping reveals precise and highly correlated estimates at late $m$, such that the final output of the analysis can simply be taken as the single estimate at maximal $m$ without further postprocessing.

\bibliography{main}

\end{document}